\documentclass{ws-ijmpa}

\usepackage{ifthen}
\newboolean{uprightparticles}
\setboolean{uprightparticles}{false} 

\usepackage{acronym}
\usepackage{amsmath,amssymb}
\usepackage{centernot}
\usepackage[super,compress]{cite}
\usepackage{graphicx}
\usepackage{hyperref}
\usepackage{siunitx}
\usepackage{subfigure}
\usepackage{tikz,pgfplots}
\usepackage{xspace}
\usepackage{xcolor}

\usepackage{bigints}

\usepackage{upgreek}

\def\lhcb {\mbox{LHCb}\xspace}
\def\atlas  {\mbox{ATLAS}\xspace}
\def\cms    {\mbox{CMS}\xspace}

\def\babar  {\mbox{BaBar}\xspace}
\def\belle  {\mbox{Belle}\xspace}
\def\cleo   {\mbox{CLEO}\xspace}

\def\lhc    {\mbox{LHC}\xspace}

\def\MagUp {\mbox{\em Mag\kern -0.05em Up}\xspace}

\ifthenelse{\boolean{uprightparticles}}%
{
 
 \def\Pgamma      {\ensuremath{\upgamma}\xspace}

 \def\Pmu         {\ensuremath{\upmu}\xspace}                 
 \def\Pnu         {\ensuremath{\upnu}\xspace}                 
                  
 \def\Ppi         {\ensuremath{\uppi}\xspace}

 \def\Ptau        {\ensuremath{\uptau}\xspace}                 
                  
 \def\Pphi        {\ensuremath{\upphi}\xspace}

 \def\Ppsi        {\ensuremath{\uppsi}\xspace}

 \def\PDelta      {\ensuremath{\Delta}\xspace}                 
 \def\PXi      {\ensuremath{\Xi}\xspace}                 
 \def\PLambda      {\ensuremath{\Lambda}\xspace}                 
 \def\PSigma      {\ensuremath{\Sigma}\xspace}                 
 \def\POmega      {\ensuremath{\Omega}\xspace}                 
 \def\PUpsilon      {\ensuremath{\Upsilon}\xspace}                 
 
 \def\PB      {\ensuremath{\mathrm{B}}\xspace}                 
                  
 \def\PD      {\ensuremath{\mathrm{D}}\xspace}

 \def\PJ      {\ensuremath{\mathrm{J}}\xspace}                 
 \def\PK      {\ensuremath{\mathrm{K}}\xspace}

 \def\Pb      {\ensuremath{\mathrm{b}}\xspace}                 
 \def\Pc      {\ensuremath{\mathrm{c}}\xspace}                 
 \def\Pd      {\ensuremath{\mathrm{d}}\xspace}                 
 \def\Pe      {\ensuremath{\mathrm{e}}\xspace}

 \def\Pi      {\ensuremath{\mathrm{i}}\xspace}

 \def\Pp      {\ensuremath{\mathrm{p}}\xspace}

 \def\Ps      {\ensuremath{\mathrm{s}}\xspace}                 
 \def\Pt      {\ensuremath{\mathrm{t}}\xspace}

}
{
 
 \def\Pgamma      {\ensuremath{\gamma}\xspace}

 \def\Pmu         {\ensuremath{\mu}\xspace}                 
 \def\Pnu         {\ensuremath{\nu}\xspace}                 
                  
 \def\Ppi         {\ensuremath{\pi}\xspace}

 \def\Ptau        {\ensuremath{\tau}\xspace}                 
                  
 \def\Pphi        {\ensuremath{\phi}\xspace}

 \def\Ppsi        {\ensuremath{\psi}\xspace}                 
                  
 \mathchardef\PDelta="7101
 \mathchardef\PXi="7104
 \mathchardef\PLambda="7103
 \mathchardef\PSigma="7106
 \mathchardef\POmega="710A
 \mathchardef\PUpsilon="7107
                  
 \def\PB      {\ensuremath{B}\xspace}                 
                  
 \def\PD      {\ensuremath{D}\xspace}

 \def\PJ      {\ensuremath{J}\xspace}                 
 \def\PK      {\ensuremath{K}\xspace}

 \def\Pb      {\ensuremath{b}\xspace}                 
 \def\Pc      {\ensuremath{c}\xspace}                 
 \def\Pd      {\ensuremath{d}\xspace}                 
 \def\Pe      {\ensuremath{e}\xspace}

 \def\Pi      {\ensuremath{i}\xspace}

 \def\Pp      {\ensuremath{p}\xspace}

 \def\Ps      {\ensuremath{s}\xspace}                 
 \def\Pt      {\ensuremath{t}\xspace}

}

\makeatother

\DeclareRobustCommand{\optbar}[1]{\shortstack{{\miniscule (\rule[.5ex]{1.25em}{.18mm})}
  \\ [-.7ex] $#1$}}

\def\en         {{\ensuremath{\Pe^-}}\xspace}   
\def\ep         {{\ensuremath{\Pe^+}}\xspace}

\def\epem       {{\ensuremath{\Pe^+\Pe^-}}\xspace}

\def\mup        {{\ensuremath{\Pmu^+}}\xspace}
\def\mun        {{\ensuremath{\Pmu^-}}\xspace} 

\def\mumu       {{\ensuremath{\Pmu^+\Pmu^-}}\xspace}

\def\taup       {{\ensuremath{\Ptau^+}}\xspace}
\def\taum       {{\ensuremath{\Ptau^-}}\xspace}

\def\ellm       {{\ensuremath{\ell^-}}\xspace}
\def\ellp       {{\ensuremath{\ell^+}}\xspace}
\def\ellell     {\ensuremath{\ell^+ \ell^-}\xspace}

\def\neu        {{\ensuremath{\Pnu}}\xspace}

\def\g      {{\ensuremath{\Pgamma}}\xspace}

\def\dquark    {{\ensuremath{\Pd}}\xspace}

\def\squark    {{\ensuremath{\Ps}}\xspace}

\def\cquark    {{\ensuremath{\Pc}}\xspace}

\def\bquark    {{\ensuremath{\Pb}}\xspace}
\def\bquarkbar {{\ensuremath{\overline \bquark}}\xspace}
\def\bbbar     {{\ensuremath{\bquark\bquarkbar}}\xspace}
\def\tquark    {{\ensuremath{\Pt}}\xspace}

\def\pion   {{\ensuremath{\Ppi}}\xspace}

\def\pip    {{\ensuremath{\pion^+}}\xspace}
\def\pim    {{\ensuremath{\pion^-}}\xspace}

\def\kaon    {{\ensuremath{\PK}}\xspace}
  \def\Kbar    {{\kern 0.2em\overline{\kern -0.2em \PK}{}}\xspace}

\def\KorKbar    {\kern 0.18em\optbar{\kern -0.18em K}{}\xspace}
\def\Kz      {{\ensuremath{\kaon^0}}\xspace}

\def\Kp      {{\ensuremath{\kaon^+}}\xspace}
\def\Km      {{\ensuremath{\kaon^-}}\xspace}

\def\KS      {{\ensuremath{\kaon^0_{\mathrm{ \scriptscriptstyle S}}}}\xspace}

\def\Kstarz  {{\ensuremath{\kaon^{*0}}}\xspace}
\def\Kstarzb {{\ensuremath{\Kbar{}^{*0}}}\xspace}
\def\Kstar   {{\ensuremath{\kaon^*}}\xspace}

\def\Kstarp  {{\ensuremath{\kaon^{*+}}}\xspace}

\newcommand{\phiz}{\ensuremath{\Pphi}\xspace}

  \def\Dbar    {{\kern 0.2em\overline{\kern -0.2em \PD}{}}\xspace}

\def\DorDbar    {\kern 0.18em\optbar{\kern -0.18em D}{}\xspace}

\def\B       {{\ensuremath{\PB}}\xspace}
\def\Bbar    {{\ensuremath{\kern 0.18em\overline{\kern -0.18em \PB}{}}}\xspace}
\def\Bb      {{\ensuremath{\Bbar}}\xspace}
\def\BorBbar    {\kern 0.18em\optbar{\kern -0.18em B}{}\xspace}
\def\Bz      {{\ensuremath{\B^0}}\xspace}
\def\Bzb     {{\ensuremath{\Bbar{}^0}}\xspace}
\def\Bu      {{\ensuremath{\B^+}}\xspace}
\def\Bub     {{\ensuremath{\B^-}}\xspace}
\def\Bp      {{\ensuremath{\Bu}}\xspace}
\def\Bm      {{\ensuremath{\Bub}}\xspace}

\def\Bd      {{\ensuremath{\B^0}}\xspace}
\def\Bs      {{\ensuremath{\B^0_\squark}}\xspace}
\def\Bsb     {{\ensuremath{\Bbar{}^0_\squark}}\xspace}

\def\jpsi     {{\ensuremath{{\PJ\mskip -3mu/\mskip -2mu\Ppsi\mskip 2mu}}}\xspace}
\def\psitwos  {{\ensuremath{\Ppsi{(2S)}}}\xspace}

  \def\Y#1S{\ensuremath{\PUpsilon{(#1S)}}\xspace}

\def\FiveS {{\Y5S}}

\def\proton      {{\ensuremath{\Pp}}\xspace}

\def\Lz          {{\ensuremath{\PLambda}}\xspace}
\def\Lbar        {{\ensuremath{\kern 0.1em\overline{\kern -0.1em\PLambda}}}\xspace}
\def\LorLbar    {\kern 0.18em\optbar{\kern -0.18em \PLambda}{}\xspace}

\def\Lb      {{\ensuremath{\Lz^0_\bquark}}\xspace}
\def\Lbbar   {{\ensuremath{\Lbar{}^0_\bquark}}\xspace}

\def\BF         {{\ensuremath{\mathcal{B}}}\xspace}

\def\BR         {\BF}
\newcommand{\decay}[2]{\ensuremath{#1\!\to #2}\xspace}         

\def\to                 {\ensuremath{\rightarrow}\xspace}

\def\order   {{\ensuremath{\mathcal{O}}}\xspace}

\def\qsq       {{\ensuremath{q^2}}\xspace}

\def\CP                {{\ensuremath{C\!P}}\xspace}

\newcommand{\ACP}{{\ensuremath{{\mathcal{A}}^{\CP}}}\xspace}

\def\AT#1     {\ensuremath{A_{\mathrm{T}}^{#1}}\xspace}           

\def\Bsmm     {\decay{\Bs}{\mup\mun}}

\def\C#1      {\ensuremath{\mathcal{C}_{#1}}\xspace}                       
\def\Cp#1     {\ensuremath{\mathcal{C}_{#1}^{'}}\xspace}                    
\def\Ceff#1   {\ensuremath{\mathcal{C}_{#1}^{\mathrm{(eff)}}}\xspace}        
\def\Cpeff#1  {\ensuremath{\mathcal{C}_{#1}^{'\mathrm{(eff)}}}\xspace}       
\def\Ope#1    {\ensuremath{\mathcal{O}_{#1}}\xspace}                       
\def\Opep#1   {\ensuremath{\mathcal{O}_{#1}^{'}}\xspace}                    

\newcommand{\bra}[1]{\ensuremath{\langle #1|}}             
\newcommand{\ket}[1]{\ensuremath{|#1\rangle}}              

\newcommand{\tev}{\ensuremath{\mathrm{\,Te\kern -0.1em V}}\xspace}
\newcommand{\gev}{\ensuremath{\mathrm{\,Ge\kern -0.1em V}}\xspace}
\newcommand{\mev}{\ensuremath{\mathrm{\,Me\kern -0.1em V}}\xspace}
\newcommand{\kev}{\ensuremath{\mathrm{\,ke\kern -0.1em V}}\xspace}
\newcommand{\ev}{\ensuremath{\mathrm{\,e\kern -0.1em V}}\xspace}
\newcommand{\gevc}{\ensuremath{{\mathrm{\,Ge\kern -0.1em V\!/}c}}\xspace}
\newcommand{\mevc}{\ensuremath{{\mathrm{\,Me\kern -0.1em V\!/}c}}\xspace}
\newcommand{\gevcc}{\ensuremath{{\mathrm{\,Ge\kern -0.1em V\!/}c^2}}\xspace}
\newcommand{\gevgevcccc}{\ensuremath{{\mathrm{\,Ge\kern -0.1em V^2\!/}c^4}}\xspace}
\newcommand{\mevcc}{\ensuremath{{\mathrm{\,Me\kern -0.1em V\!/}c^2}}\xspace}

\def\invfb   {\ensuremath{\mbox{\,fb}^{-1}}\xspace}

\def\invab   {\ensuremath{\mbox{\,ab}^{-1}}\xspace}

\def\ps   {\ensuremath{{\mathrm{ \,ps}}}\xspace}

\def\order{{\ensuremath{\mathcal{O}}}\xspace}
\newcommand{\chisq}{\ensuremath{\chi^2}\xspace}

\def\gsim{{~\raise.15em\hbox{$>$}\kern-.85em
          \lower.35em\hbox{$\sim$}~}\xspace}
\def\lsim{{~\raise.15em\hbox{$<$}\kern-.85em
          \lower.35em\hbox{$\sim$}~}\xspace}

\newcommand{\Real}{\ensuremath{\mathcal{R}e}\xspace}
\newcommand{\Imag}{\ensuremath{\mathcal{I}m}\xspace}

\def\tell1  {TELL1\xspace}
\def\ukl1   {UKL1\xspace}

\newcommand{\eg}{\mbox{\itshape e.g.}\xspace}

\newcommand{\refeq}[1]{Eq.~(\ref{eq:#1})}

\newcommand{\reffig}[1]{Fig.~\ref{fig:#1}}

\newcommand{\refsec}[1]{Section~\ref{sec:#1}}

\makeatletter
\def\ja{\@ifstar\@@ja\@ja}
\newcommand{\@ja}[1]{\textcolor{green}{[\textbf{JA:} #1]}}
\newcommand{\@@ja}[1]{\textcolor{green}{#1}}
\def\sr{\@ifstar\@@sr\@sr}
\newcommand{\@sr}[1]{\textcolor{blue}{[\textbf{SR:} #1]}}
\newcommand{\@@sr}[1]{\textcolor{blue}{#1}}
\def\dvd{\@ifstar\@@dvd\@dvd}
\newcommand{\@dvd}[1]{\textcolor{purple}{[\textbf{DvD:} #1]}}
\newcommand{\@@dvd}[1]{\textcolor{purple}{#1}}
\makeatother

\newcommand{\dd}[1][]{\mathrm{d}^{#1}}

\newcommand{\GeV}{\si{\giga\electronvolt}}
\newcommand{\MeV}{\si{\mega\electronvolt}}

\acrodef{NP}{New Physics}\newcommand{\NP}{\ac{NP}\xspace}
\acrodef{CKM}{Cabibbo-Kobayashi-Maskawa}\newcommand{\CKM}{\ac{CKM}\xspace}
\acrodef{EFT}{Effective Field Theory}\newcommand{\EFT}{\ac{EFT}\xspace}
\acrodef{OPE}{Operator Product Expansion}\newcommand{\OPE}{\ac{OPE}\xspace}
\acrodef{SM}{Standard Model}\newcommand{\SM}{\ac{SM}\xspace}
\acrodef{WC}{Wilson coefficients}\newcommand{\WC}{\ac{WC}\xspace}

\newcommand{\GF}{\ensuremath{G_\text{F}}\xspace}
\newcommand{\Ci}[1]{\ensuremath{ {C_{#1}} } \xspace}
\newcommand{\Oi}[1]{\ensuremath{ {O_{#1}} } \xspace}
\newcommand{\Cnine}{\ensuremath{   {C^{\rm NP}_9}^{\mu\mu}  }   \xspace}
\newcommand{\Cten}{\ensuremath{ {C^{\rm NP}_{10}}^{\mu\mu}   }   \xspace}

\newcommand{\Cninep}{\ensuremath{ {C^{\prime}_{9}}^{\mu\mu}  }   \xspace}
\newcommand{\Ctenp}{\ensuremath{  {C^{\prime}_{10}}^{\mu\mu}   }   \xspace}
\newcommand{\ReCsev}{\ensuremath{ \Real \left ( C^{\rm NP }_7\right )}\xspace}
\newcommand{\ImCsev}{\ensuremath{ \Imag \left ( C^{\rm NP }_7 \right )}\xspace}

\newcommand{\Rk}{\ensuremath{R(\kaon)}\xspace}
\newcommand{\Rkst}{\ensuremath{R(\kaon^*)}\xspace}

\def \btosll     {\decay{\bquark}{\squark\ellell}}
\def \btosmumu   {\decay{\bquark}{\squark\mumu}}
\def \btosee     {\decay{\bquark}{\squark\ep\en}}
\def \btodll     {\decay{\bquark}{\dquark\ellell}}
\def \btoll     {\decay{\B^0_{(\squark)}}{\ellell}}
\def \btosgamma     {\decay{\bquark}{\squark\g}}

\def \Lbpmunu    {\decay{\Lb}{\proton \mun \neu}}
\def \Lblmumu    {\decay{\Lb}{\Lz \mup \mun}}
\def \Lblll    {\decay{\Lb}{\Lz \ellell}}
\def \Lblsll    {\decay{\Lb}{\Lz^{(*)} \ellell}}
\def \Lblsmumu    {\decay{\Lb}{\Lz^{(*)} \ellell}}

\def \Bsmumu     {\decay{\Bs}{\mup\mun}}
\def \Bsbmumu     {\decay{\Bsb}{\mup\mun}}
\def \Bmumu     {\decay{\B^0_{(\squark)}}{\mup\mun}}
\def \Bsll     {\decay{\Bs}{\ell^+ \ell^-}}
\def \Bstautau     {\decay{\Bs}{\taup\taum}}
\def \Bsee     {\decay{\Bs}{\ep\en}}

\def \Bdmumu     {\decay{\Bd}{\mup\mun}}
\def \Bdtautau     {\decay{\Bd}{\taup\taum}}
\def \Bdee     {\decay{\Bd}{\ep\en}}

\def \Bsdtoee     {\decay{\B_{(\squark)}}{\ep\en}}
\def \Bsdtotautau     {\decay{\B_{(\squark)}}{\taup\taum}}
\def \Bstophigamma     {\decay{\Bs}{\phi\g}}
\def \Bztophigamma     {\decay{\Bz}{\phi\g}}

\def \BtoKstmumu      {\decay{\Bz}{\Kstarz\mup\mun}}
\def \BtoKstJpsi      {\decay{\Bz}{\Kstarz\jpsi}}
\def \BtoKstpsip      {\decay{\Bz}{\Kstarz\psi(2S)}}
\def \BToKpipimumu {\decay{\Bp}{\Kp\pip\pim\mup\mun}}
\def \BTophiKmumu {\decay{\Bp}{\phi\Kp\mup\mun}}
\def \BToKpmumu {\decay{\Bp}{\Kp\mup\mun}}
\def \BToKzmumu {\decay{\Bz}{\Kz\mup\mun}}
\def \BToKsmumu {\decay{\Bz}{\KS\mup\mun}}
\def \BToKstpmumu {\decay{\Bp}{\Kstarp\mup\mun}}
\def \BToKstzmumu {\decay{\Bz}{\Kstarz\mup\mun}}
\def \BbToKstzbmumu {\decay{\Bzb}{\Kstarzb\mup\mun}}
\def \BToKstzee {\decay{\Bz}{\Kstarz\ep\en}}
\def \BToKstzll {\decay{\Bz}{\Kstarz\ell^+\ell^-}}
\def \BToKstll {\decay{\Bp}{\Kstarp\ell^+\ell^-}}
\def \BtoKll {\decay{\Bp}{\kaon^+\ell^+\ell^-}}
\def \BToJpsiKp {\decay{\Bp}{\Kp\jpsi}}
\def \BtoKstee      {\decay{\Bz}{\Kstarz\ep\en}}
\def \BtoKmumu      {\decay{\Bp}{\Kp\mup\mun}}
\def \BtoKee      {\decay{\Bp}{\Kp\ep\en}}
\def \Bstophimumu      {\decay{\Bs}{\phi\mup\mun}}

\def \BtoKstpgamma	{\decay{\Bp}{\Kstarp\g}}
\def \BtoKstzgamma	{\decay{\Bz}{\Kstarz\g}}
\def \BtoKstgamma	{\decay{\B}{\Kstar\g}}

\def \BXsll      {\decay{\B}{X_s\ellell}}
\def \bbbar	     {\bquark\bquarkbar}

\def \Butopimumu      {\decay{\Bp}{\pip\mup\mun}}
\def \BstoKstmumu      {\decay{\Bs}{\Kstarzb\mup\mun}}

\newcommand{\Pone}{\ensuremath{P_1} \xspace}
\newcommand{\Ptwo}{\ensuremath{P_2} \xspace}
\newcommand{\Pfourp}{\ensuremath{P^{\prime}_4} \xspace}
\newcommand{\Pfivep}{\ensuremath{P^{\prime}_5} \xspace}

\begin{document} 
\markboth{Johannes Albrecht, Stefanie Reichert, Danny van Dyk}{Status of rare exclusive \B hadron decays in 2018}

%
\catchline{}{}{}{}{}
%

\title{Status of rare exclusive \B meson decays in 2018}

\author{Johannes Albrecht}

\address{Fakult\"at Physik, Technische Universit\"at Dortmund, Otto-Hahn-Stra\ss{}e 4a, 44227 Dortmund, Germany\\
johannes.albrecht@tu-dortmund.de}

\author{Stefanie Reichert}

\address{Fakult\"at Physik, Technische Universit\"at Dortmund, Otto-Hahn-Stra\ss{}e 4a, 44227 Dortmund, Germany\\
stefanie.reichert@cern.ch}

\author{Danny van Dyk\footnote{TUM-HEP-1146/18}}

\address{Physik Department, Technische Universit\"at M\"unchen, James-Franck-Stra\ss{}e 1, 85748 Garching, Germany\\
danny.van.dyk@gmail.com}

\maketitle

\begin{history}
\received{Day Month Year}
\revised{Day Month Year}
\end{history}

\begin{abstract}
This review discusses the present experimental and theoretical status of rare
flavour-changing neutral current \bquark-quark decays at the beginning of 2018. It
includes a discussion of the experimental situation and details of the
currently observed anomalies in measurements of flavour observables, including
lepton flavour universality. Progress on the theory side, within and beyond the
Standard Model theory is also discussed, together with potential New Physics
interpretations of the present measurements.
\end{abstract}

\ccode{PACS numbers: 13.20.He 13.30.Ce 12.15.Mm 12.60.Jv }

\tableofcontents

\section{Introduction}
\label{sec:intro}

The year 2018 marks a special point for the field of particle physics as it is not only the end of second period of data-taking at the Large Hadron Collider(LHC) but also the
start-up of the \belle II experiment. In the field of flavour physics, a number of
anomalies have been observed when comparing measurements to the
Standard Model (\SM) expectation; from which a coherent and conclusive
picture emerges. Those anomalies could, if they persist,  point
towards certain \NP models such as leptoquarks or $Z^{\prime}$
bosons. In this review, we will give a comprehensive overview of the
current state-of-the-art of exclusive rare $b$ decays from both experimental and theoretical perspective. We will present the current experimental landscape with a focus of the observed anomalies and discuss potential new physics interpretations.  

The \lhcb experiment will continue to run for the next two decades and
hence will be able to either confirm or rule out many of the
present-day anomalies. With
the imminent start-up of the data-taking at the \belle II experiment,
these tensions will be independently cross-checked and further
complementary measurements will be accessible.

As the progress in inclusive \btosll decays was limited after the end
of the \B factories, this review focusses on exclusive measurements of
rare decays of \B hadrons as these are excellent probes for many new
physics scenarios. These rare decay processes are CKM-, GIM- and
loop-suppressed, wherefore potential new physics effects can be large
compared to the \SM amplitude. Such indirect searches allow to probe \NP models at much higher mass scales as are currently 
accessible through direct searches. In the past, exclusive decays with
muons in the final state have been measured 
extensively by the \lhc experiments, most notably by \lhcb, as well as
the \B factories. We review the status of these measurements, discuss
their theoretical 
description in the \SM and model-independently in the presence of new physics. A special focus is set on the anomalies
seen in angular distributions of \BtoKstmumu and in the branching
fraction measurements of all exclusive \btosll decays as well as to the
tensions observed in tests of lepton flavour universality in this class of
decays. Measurements of the type $\decay{\bquark}{\cquark \ell \nu}$
are not covered in this review. 

On the theoretical side, the current state of the \SM predictions 
for the discussed observables is presented. The anatomy of the amplitudes, and their dependence on various local and non-local hadronic matrix elements is discussed. Particular focus is set on recent developments for the non-local matrix elements. The model-independent
interpretation of the \btosll measurements is examined in the
framework of the usual Effective Field Theory.

This review is structured as follows:~\refsec{amplitudes}
introduces the amplitudes of \btosll decays in an effective field theory approach.~\refsec{exp} discusses the experimental landscape with a focus on 
semileptonic \btosll decays, which is followed by a brief discussion of \btodll decays, purely
leptonic \btoll and \btosgamma decays. The experimental part of the review
closes with a brief discussion of \btosgamma decays. 

The second part of the review,~\refsec{th-had}, examines
the foundation for the \SM predictions required in the first part of the review.~\refsec{ph-np} discusses then a
combined interpretation of all presented measurements in the framework
of global fits and possible interpretations of the observed
patterns in \NP frameworks. 

The review closes with a discussion of the experimental outlook in~\refsec{exp-outlook}.    


\section{Anatomy of the amplitudes}
\label{sec:amplitudes}

Within the \SM and at the Born level there are no flavor-changing neutral currents
(FCNC).  As a consequence, any FCNC-mediated quark-flavor transitions, such as
$b\to s$, emerge only from virtual loop corrections, in which a $W$ is first
emitted and subsequently re-absorbed.  The emergence at the loop level
naturally suppresses the rate at which these processes occur.  In the \SM, the
unitarity of the \CKM matrix introduces an additional suppression mechanism that renders \btosll
transitions \emph{very rare}~\cite{Buchalla:1995vs}. Moreover, descriptions of these rare
\bquark decays in the \SM are further complicated since they pose a multi-scale problem:
as weak decays they are mediated through the exchange of $W$ and $Z$ bosons, which are
much heavier than the remaining particles (with the exception of the \tquark quark).
The use of an \EFT{} helps our understanding of such
multi-scale dynamics. In the following, we will discuss the \EFT{} used for predictions of rare $b$ decays
below the electroweak breaking scale $\Lambda_\text{EW} \simeq
80\,\GeV$. Within the \SM, this \EFT{} captures the effects of the $t$ quark
as well as the $W$ and $Z$ bosons, which are no-longer dynamical degrees of freedom
for scales $\mu < \Lambda_\text{EW}$. 

The effective Lagrangian reads (see \eg~\cite{Buchalla:1995vs})
\begin{equation}
    \label{eq:Leff}
    \mathcal{L}_{\btosll}
        = \mathcal{L}_\text{QED} + \mathcal{L}_\text{QCD,5}
        + \frac{4 \GF}{\sqrt{2}} V_{tb} V_{ts}^* \left[\sum_i \Ci{i}(\mu) \Oi{i}\right]
        + \order\left(V_{ub} V_{us}^*\right)\,,
\end{equation}
where $\mathcal{L}_\text{QED}$ and $\mathcal{L}_\text{QCD,5}$ denote the
Lagrangians of the electromagnetic and strong interactions (with five quark flavours after integrating out the top quark), \GF{} refers to the
Fermi constant, and the last term captures the local effective operators \Oi{i}
with their effective couplings - or \WC - \Ci{i} at the
renormalisation scale $\mu$.  For convenience, the product of \CKM matrix
elements $V_{tb} V_{ts}^*$ has been extracted from the definition of the \WC.

The matrix elements of the effective operators $\Oi{i}$ need to be evaluated
at a low scale $\mu_b = m_b \simeq 4.2\,\GeV$, which minimises logarithms in the perturbative expansion of the matrix elements. Consequently, one requires
the \WC evaluated at $\mu_b$. Problems arising from large logarithms
$\ln(M_W / \mu_b)$ are resummed through Renormalisation Group (RG) improved
running. Contemporary analyses use resummation of QCD-induced large logarithms
up to Next-to-Next-to-Leading-Logarithm (NNLL).
This requires knowledge of the relevant anomalous dimensions at the four-loop level, and
the matching conditions at the three-loop level~\cite{Bobeth:1999mk,Misiak:2004ew,Gorbahn:2004my,Gorbahn:2005sa,Czakon:2006ss}.
The electroweak effects to NLL, which are numerically sub-leading, are known only for a subset of the \WC~\cite{Bobeth:2013tba}.

For a consistent treatment at the leading order in the electromagnetic coupling
$e$, all operators of the following set - usually called the \SM
basis - are required (see \eg~\cite{Bobeth:1999mk}). The \SM basis consists of the current-current operators ($q=u,c$)
\begin{align}
    \Oi{1q} & = \left[\bar{s} \gamma^\mu T^A P_L q\right]\, \left[\bar{q} \gamma_\mu T^A P_L b\right]\,,  &
    \Oi{2q} & = \left[\bar{s} \gamma^\mu     P_L q\right]\, \left[\bar{q} \gamma_\mu     P_L b\right]\,;
\end{align}
the QCD-penguin operators
\begin{equation}
\begin{aligned}
    \Oi{3}  & = \left[\bar{s} \gamma^\mu     P_L b\right]\, \sum_q \left[\bar{q} \gamma_\mu     q\right]\,, &
    \Oi{4}  & = \left[\bar{s} \gamma^\mu T^A P_L b\right]\, \sum_q \left[\bar{q} \gamma_\mu T^A q\right]\,, \\
    \Oi{5}  & = \left[\bar{s} \gamma^{\mu\nu\rho}     P_L b\right]\, \sum_q \left[\bar{q} \gamma_{\mu\nu\rho}     q\right]\,, &
    \Oi{6}  & = \left[\bar{s} \gamma^{\mu\nu\rho} T^A P_L b\right]\, \sum_q \left[\bar{q} \gamma_{\mu\nu\rho} T^A q\right]\,;
\end{aligned}
\end{equation}
the electromagnetic and chromomagnetic dipole operators
\begin{align}
    \Oi{7}  & = \frac{e}{16 \pi^2} \overline{m}_b \left[\bar{s} \sigma^{\mu\nu}     P_R b\right]\, F_{\mu\nu}\,,   &
    \Oi{8}  & = \frac{g_s}{16\pi^2}\overline{m}_b \left[\bar{s} \sigma^{\mu\nu} T^A P_R b\right]\, G_{\mu\nu}^A\,;
\end{align}
and the semileptonic operators
\begin{align}
    \Oi{ 9} & = \frac{e^2}{16\pi^2} \left[\bar{s} \gamma^\mu P_L b\right]\,\left[\bar{\ell} \gamma_\mu          \ell\right]\,, &
    \Oi{10} & = \frac{e^2}{16\pi^2} \left[\bar{s} \gamma^\mu P_L b\right]\,\left[\bar{\ell} \gamma_\mu \gamma_5 \ell\right]\,.
\end{align}
In the above, we abbreviate $\gamma^{\mu\nu\rho}\equiv
\gamma^\mu\gamma^\nu\gamma^\rho$, $P_{L(R)} \equiv (1 \mp \gamma_5)/2$, $e^2 = 4\pi \alpha_e$, and
sums over $q$ run over all active quark flavours $u$, $d$, $s$, $c$, and $b$.
Through using $\alpha_e = \alpha_e(\mu_b)$, universal QED corrections at NLO are
taken care of~\cite{Bobeth:2003at,Huber:2005ig}.

Searches for \NP effects at energies smaller than $\mu \approx M_W$ have so far not
discovered either new interactions or new particles.  Assuming that no such
low-mass fields exist, the \EFT{} framework, which is necessary for accurate
theory predictions, can also be used to systematically describe \NP effects.
To this end, the effective Lagrangian~\refeq{Leff} has to be modified in the following
way~\cite{Ali:1994bf}:
\begin{enumerate}[]
    \item the set of effective field operators is enlarged to include all
        operators allowed by field content and Lorentz symmetry up to a given
        mass dimension (typically mass dimension six),
    \item their \WC are assumed to be independent and uncorrelated parameters.
\end{enumerate}
When limiting this enlarged set to only semileptonic operators, a basis of all operators up to
and including mass dimension six can be formed by also including the chirality-flipped
operators,
\begin{align}
    \Oi{9'}  & = \frac{e^2}{16\pi^2} \left[\bar{s} \gamma^\mu P_R b\right]\,\left[\bar{\ell} \gamma_\mu          \ell\right]\,, &
    \Oi{10'} & = \frac{e^2}{16\pi^2} \left[\bar{s} \gamma^\mu P_R b\right]\,\left[\bar{\ell} \gamma_\mu \gamma_5 \ell\right]\,;
\end{align}
the (pseudo)scalar operators
\begin{equation}
\begin{aligned}
    \Oi{ S} & = \frac{e^2}{16\pi^2} \left[\bar{s} \gamma^\mu P_R b\right]\,\left[\bar{\ell}          \ell\right]\,, &
    \Oi{ P} & = \frac{e^2}{16\pi^2} \left[\bar{s} \gamma^\mu P_R b\right]\,\left[\bar{\ell} \gamma_5 \ell\right]\,, \\
    \Oi{S'} & = \frac{e^2}{16\pi^2} \left[\bar{s} \gamma^\mu P_L b\right]\,\left[\bar{\ell}          \ell\right]\,, &
    \Oi{P'} & = \frac{e^2}{16\pi^2} \left[\bar{s} \gamma^\mu P_L b\right]\,\left[\bar{\ell} \gamma_5 \ell\right]\,;
\end{aligned}
\end{equation}
and the tensor operators
\begin{align}
    \Oi{ T} & = \frac{e^2}{16\pi^2} \left[\bar{s} \sigma^{\mu\nu} b\right]\,\left[\bar{\ell} \sigma_{\mu\nu}          \ell\right]\,, &
    \Oi{T5} & = \frac{e^2}{16\pi^2} \left[\bar{s} \sigma^{\mu\nu} b\right]\,\left[\bar{\ell} \sigma_{\mu\nu} \gamma_5 \ell\right]\,.
\end{align}
A complete and non-redundant set of dimension-six operators, including their one-loop anomalous dimensions
in both QCD and QED is compiled in~\cite{Aebischer:2017gaw}.\\

This \emph{bottom up} approach probes model-independently for deviations from
the \SM:~\cite{%
Beaujean:2013soa,Beaujean:2015gba,Descotes-Genon:2015uva,Hurth:2016fbr%
}the WCs are sensitive to \NP effects of new particles
that are too heavy to be produced directly. Rare semileptonic $b$ decays therefore
provide information that is complementary to the ``direct'' searches for new
interactions and particles that are carried out at the Large Hadron Collider.\\

The above basis is further reduced when one assumes a manifest invariance of \NP effects
under the \SM gauge group within a \SM-like (i.e. linear $\sigma$) Higgs model.
For the \btosll operators, this was explicitly
demonstrated in~\cite{Alonso:2014csa}.\footnote{
Note, however, that a nonlinear representation
fully restores the basis of dimension six operators to the set introduced above~\cite{Cata:2015lta}.
}
The effective Lagrangian in~\refeq{Leff} can be matched onto the \SM Effective Field Theory (SMEFT)
(see \eg~\cite{Buchmuller:1985jz}),
whose operators are manifestly invariant under the \SM gauge group. The required matching
formulas are compiled in~\cite{Aebischer:2015fzz} to leading non-trivial loop level.
Matching, basis transformations and Renormalization-Group-Equation (RGE) evolution can be conveniently
achieved through a variety of computational tools, including but not limited to:
the ``DsixTools'' Mathematica package~\cite{Celis:2017hod};
the ``Wilson Coefficient Exchange Format'' and its reference Python implementation~\cite{Aebischer:2017ugx};
and the ``Wilson'' Python package~\cite{Aebischer:2018bkb}.\\

Schematically, the matrix elements for all exclusive \btosll processes can now
be expressed as
\begin{equation}
    \mathcal{A} \sim C_{10} \mathcal{F}_{10} + \left[C_9 \mathcal{F}_9 - \frac{2m_b M_B}{q^2} C_7 \mathcal{F}_{7}\right] - \frac{32 \pi^2 M_B^2}{q^2} \mathcal{H}
        + \order\left(\alpha_e\right)\,,
\end{equation}

where \qsq refers to the invariant mass squared of the di-lepton pair. The $\mathcal{F}_i$ refer to hadronic matrix elements of local $\bar{s} b$
currents as induced by the operators \Oi{7,9,10}, while $\mathcal{H}$ denotes
matrix elements of time-ordered products involving four-quark operators
\Oi{1c,2c,3,\dots,6} and the chromomagnetic operator \Oi{8} together with the
electromagnetic current.  In the presence of \NP effects in the semileptonic and radiative operators,
the hadronic matrix elements remain unchanged. However, the coefficients multiplying the latter are then
modified. For the complete anatomy of the amplitudes of $B\to K^{(*)}\ell^+\ell^-$ decays in the presence
of \NP operators of mass dimension six we refer to~\cite{Bobeth:2012vn,Gratrex:2015hna}. A similar study for \Lblll is presented in~\cite{Das:2018iap}.

\section{Experimental measurements}
\label{sec:exp}

In this section, we draw a picture of the current experimental status of semi-leptonic \btosll (including a brief discussion of \btodll) decays, purely leptonic \btoll and radiative \btosgamma transitions. Hereby, we will discuss measurements of the various hadronic final states, \eg kaon and \Kstar states, giving access to a broad range of observables such as branching fractions, \CP and isospin asymmetries as well as angular observables. Within this section, the \lhcb measurements refer to the full  Run 1 dataset corresponding to 3\invfb collected during the years 2011 and 2012, the \babar results were obtained on their full dataset of 424\invfb and the \belle publications exploit their complete dataset of 711\invfb if not otherwise stated. Presented results integrated over the whole \qsq range have been obtained by vetoing the charmonium resonances and interpolating over this vetoed \qsq region. It should be noted that the exact range to veto the charmonium resonances depends on the respective analysis. 

\subsection{Semileptonic \btosll decays}
\label{sec:btosll}

Particular interest was raised throughout the past years by various
measurements of semileptonic \btosll decays, in which several
tensions between the \SM predictions and experimental measurements
have been observed. These deviations are mostly in the range of two to
three standard deviations and show a consistent pattern, which can be explained
by lowering the \WC
$C_9$ with respect to its \SM value~\cite{Descotes-Genon:2013wba,Descotes-Genon:2015uva,Altmannshofer:2017fio} as detailed
further in~\refsec{global}. 

\subsubsection{$B \to \kaon \ellell$ decays}

\begin{figure}[h]
\begin{minipage}[t]{0.49\linewidth}
\centering
\includegraphics[trim={8cm 6cm 0 0}, clip, width=1.1\linewidth]{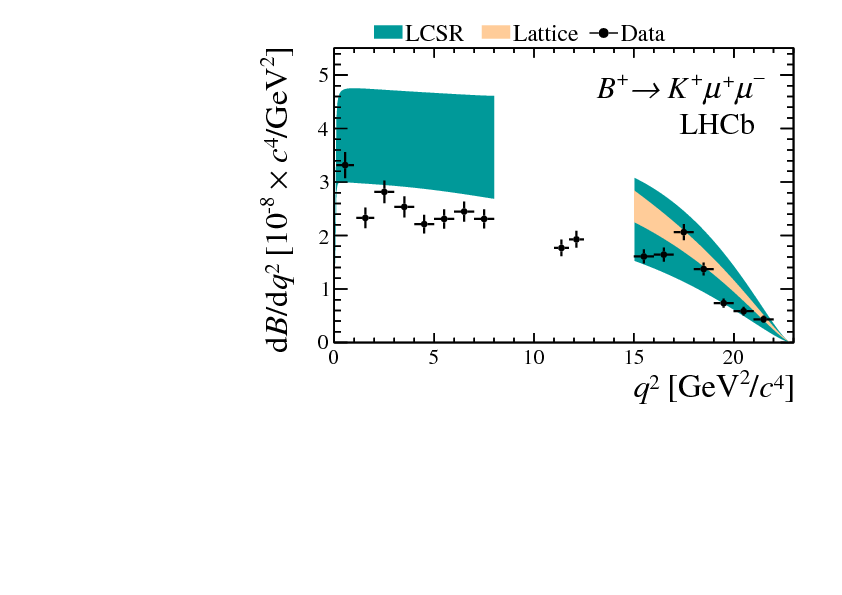}
\end{minipage}
\hspace{\fill}
\begin{minipage}[t]{0.49\linewidth}
\centering
\includegraphics[trim={8cm 6cm 0 0}, clip, width=1.1\linewidth]{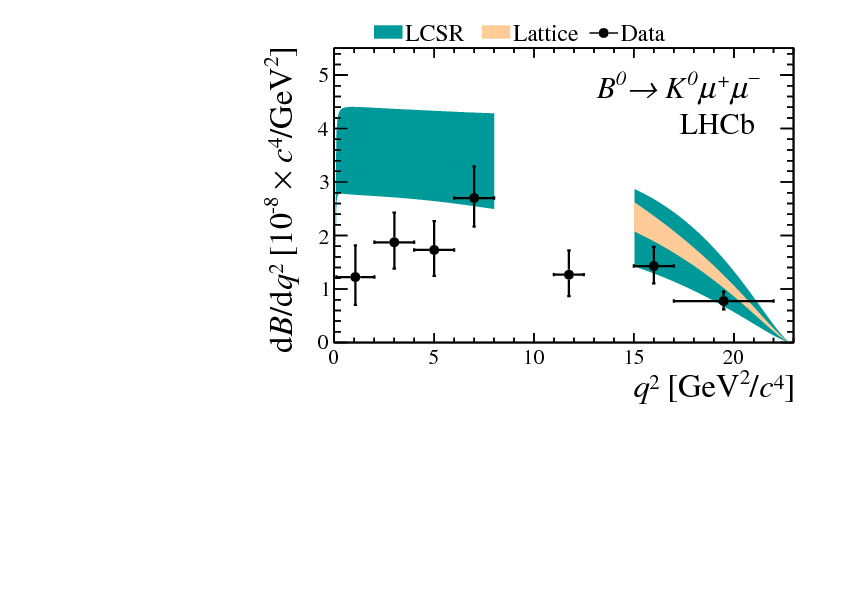}
\end{minipage}
\begin{minipage}[t]{\linewidth}
\centering
\includegraphics[trim={8cm 6cm 0 0}, clip, width=0.55\linewidth]{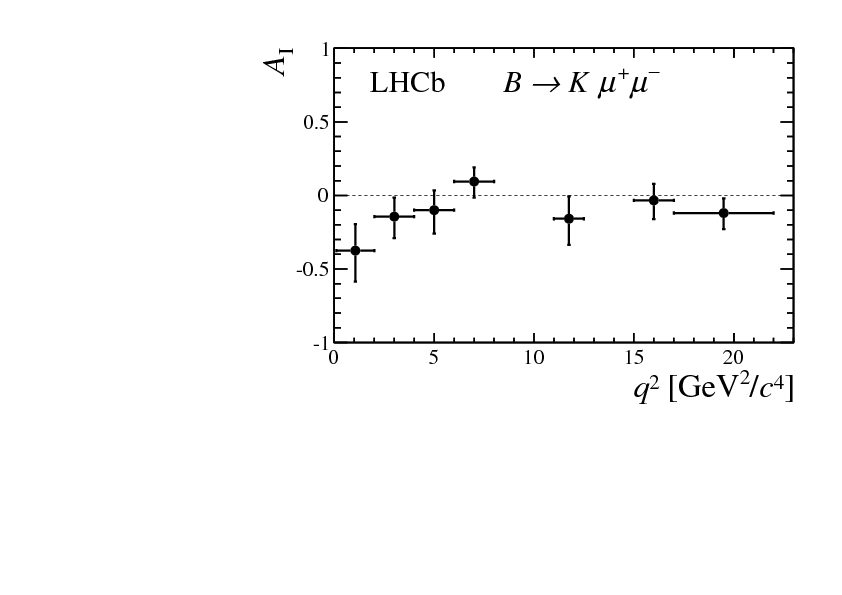}
\end{minipage}
\caption{Differential branching fraction results for \BToKpmumu (top left) and \BToKzmumu (top right) decays with theory predictions overlaid~\cite{Bobeth:2011nj} and the \CP-averaged isospin asymmetry (bottom). Figures from~\cite{Aaij:2014pli}.}
 \label{fig:BF_AI_kaon}
\end{figure}

The differential branching fractions of the decays \BToKpmumu and
\BToKzmumu versus the invariant mass squared of the dimuon pair, \qsq, were determined by the \lhcb collaboration with the \CP-averaged isospin asymmetry
$A_\text{I}$~\cite{Aaij:2014pli} defined as 
\begin{align}
\label{eq:AI}
A_\text{I} &= \frac{\Gamma(\decay{\Bz}{\kaon^{(*)0} \mup\mun}) - \Gamma(\decay{\Bp}{\kaon^{(*)+} \mup\mun})}{\Gamma(\decay{\Bz}{\kaon^{(*)0} \mup\mun}) + \Gamma(\decay{\Bp}{\kaon^{(*)+} \mup\mun})},\\
&= \frac{\mathcal{B}(\decay{\Bz}{\kaon^{(*)0} \mup\mun}) - (\tau_0/\tau_+)\cdot \mathcal{B}(\decay{\Bp}{\kaon^{(*)+} \mup\mun})}{\mathcal{B}(\decay{\Bz}{\kaon^{(*)0} \mup\mun}) + (\tau_0/\tau_+)\cdot \mathcal{B}(\decay{\Bp}{\kaon^{(*)+} \mup\mun})},
\end{align}
with the partial widths $\Gamma$ and branching fractions $\mathcal{B}$ of the corresponding decay channels and the ratio of \Bz to \Bp lifetimes $\tau_0/\tau_+$. The analysis finds values individually compatible with the \SM prediction. However, the entity of measurements lies systematically below the predictions as shown in~\reffig{BF_AI_kaon}. In the same analysis, the \CP-averaged isospin asymmetry has been measured and is depicted in~\reffig{BF_AI_kaon}. In a similar analysis, the \CP asymmetry 
\begin{align}
\label{eq:ACP}
\mathcal{A}_{\CP} &= \frac{\Gamma(\decay{\bar{\B}}{\bar{\kaon}^{(*)} \mup\mun}) - \Gamma(\decay{\bar{\B}}{\bar{\kaon}^{(*)} \mup\mun})}{\Gamma(\decay{\bar{\B}}{\bar{\kaon}^{(*)} \mup\mun}) + \Gamma(\decay{\bar{\B}}{\bar{\kaon}^{(*)} \mup\mun})},
\end{align}
was determined under the assumption of no direct \CP violation in the control mode \BToJpsiKp to be $\mathcal{A}_{\CP}(\BToKpmumu) = 0.012 \pm 0.017 \pm 0.001$, where the uncertainties are statistical and systematic, respectively~\cite{Aaij:2014bsa}, and this measurement is consistent with the \SM prediction~\cite{Altmannshofer:2008dz}.\\

The double differential decay rate of \BToKpmumu decays~\cite{Bobeth:2007dw} is given by

\begin{align}
\frac{1}{\Gamma} \frac{d\Gamma}{d\cos\theta_l} = \frac{3}{4}(1-F_{\text{H}})(1-\cos^2\theta_l) + \frac{1}{2}F_{\text{H}} + A_{\text{FB}}\cos\theta_l,
\end{align}

where the angle between the \mun (\mup) and the oppositely charged kaon \Kp (\Km) of the \Bp (\Bm) decay is denoted by $\theta_l$, $F_{\text{H}}$ is the so-called flat term, and $A_{\text{FB}}$ is the forward-backward asymmetry of the dimuon system.

\begin{figure}[h]
\begin{minipage}[t]{0.49\linewidth}
\centering
\includegraphics[trim={8cm 6cm 0 0}, clip, width=1.1\linewidth]{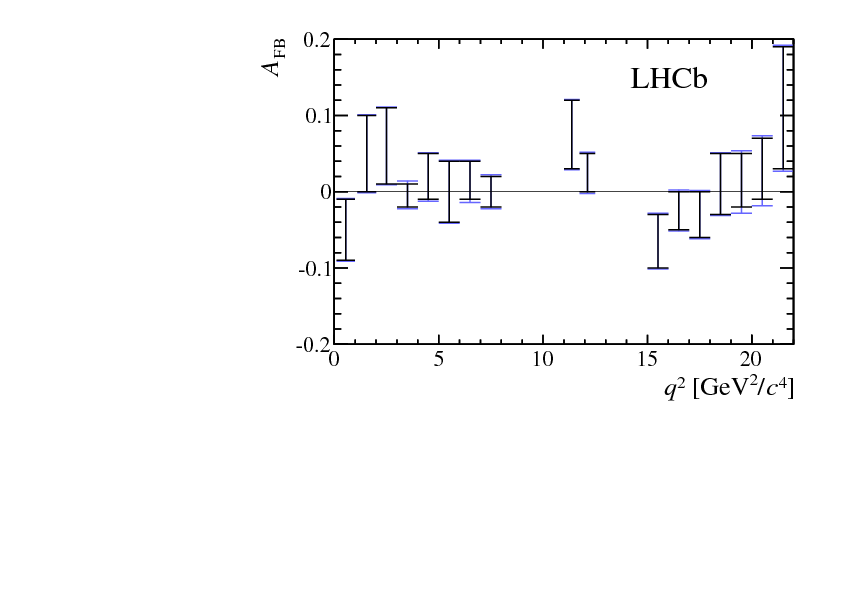}
\end{minipage}
\hspace{\fill}
\begin{minipage}[t]{0.49\linewidth}
\centering
\includegraphics[trim={8cm 6cm 0 0}, clip, width=1.1\linewidth]{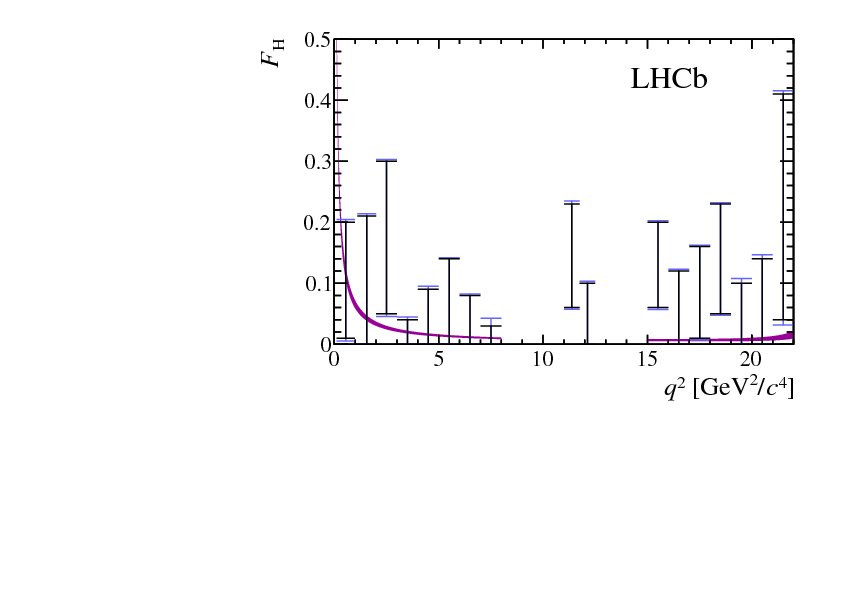}
\end{minipage}
\begin{minipage}[t]{\linewidth}
\centering
\includegraphics[trim={8cm 6cm 0 0}, clip, width=0.55\linewidth]{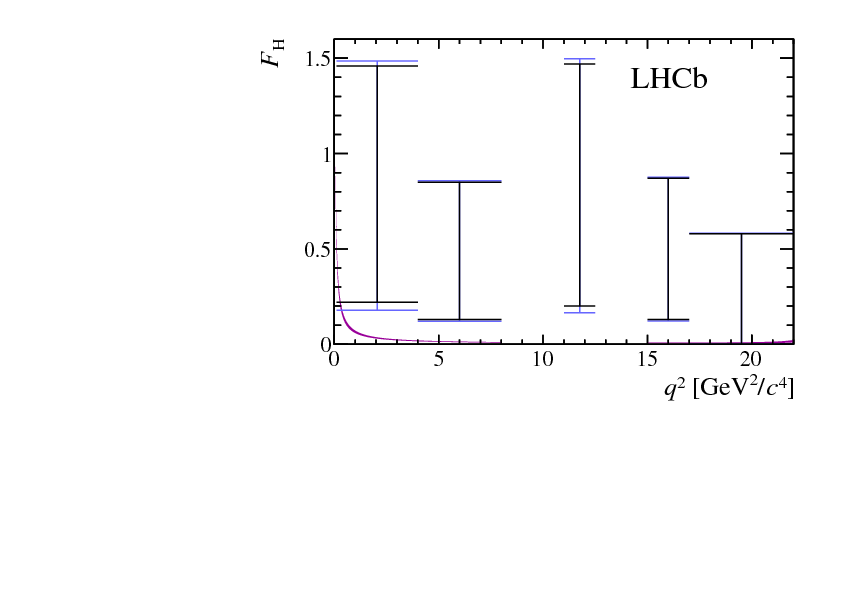}
\end{minipage}
\caption{Results for $A_{\text{FB}}$ (left) and $F_{\text{H}}$ in \BToKpmumu decays and for $F_{\text{H}}$ in \BToKsmumu decays (bottom).  The inner horizontal bars indicate the one-dimensional $68\%$ confidence intervals, whereas the outer vertical bars include contributions from systematic uncertainties. The \SM predictions are overlaid as band. Figures from~\cite{Aaij:2014tfa}.}
 \label{fig:FH_AFB_kaon}
\end{figure}

As the flavour of the neutral \B meson cannot be determined for the self-conjugate final state $\KS\mup\mun$, the double differential decay rate is extracted as a function of the absolute value of $\cos \theta_l$ as
\begin{align}
\frac{1}{\Gamma} \frac{d\Gamma}{d|\cos\theta_l|} = \frac{3}{2}(1-F_{\text{H}})(1-|\cos\theta_l|^2) + F_{\text{H}},
\end{align}
with the constraint of $0 \leq F_{\text{H}} \leq 3$ enforcing the expression to be positive-definite for all values of $|\cos \theta_l|$~\cite{Aaij:2014tfa}. The results of $A_{\text{FB}}$ and $F_{\text{H}}$ are compatible with the \SM predictions~\cite{Bobeth:2011nj} and are shown in~\reffig{FH_AFB_kaon}.\\

Interference effects between the short- and long-distance contributions were studied in \BToKpmumu decays at low recoil~\cite{Aaij:2013pta}. The \lhcb collaboration reported the first observation of the decays $\decay{\Bp}{\Psi(4160)\Kp}$ and the subsequent decay $\decay{\Psi(4160)}{\mup\mun}$ with $\mathcal{B}(\decay{\Bp}{\Psi(4160)\Kp}) = (5.1^{+1.3}_{-1.2} \pm 3.0)\cdot 10^{-4}$, which is determined under the assumption of lepton flavour universality and hence the second uncertainty stems from the known $\decay{\Psi(4160)}{\ep\en}$ branching fraction. No significant signal was observed for the $\Psi(4040)$ resonance, and an upper limit $\mathcal{B}(\decay{\Bp}{\Psi(4040)\Kp}) < 1.3 \,(1.7) \cdot 10^{-4}$ at $90 \, (95)\%$ C.L. is set. The mean and the width of the $\Psi(4160)$ resonance were measured to be $4191^{+9}_{-8} \mevcc$ and $65^{+22}_{-16} \mevcc$, respectively, where the uncertainties comprise both statistical and systematic sources. The interference between the observed resonant $\decay{\Bp}{\Psi(4160)\Kp}$ and the non-resonant \BToKpmumu decay in the large \qsq region amounts to $20\%$~\cite{Aaij:2013pta} leading to renewed interest in previous estimates of quark-hadron duality violation for these processes~\cite{Beylich:2011aq} as discussed in~\refsec{th-had:non-local}.

\begin{figure}
    \centering
    \includegraphics[trim={9cm 4cm 0 0}, clip,width=0.7\textwidth]{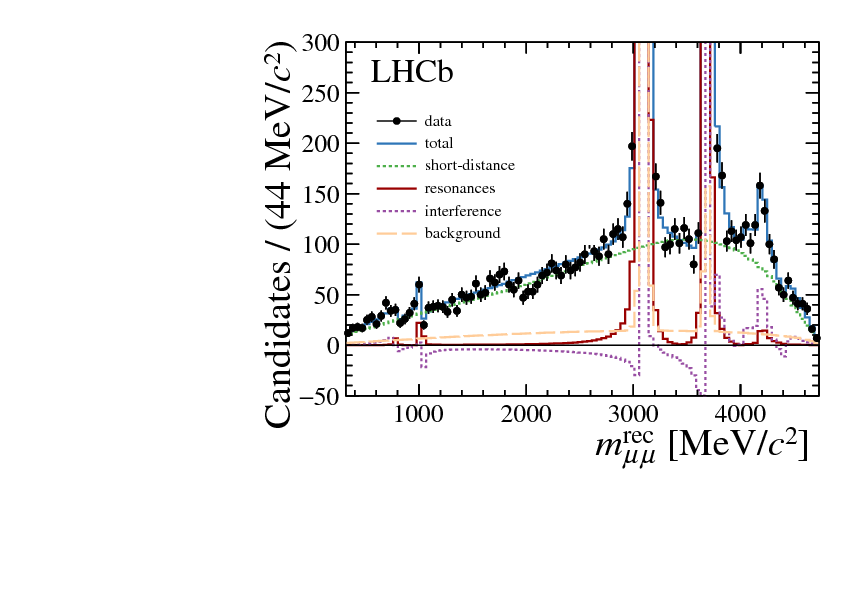}
    \caption{Fit to the mass of the dimuon pair for the favoured case where both \jpsi and \psitwos phases are negative. Figure from~\cite{Aaij:2016cbx}.}
        \label{fig:phasediff}
\end{figure}

By assuming a model of hadronic resonances, the \lhcb collaboration fits the differential branching fraction of \BToKpmumu to data. The branching fraction of the short-distance component is determined by setting the \jpsi and \psitwos resonance amplitudes to zero, and is found to be $\mathcal{B}(\BToKpmumu) = (4.37 \pm 0.15 \pm 0.23)\cdot 10^{-7}$~\cite{Aaij:2016cbx}, where the uncertainties are statistical (including the form factor uncertainties) and systematic. In addition, the phase difference between the short-distance and the narrow-resonance amplitudes in \BToKpmumu decays was determined by performing a fit to the mass of the dimuon pair; of which one possible solution is illustrated in~\reffig{phasediff}. As the values of the \jpsi phases are compatible with $\pm \pi/2$, the interference with the short-distance contributions far from the pole masses is small.

\subsubsection{$B \to \Kstar \ellell$ decays}
\label{sec:Kstll}


In addition to the aforementioned studies of \Kz and \Kp states, channels with a pair of \Kp\pim mesons in the final
state have received particular attention over the past years as those decays show a rich phenomenology. Amongst these decays, events with a \Kp\pim invariant mass
close to the vector resonance \Kstar(892) are the subject of numerous studies . Here and in the following, we discuss
treating the \Kstar(892) as a quasi-stable particle, and decays \BToKstzmumu should be interpreted as the decay chain
$\decay{\B}{\Kstar(892)(\to \kaon \pi) \mup \mun}$ if not otherwise
stated.\\

The \CP asymmetry defined in~\refeq{ACP} was measured by the
\lhcb collaboration to be $\mathcal{A}_{\CP}(\BToKstzmumu)
= -0.035 \pm 0.024 \pm 0.003$, where the uncertainties are statistical
and systematic, respectively~\cite{Aaij:2014bsa}, and the measurement
is found to be consistent with the \SM expectation of a small \CP
asymmetry in this decay.\\ 

\begin{figure}[h]
\begin{minipage}[t]{0.49\linewidth}
\centering
\includegraphics[trim={8cm 6cm 0 0}, clip, width=1.1\linewidth]{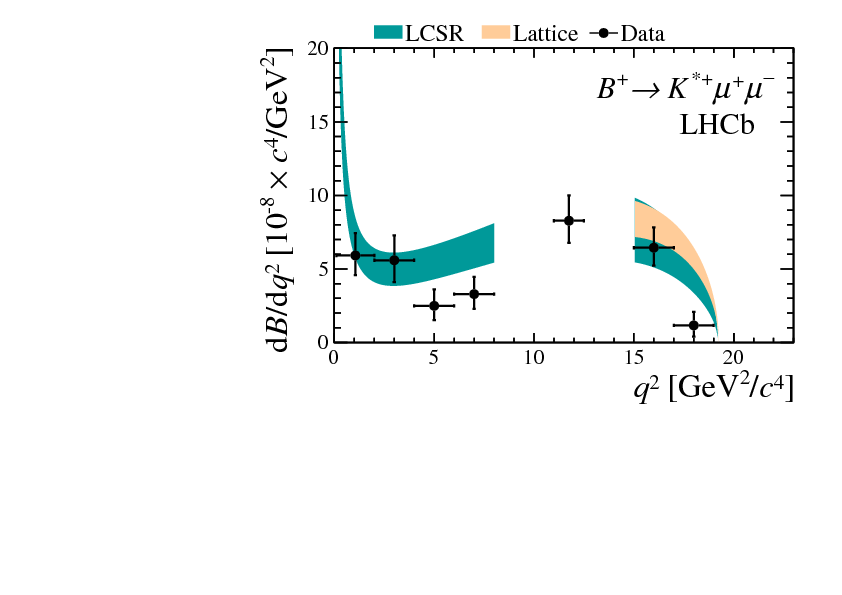}
\end{minipage}
\hspace{\fill}
\begin{minipage}[t]{0.49\linewidth}
\centering
\includegraphics[trim={8cm 6cm 0 0}, clip, width=1.1\linewidth]{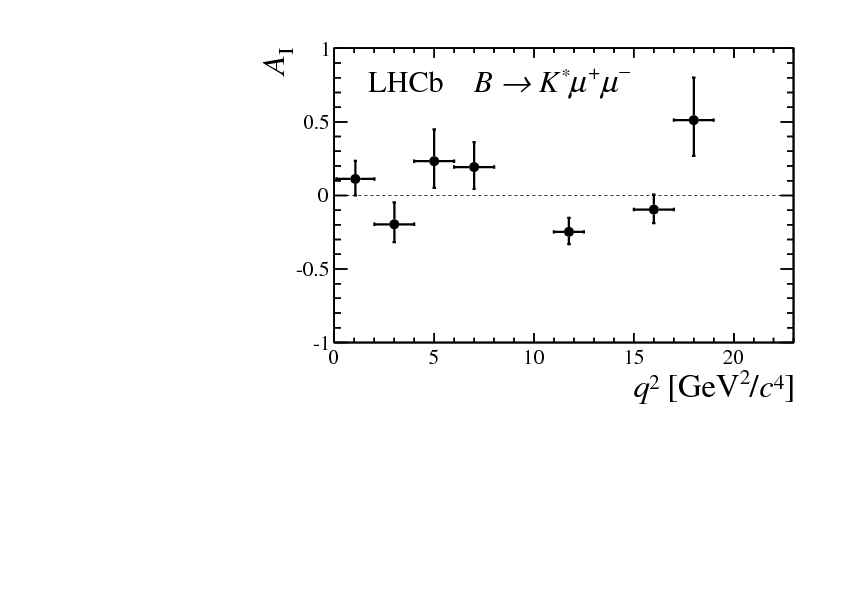}
\end{minipage}
\caption{Differential branching fraction results for \BToKstpmumu (left) decays with theory predictions overlaid~\cite{Bobeth:2011gi,Bobeth:2011nj} and the \CP-averaged isospin asymmetry (right). Figures from~\cite{Aaij:2014pli}.}
 \label{fig:BF_AI_kstar}
\end{figure}

\begin{figure}
\begin{minipage}[t]{0.49\linewidth}
\centering
    \includegraphics[width=0.9\textwidth]{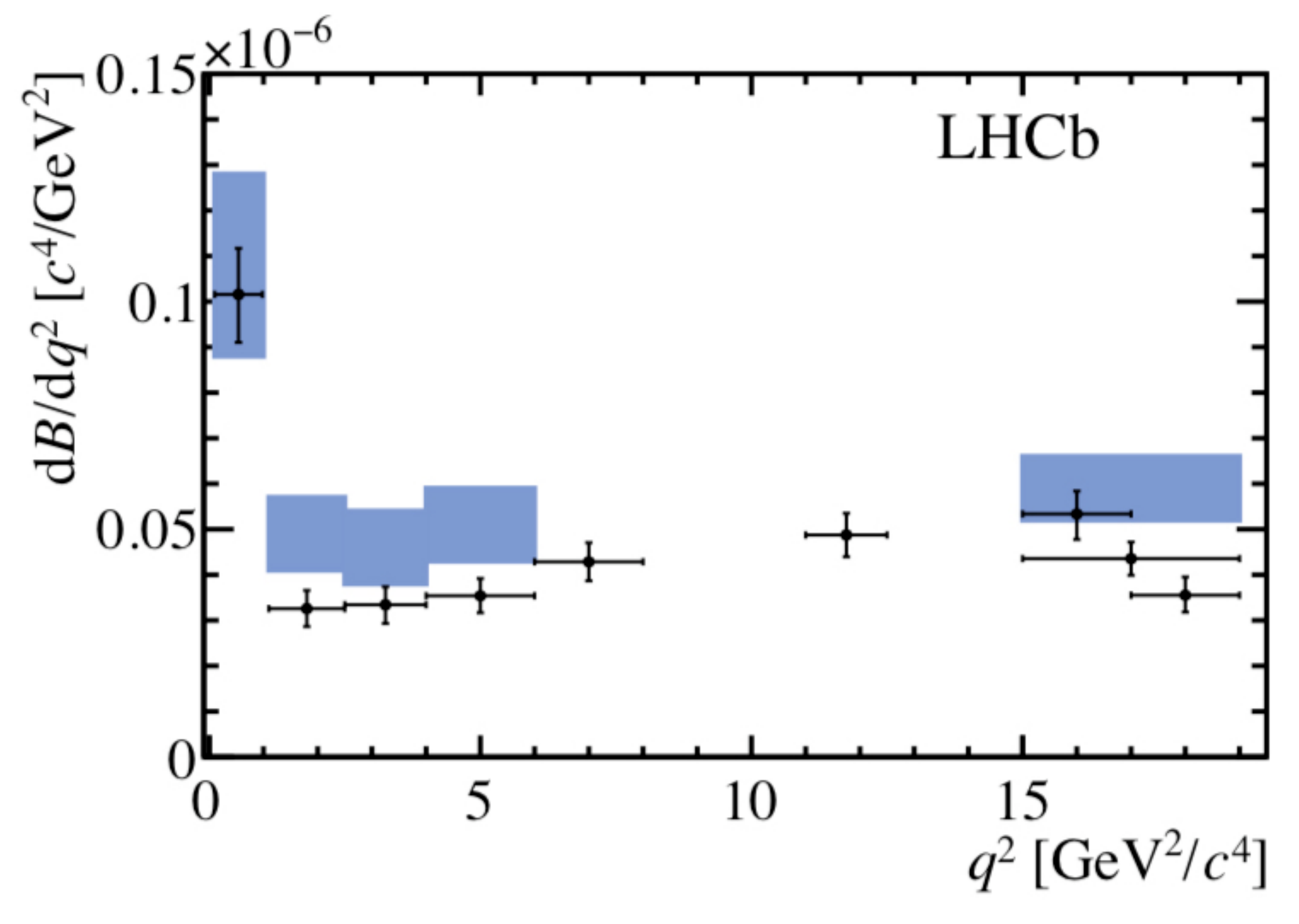}
    \end{minipage}
\hspace{\fill}
\begin{minipage}[t]{0.49\linewidth}
    \centering
\includegraphics[trim={6cm 6cm 0 0}, clip, width=1.05\textwidth]{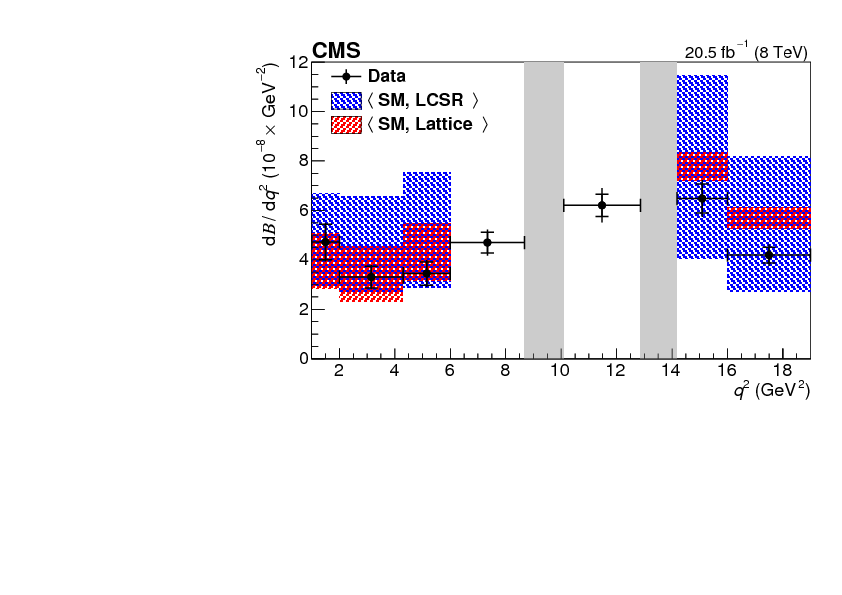}
\end{minipage}
\caption{Differential branching fraction results for the P-wave component of \BToKstzmumu decays with theory predictions overlaid~\cite{Horgan:2013hoa,Straub:2015ica}, which are purely made for the resonant P-wave part of the \Kp\pim system (left)~\cite{Aaij:2016flj} and for the S- and P-wave components (right)~\cite{Khachatryan:2015isa}.}
    \label{fig:kstzmumu_BF}
\end{figure}

The differential branching fraction of a charged \B meson into a
\Kstarp final state, \BToKstpmumu, has been measured by the \lhcb
collaboration; the results are shown in~\reffig{BF_AI_kstar} in several bins of \qsq. A study of the
neutral decay mode was performed, which resulted in the measurement of
the isospin asymmetry as defined in~\refeq{AI} illustrated in~\reffig{BF_AI_kstar}. As the differential branching fractions of
the decay \BToKstzmumu had been previously reported in~\cite{Aaij:2013iag}, the values were not updated in~\cite{Aaij:2014pli} until later in~\cite{Aaij:2016flj}. The
latter results are as well depicted in~\reffig{kstzmumu_BF}, and the
analysis yields the most precise measurement to date of the \qsq-averaged branching ratio
\begin{align}
\mathcal{B}(\BToKstzmumu)/\delta\qsq = (0.342^{+0.17}_{-0.17} \pm 0.009 \pm 0.023)\cdot 10^{-7} \frac{c^4}{\gev^2},
\end{align}
with $\delta\qsq = 4.9 \gevgevcccc$ (in the bin $1.1 \gevgevcccc < \qsq <  6.0 \gevgevcccc $) where the uncertainties are statistical, systematic and from the uncertainty on the branching fraction of the normalisation channel \BtoKstJpsi. In the region $1.1  \gevgevcccc < \qsq < 6.0 \gevgevcccc$ and for the \Kp\pim invariant mass range $796 \mevcc < m(\kaon\pi) < 996 \mevcc$, the S-wave component is measured to be $F_S = 0.101 \pm 0.017 \pm 0.009$~\cite{Aaij:2016flj}, where the uncertainties are of statistical and systematic origin. Assuming the absence of high-order waves \eg D- and F-waves, the pure P-wave component of the differential branching fraction has been determined for the first time. However, in previous analyses, the S-wave fraction was not taken into account as is the case in a measurement of the differential branching fraction published by the \cms collaboration~\cite{Khachatryan:2015isa}, whose results are shown in~\reffig{kstzmumu_BF}. 

The four-dimensional differential decay rate of \BToKstzmumu (\BbToKstzbmumu) decays is expressed as a function of the invariant mass of the lepton pair $q$, and the angles $\theta_{\ell}$, $\theta_{\Kstar}$ and $\phi$, where $\theta_{\ell}$ refers to the angle between the \mun and the \Kstarz (\Kstarzb) flight directions in the di-lepton rest frame, and $\theta_{\Kstar}$ to the angle between the \Kstarz (\Kstarzb) and the \Kp (\Km) flight directions in the \Kstar rest frame and $\phi$ corresponds to the angle between the planes defined by the di-muon and the kaon and pion in the \B rest frame~\cite{Kruger:1999xa, Kruger:2005ep}. The fully differential decay rate is then given by
\begin{align}
\label{eq:fullydiffdecayrate}
\frac{\mathrm{d}^4\bar{\Gamma}(\BToKstzmumu)}{\mathrm{d}q \, \mathrm{d cos \,}{\theta_{\ell}} \, \mathrm{d cos \,} {\theta_{\Kstar}} \, \mathrm{d cos \,}{\phi} } &= \frac{9}{32\pi} \sum\limits \bar{I_i}(q, \theta_{\Kstar}) f_i (\theta_{\ell}, \phi), \\
\frac{\mathrm{d}^4\Gamma(\BbToKstzbmumu)}{\mathrm{d}q \, \mathrm{d cos \,}{\theta_{\ell}} \, \mathrm{d cos \,}{\theta_{\Kstar}} \, \mathrm{d cos \,}{\phi} } &= \frac{9}{32\pi} \sum\limits I_i(q, \theta_{\Kstar}) f_i (\theta_{\ell}, \phi),
\end{align}
where the $I_i$ ($\bar{I_i}$) terms depend on products of the \Kstar spin amplitudes and the $f_i$ functions are the corresponding angular distribution functions\footnote{The differential decay rate taking the full basis including scalar and tensor operators has been derived in~\cite{Bobeth:2012vn}.} assuming an on-shell \Kstar meson~\cite{Altmannshofer:2008dz}. As the theoretical calculations define $\theta_{\ell}$ with respect to the negatively charged muon in both \Bz and \Bzb decays, the angular distributions between theory and experiment differ. A translation scheme between the two conventions is given in~\cite{Gratrex:2015hna}. More common than expressing the differential decay rates as functions of $I_i$ and $\bar{I_i}$ is using the \CP-averaged and \CP-asymmetric observables $S_i$ and $A_i$ as introduced in~\cite{Altmannshofer:2008dz}.
\begin{align}
S_i &= (I_i + \bar{I_i}) \left / \right. \left ( \frac{\mathrm{d}\Gamma}{\mathrm{d}\qsq} + \frac{\mathrm{d}\bar{\Gamma}}{\mathrm{d}\qsq} \right ), \\
A_i &= (I_i - \bar{I_i}) \left / \right. \left ( \frac{\mathrm{d}\Gamma}{\mathrm{d}\qsq} + \frac{\mathrm{d}\bar{\Gamma}}{\mathrm{d}\qsq} \right ).
\end{align}
One can construct a complete set of observables $P^{(\prime)}_i$ with a reduced $\decay{\Bz}{\Kstar}$ form-factor dependence~\cite{DescotesGenon:2012zf,Matias:2012qz,Matias:2012xw} in the large energy limit. They emerge from combinations of $F_L$ and $S_3-S_9$, where $\Pfivep = S_5/\sqrt{F_L(1-F_L)}$ is the most notable and hence relevant for further discussions in this review.\\

\begin{figure}
\centering
\begin{minipage}[t]{0.49\linewidth}
\centering
\includegraphics[trim={8cm 6cm 0 0}, clip, width=1.\linewidth]{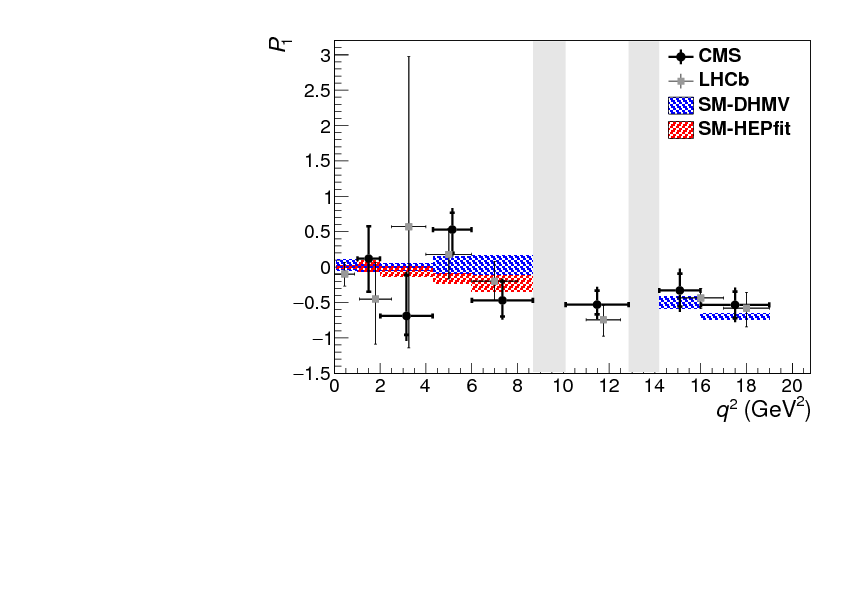}
\end{minipage}
\hspace{\fill}
\begin{minipage}[t]{0.49\linewidth}
\centering
\includegraphics[trim={8cm 6cm 0 0}, clip, width=1.\linewidth]{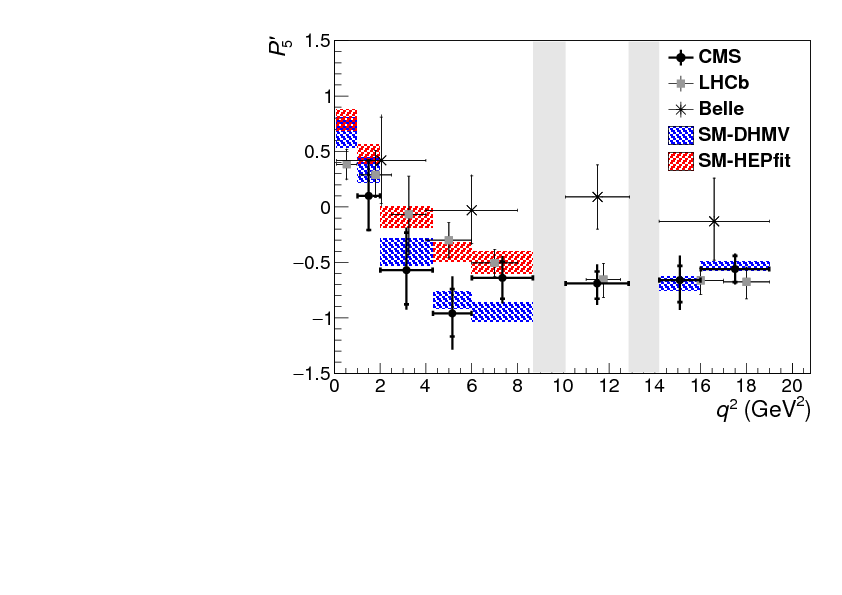}
\end{minipage}    
 \caption{Angular observables \Pone (left) and \Pfivep (right) of
   \BToKstzmumu decays depending on \qsq compared to the
   \lhcb~\cite{Aaij:2015oid} and \belle~\cite{Wehle:2016yoi}
   measurements and with theory predictions overlaid
   (DHMV~\cite{DescotesGenon:2012zf,Descotes-Genon:2013vna},
   HEPfit~\cite{Ciuchini:2015qxb,Ciuchini:2016weo}). The inner bars
   represent the statistical uncertainty, and the total uncertainty is
   illustrated by the outer vertical bars. The bin widths are
   indicated by the horizontal bars. Figures from~\cite{Sirunyan:2017dhj}.} 
    \label{fig:kstzmumu_ang_CMS}
\end{figure}

In the angular analysis of \BToKstzmumu decays by the \cms collaboration on a dataset of $20.5\invfb$ recorded in 2012, both \Pfivep and $\Pone = 2S_3/(1-F_{\text{L}})$ were determined~\cite{Sirunyan:2017dhj}, by using a folded differential decay rate - an approach originally developed by the \lhcb collaboration~\cite{Aaij:2013iag,Aaij:2013qta}. After the folding, a multi-dimensional fit is performed, with the parameters of interest \Pone, \Pfivep and $A_{\text{S}}^5$ along with signal and background yields, whereas the values of the longitudinal polarisation of the \Kstar meson, $F_{\text{L}}$, $F_{\text{S}}$ and and the interference between S- and P-wave, $A_{\text{S}}$, have been fixed to values determined in a previous analysis performed on the same dataset~\cite{Khachatryan:2015isa}. The results of \Pone and \Pfivep are shown in~\reffig{kstzmumu_ang_CMS} and are consistent with the \SM predictions and previous measurements.

\begin{figure}
\centering
\begin{minipage}[t]{0.49\linewidth}
\centering
\includegraphics[width=1.\linewidth]{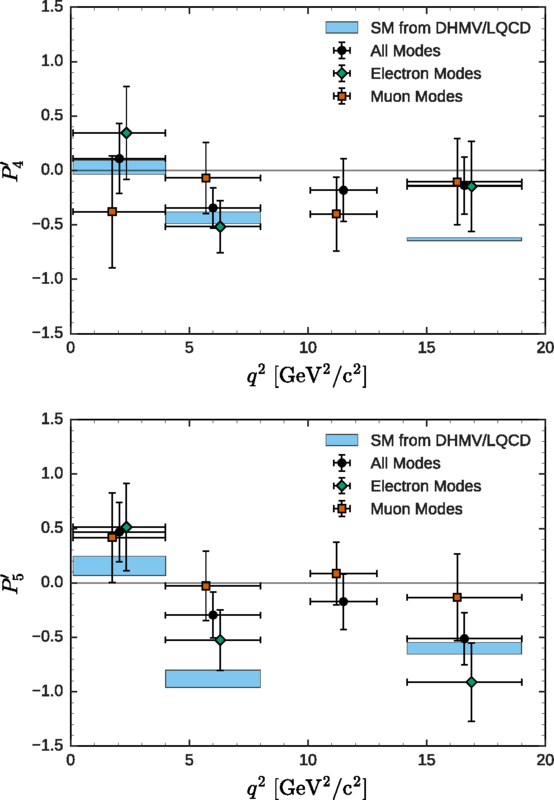}
\end{minipage}
\hspace{\fill}
\begin{minipage}[t]{0.49\linewidth}
\centering
\includegraphics[width=1.\linewidth]{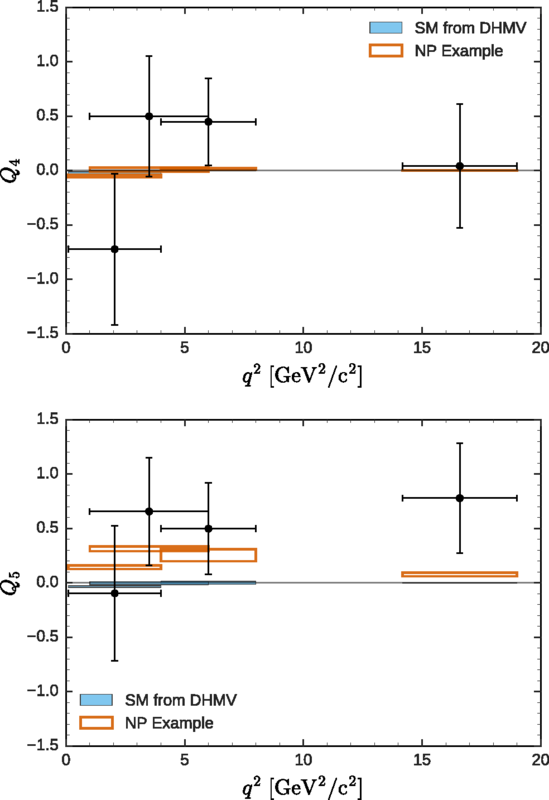}
\end{minipage}    
 \caption{Belle measurement of angular observables \Pfourp (top left) and \Pfivep (bottom
   left) depending on \qsq for \BToKstzll with $\ell=e,\mu$ decays
   with theory predictions overlaid from
   DHMV~\cite{Descotes-Genon:2013vna} and lattice
   QCD~\cite{Horgan:2015vla}; the measured values of $Q_4$ (top right)
   and $Q_5$ (bottom right) are compared to the DHMV prediction and an
   arbitrary new physics scenario (scenario 1 from
   ~\cite{Descotes-Genon:2013vna}). Figures
   from~\cite{Wehle:2016yoi}.} 
    \label{fig:kstzmumu_ang_Belle}
\end{figure}

The aforementioned folding technique has also been employed by the \belle collaboration to extract \Pfivep and $\Pfourp=S_4/\sqrt{F_{\text{L}}(1-F_{\text{L}})}$ from the full dataset for both \BToKstzmumu and \BToKstzee decays, as well as the combination of both leptonic final states~\cite{Wehle:2016yoi}. The results on \Pfourp and \Pfivep are shown in~\reffig{kstzmumu_ang_Belle}. Tensions between measurement and \SM prediction~\cite{Descotes-Genon:2013vna,Horgan:2015vla} are observed for \Pfivep for the muon final state in the region $4 \gevgevcccc< \qsq < 8 \gevgevcccc$ of $2.6\sigma$, whereas the electron mode differs by $1.3\sigma$ in the very same region, leading to a combined tension of $2.5\sigma$. In addition to the $P^{\prime}_{4,5}$ observables, the so-called $Q_{4,5} = P^{\prime, \mu}_{4,5} - P^{\prime, e}_{4,5} $ observables are determined for the first time; any deviation of these observables from zero would be a clear sign for new physics~\cite{Capdevila:2016ivx}. However in the \belle analysis, no deviation from zero is observed as can be seen in~\reffig{kstzmumu_ang_Belle}.

\begin{figure}
\centering
\begin{minipage}[t]{0.49\linewidth}
\centering
\includegraphics[trim={8cm 6cm 0 0}, clip, width=1.\linewidth]{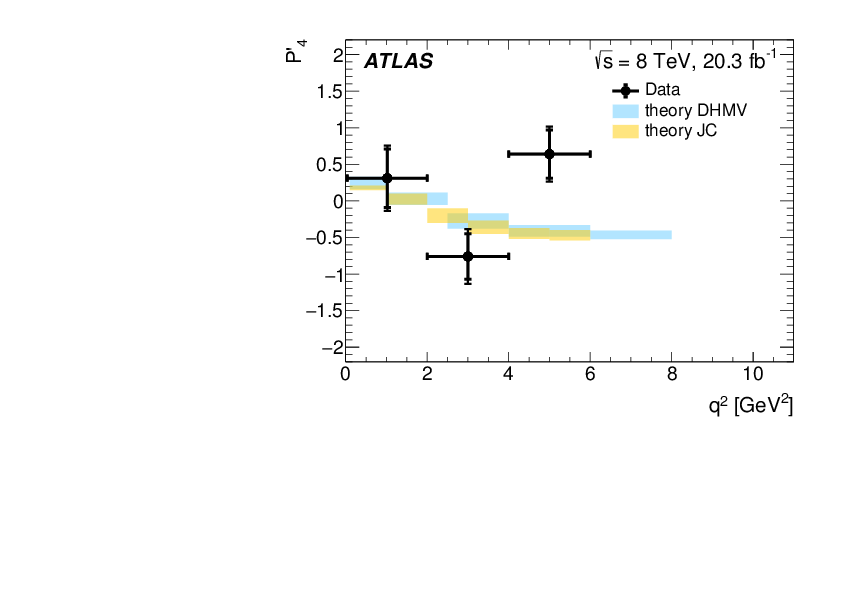}
\end{minipage}
\hspace{\fill}
\begin{minipage}[t]{0.49\linewidth}
\centering
\includegraphics[trim={8cm 6cm 0 0}, clip, width=1.\linewidth]{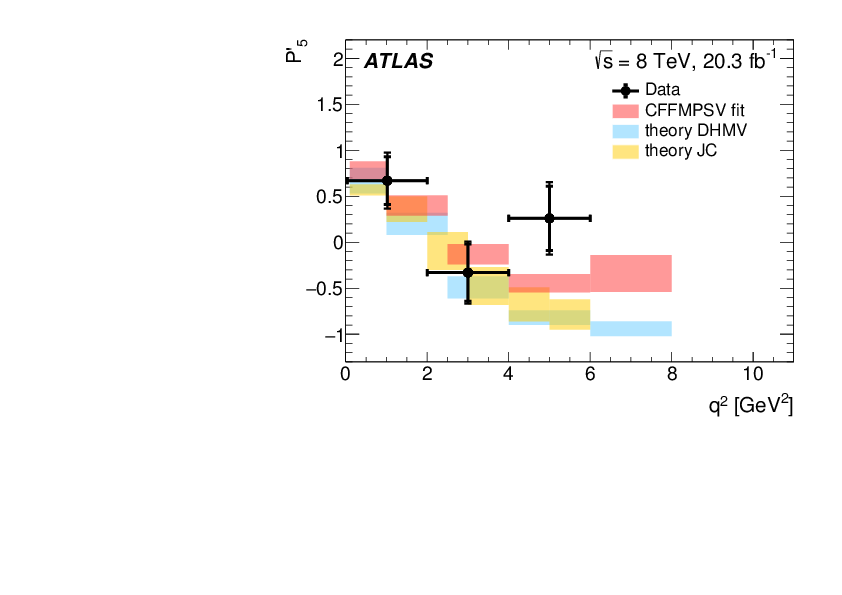}
\end{minipage}    
 \caption{Angular observables \Pfourp (left) and \Pfivep (right) depending on \qsq for \BToKstzmumu decays with theory predictions overlaid from DHMV~\cite{Descotes-Genon:2013vna} and JC~\cite{Jager:2012uw,Jager:2014rwa}. Figures from~\cite{Aaboud:2018krd}.}
    \label{fig:kstzmumu_ang_ATLAS}
\end{figure}

A more complete angular analysis was performed by the \atlas collaboration on its 20.3\invfb dataset from 2012 by exploiting four different folding schemes allowing to extract a set of four observables for each scheme~\cite{Aaboud:2018krd}. However neither $S_9$ nor $A_{\text{FB}}$ can be measured with the chosen approach. The values for $F_{\text{L}}$ and $S_3$, which are common to each folding scheme, have been compared between the four fits and have found to be consistent. The publication comprises results on $F_{\text{L}}$, $S_{3,4,5,7,8}$, $P_1$ and $P^{\prime}_{4,5,6,8}$, from which \Pfourp and \Pfivep are shown in~\reffig{kstzmumu_ang_ATLAS}. The obtained results are consistent with the different \SM predictions within less than three standard deviations. 

\begin{figure}
\centering
\begin{minipage}[t]{0.49\linewidth}
\centering
\includegraphics[trim={8cm 6cm 0 0}, clip, width=1.1\linewidth]{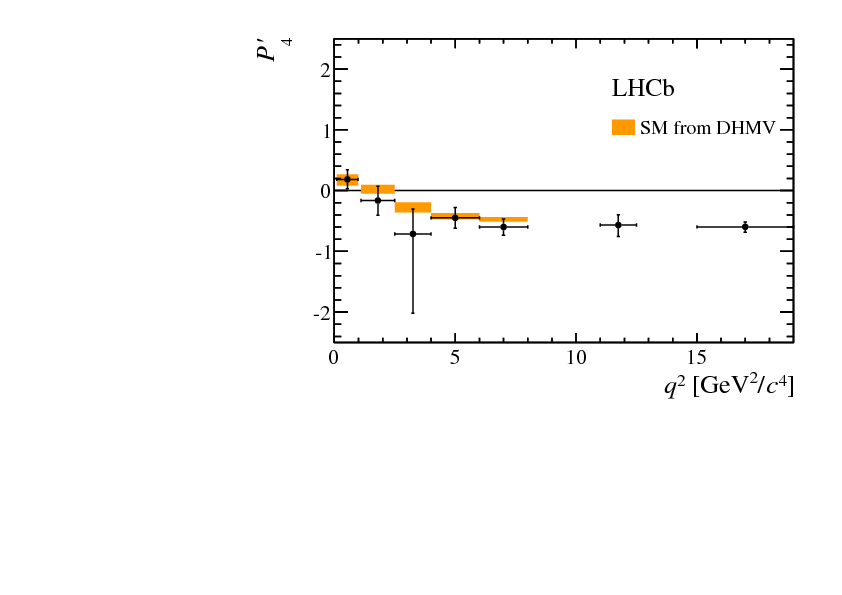}
\end{minipage}
\hspace{\fill}
\begin{minipage}[t]{0.49\linewidth}
\centering
\includegraphics[trim={8cm 6cm 0 0}, clip, width=1.1\linewidth]{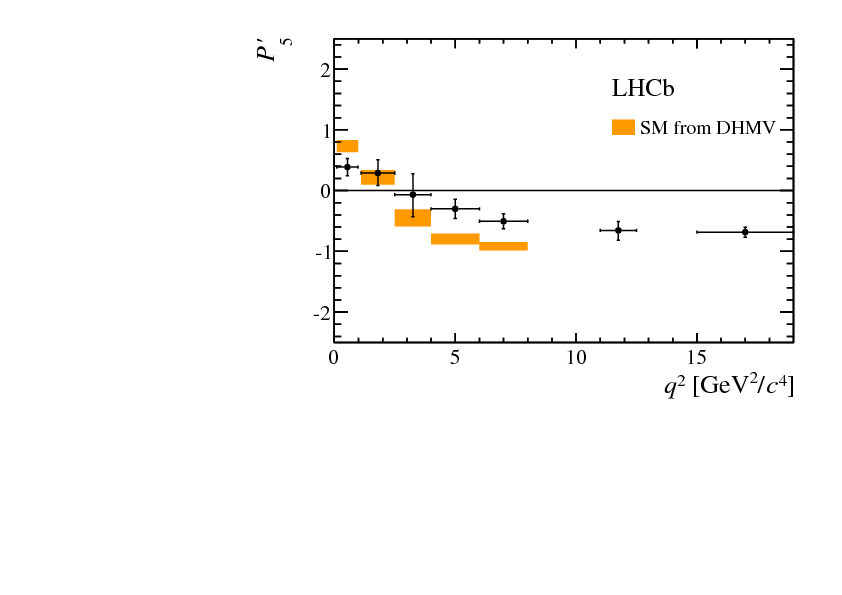}
\end{minipage}    
 \caption{Angular observables \Pfourp (left) and \Pfivep (right) of \BToKstzmumu decays depending on \qsq as extracted from a maximum likelihood fit with theory predictions overlaid~\cite{Descotes-Genon:2014uoa}. Figures from~\cite{Aaij:2015oid}.}
    \label{fig:kstzmumu_ang_LHCb}
\end{figure}

In contrast to previous analyses, the \lhcb collaboration has performed a full angular analysis, and has determined the set of angular observables by employing three different methods: by a maximum likelihood fit, by exploiting the angular principal moments~\cite{Beaujean:2015xea}, and in addition the zero-crossing points of $S_{4,5}$ and $A_{FB}$ were determined from a fit to the decay amplitudes~\cite{Aaij:2015oid}. The full angular analysis gives access to the correlation matrices, which provide crucial input for global fits to theoretical models. The results of the maximum likelihood and the method of angular principal moments are found to be compatible. As the maximum likelihood yields the most precise results, we restrict the discussion to this approach. However, we note in passing that the angular principal moments allow for smaller bin widths in \qsq and therefore provide more information of the \qsq shape of the angular observables compared to the maximum likelihood method. In this analysis performed by the \lhcb collaboration, the fraction of the S-wave component was taken into account. Neglecting theoretical correlations, the $S_i$ observables appear to be compatible with the \SM predictions. The theory correlations can be displayed best in the space of the $P^{(\prime)}$ observables with the most notable deviation of the \SM in \Pfivep as can be seen from~\reffig{kstzmumu_ang_LHCb}: in the regions $4 \gevgevcccc < \qsq < 6 \gevgevcccc $ and $6 \gevgevcccc < \qsq < 8 \gevgevcccc$, deviations of the measured values from the \SM prediction~\cite{Descotes-Genon:2014uoa} of $2.8\sigma$ and $3.0\sigma$ can be observed, respectively. This confirms the tension seen in a previous \lhcb analysis~\cite{Aaij:2013qta} in the region $4.30 \gevgevcccc < \qsq < 8.68 \gevgevcccc$, which had a local significance of $3.7\sigma$.

\begin{figure}
\centering
\includegraphics[trim={8cm 6cm 0 0}, clip, width=0.7\linewidth]{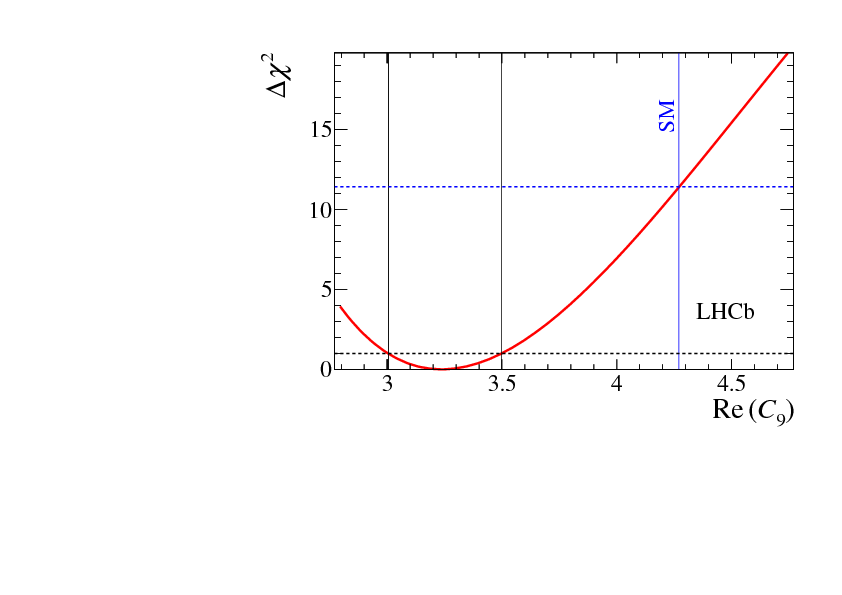}
 \caption{The $\Delta \chisq$ distribution for \Real \Cnine from a
   global fit to the full set of \CP-averaged observables. The best
   fit point lies at $\Real \Delta C_9 = -1.04 \pm 0.25$, whereas the
   central value of the \SM prediction is at $\Real C_9 =
   4.27$~\cite{Beaujean:2013soa}, hence indicating a tension of the
   measurement with respect to the \SM prediction. Figures
   from~\cite{Aaij:2015oid} using the EOS software
   package~\cite{Bobeth:2010wg,EOS}.} 
    \label{fig:kstzmumu_ang_LHCb_WC}
\end{figure}

A fit to the complete set of \CP-averaged observables, namely $F_{\text{L}}$, $A_{\text{FB}}$ and $S_3-S_9$, as determined from the maximum likelihood fit is performed by the \lhcb collaboration~\cite{Aaij:2015oid} using the EOS software package~\cite{Bobeth:2010wg,EOS} for observables in the \qsq ranges below $8 \gevgevcccc$ and $15.0  \gevgevcccc < \qsq < 19.0 \gevgevcccc$. The fit result illustrated in~\reffig{kstzmumu_ang_LHCb_WC} indicates a tension of $3.4\sigma$ between the measurements and the \SM prediction of \BToKstzmumu alone.

\begin{figure}[h]
\begin{minipage}[t]{\linewidth}
\centering
\includegraphics[width=1.\linewidth]{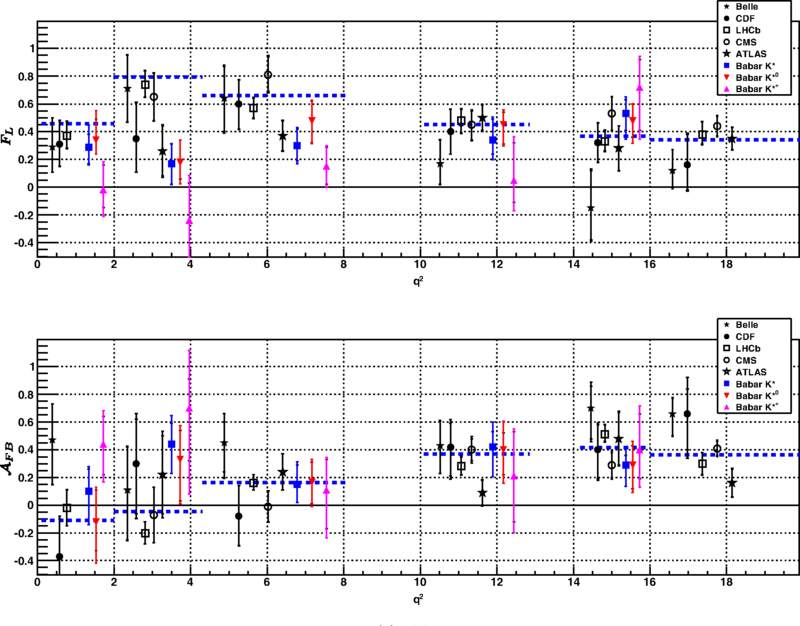}
\end{minipage}
\hspace{\fill}
\begin{minipage}[t]{\linewidth}
\centering
\includegraphics[width=1.\linewidth]{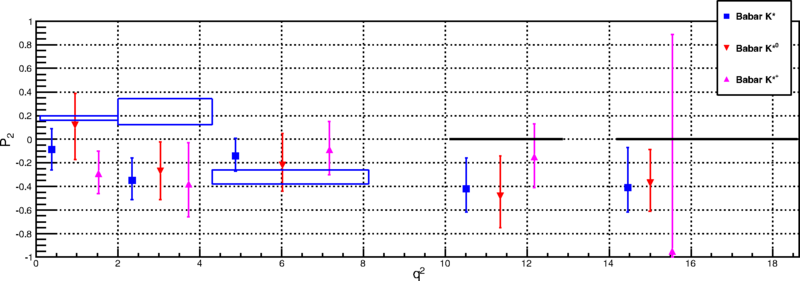}
\end{minipage}
\caption[Results for $F_{\text{L}}$ (top), $A_{\text{FB}}$ (middle) and $P_2$ (bottom) for different bins in \qsq. The results on $F_{\text{L}}$ and $A_{\text{FB}}$ are compared to previous results from \belle~\cite{Wei:2009zv}, CDF~\cite{Aaltonen:2011ja}, \lhcb~\cite{Aaij:2013iag}, \cms~\cite{Chatrchyan:2013cda} and \atlas~\cite{ATLAS:2013ola} and the \SM prediction~\cite{Descotes-Genon:2014uoa} (blue dashed line). For \Ptwo, the \SM prediction~\cite{Descotes-Genon:2014uoa} is indicated by the blue boxes, which is only available in the low \qsq region, and hence in the high \qsq regions, the bin size is illustrated by the black lines. Figures from~\cite{Lees:2015ymt}.]{Results for $F_{\text{L}}$ (top), $A_{\text{FB}}$ (middle) and $P_2$ (bottom) for different bins in \qsq. The results on $F_{\text{L}}$ and $A_{\text{FB}}$ are compared to previous results from \belle~\cite{Wei:2009zv}, CDF~\cite{Aaltonen:2011ja}, \lhcb~\cite{Aaij:2013iag}, \cms~\cite{Chatrchyan:2013cda} and \atlas~\cite{ATLAS:2013ola} \footnotemark  and the \SM prediction~\cite{Descotes-Genon:2014uoa} (blue dashed line). For \Ptwo, the \SM prediction~\cite{Descotes-Genon:2014uoa} is indicated by the blue boxes, which is only available in the low \qsq region, and hence in the high \qsq regions, the bin size is illustrated by the black lines. Figures from~\cite{Lees:2015ymt}.}
 \label{fig:kstzmumu_ang_Babar}
\end{figure}

When the fully differential decay rate (see~\refeq{fullydiffdecayrate}) is integrated over $\theta_{\ell}$ and $\phi$, the resulting expression depends on $F_{\text{L}}$ and $A_{\text{FB}}$, which is what the \babar collaboration exploited to measure those parameters~\cite{Lees:2015ymt} on their full dataset and to extract $\Ptwo=-2A_{\text{FB}}/(3[1-F_{\text{L}}])$. The studied decay channels include not only \BToKstzmumu and \BToKstzee decays but also their charged counterparts. The results are summarised in~\reffig{kstzmumu_ang_Babar} and most results are compatible amongst each other and with the \SM prediction. In the low \qsq region, a tension is observed in $F_L$ between \BToKstll and \BToKstzll decays as well as the \SM prediction. In the same region, there appears to be a small tension for \Ptwo.\\

\footnotetext{This result was later retracted in 2015.}

Although most previous analyses focus on the muonic final states, several analyses measure \eg. the branching fraction or angular observables for \BToKstzee decays. On the 2011 \lhcb dataset corresponding to 1\invfb, the branching fraction of the electron mode has been determined in the dielectron mass range $30-1000 \mevcc$ to be $\mathcal{B}(\BToKstzee)^{30-1000 \mevcc}  =(3.1^{+0.9}_{-0.8}\,^{+0.2}_{-0.3} \pm 0.2)\cdot 10^{-7}$~\cite{Aaij:2013hha}, where the uncertainties are statistical, systematic and from the normalisation mode. This result is found to agree with \SM predictions. On the 3\invfb dataset, an angular analysis has been performed~\cite{Aaij:2015dea} in the effective range $0.002 \gevgevcccc < \qsq < 1.120 \gevgevcccc$ yielding
\begin{align}
F_{\text{L}} &= +0.16 \pm 0.06 \pm 0.03, \\
P_1 = A_{\text{T}}^{(2)} &= -0.23 \pm 0.23 \pm 0.05, \\
A_{\text{T}}^{\Imag} &= +0.14 \pm 0.22 \pm 0.05, \\
A_{\text{T}}^{\Real} &= +0.10 \pm 0.18 \pm 0.05, 
\end{align}
where the uncertainties are statistical and systematic, respectively, and these results are in agreement with the \SM expectations.

\subsubsection{$B \to \Kstar \ellell$ decays beyond the ground state}

\begin{figure}[h]
\begin{minipage}[b]{0.49\linewidth}
\centering
\includegraphics[trim={8cm 6cm 0 0}, clip, width=1.1\linewidth]{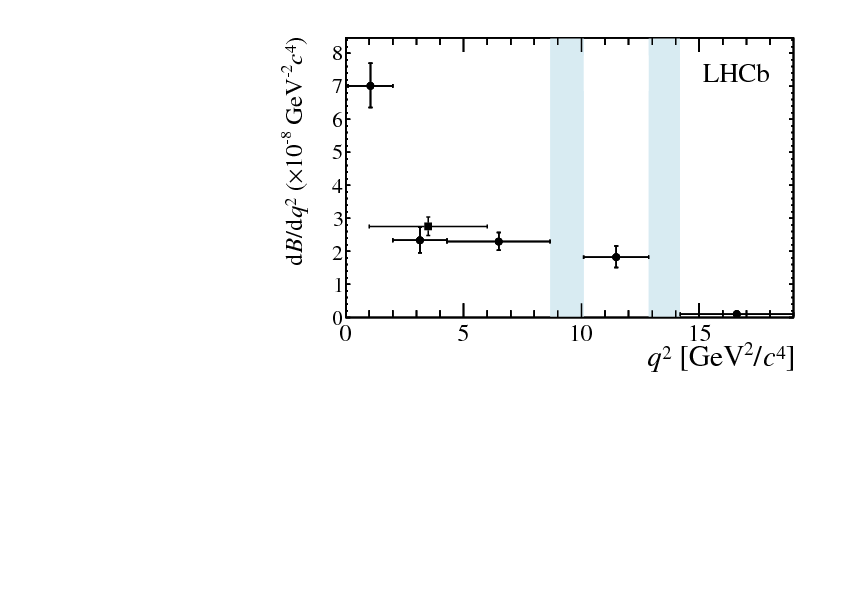}
\end{minipage}
\hspace{\fill}
\begin{minipage}[t]{0.49\linewidth}
\centering
\includegraphics[trim={0cm -1.8cm 0 0}, clip, width=1.\linewidth]{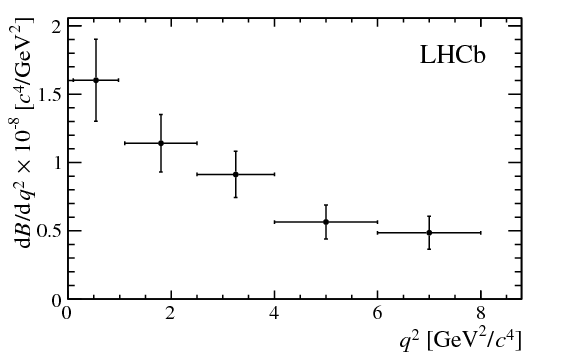}
\end{minipage}
\caption{Differential branching fraction results depending on \qsq for \BToKpipimumu decays (left - figure from~\cite{Aaij:2014kwa}) and for $\decay{\Bz}{\Kp\pim\mup\mun}$ decays in the region $1330 \mevcc < m(\Kp\pim) < 1530 \mevcc$ (right - figure from~\cite{Aaij:2016kqt}).}
 \label{fig:kpipimumu_LHCb}
\end{figure}

Apart from kaon and \Kstar states, one can consider higher \Kstar resonances above the $\Kstarz(892)$ mass range. The \lhcb collaboration reported the first observation of the decay channels \BToKpipimumu and \BTophiKmumu, whose branching fractions are
\begin{align}
\mathcal{B}(\BToKpipimumu) &= (4.36^{+0.29}_{-0.27} \pm 0.21 \pm 0.18)\cdot 10^{-7}, \\
\mathcal{B}(\BTophiKmumu) &= (0.82^{+0.19}_{-0.17}\,^{+0.10}_{-0.04} \pm 0.27)\cdot 10^{-7},
\end{align}
where the uncertainties are statistical, systematic and originating from the normalisation mode~\cite{Aaij:2014kwa}. For the \BToKpipimumu channel, the differential branching fractions illustrated in~\reffig{kpipimumu_LHCb} could be measured.

In the region above the $\Kstarz(892)$ mass around $1430 \mevcc$, the following resonances decaying to a $\Kp\pim$ final state contribute to the spectrum: the S-wave $\kaon_0^*(1430)^{0}$, the P-waves $\Kstar(1410)^{0}$ and $\Kstar(1680)^0$ and the D-wave $\kaon_2^*(1430)^{0}$. The \lhcb collaboration has, for the first time, studied the differential branching fraction of the decay $\decay{\Bz}{\Kp\pim\mup\mun}$ and performed a full angular analysis including S-, P-, and D-wave contributions in the region $1330 \mevcc < m(\Kp\pim) < 1530 \mevcc$~\cite{Aaij:2016kqt}. The former results are shown in~\reffig{kpipimumu_LHCb}; the latter results on the angular observables point towards large interference effects between the S-, P- and D-wave contributions. The fraction of the D-wave was estimated to be smaller than $F_D < 0.29$ at $95\%$ C. L.~\cite{Aaij:2016kqt}.

\subsubsection{$\Bs \to \phiz \ellell$ decays}

\begin{figure}[h]
\begin{minipage}[t]{0.49\linewidth}
\centering
\includegraphics[width=0.98\linewidth]{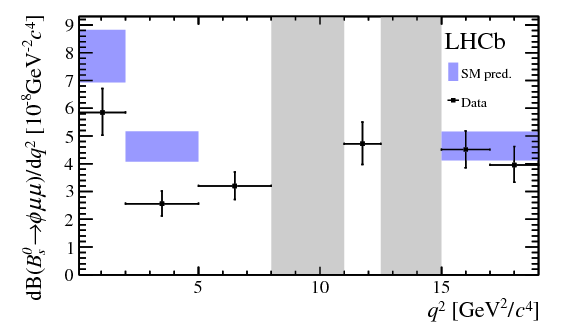}
\end{minipage}
\hspace{\fill}
\begin{minipage}[t]{0.49\linewidth}
\centering
\includegraphics[trim={0cm 0.5cm 0 0}, clip, width=1.025\linewidth]{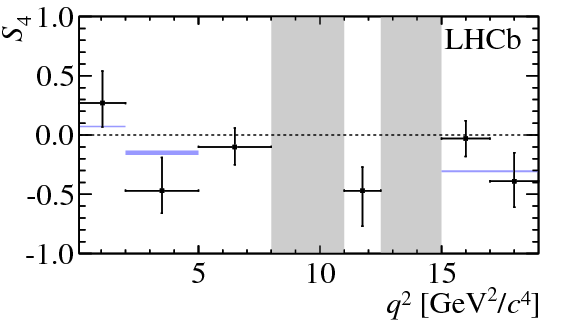}
\end{minipage}
\caption{Differential branching fraction results for \Bstophimumu
  decays depending on \qsq (left) and the angular observable $S_4$
  (right) both with \SM predictions
  overlaid~\cite{Altmannshofer:2014rta,Straub:2015ica}. Figures
  from~\cite{Aaij:2015esa}.} 
 \label{fig:phimumu_LHCb}
\end{figure}

To gain further insight into the observed tensions in \btosll decays, \Bs decays were studied to complement studies of \Bd decays. In the decay channel \Bstophimumu, a full time-integrated angular analysis in line with the previously mentioned \BToKstzmumu study has been performed. In contrast to the self-tagging \BToKstzmumu decays, the $\phiz$ meson decays into a pair of oppositely charged kaons, and hence is not flavour-specific, wherefore the accessible angular observables are $F_{\text{L}}$, the \CP-averaged observables $S_{3,4,7}$ and the \CP asymmetries $A_{5,6,8,9}$. For the first time, the observables $S_4$ (shown in~\reffig{phimumu_LHCb}) and $S_7$ have been measured. This \lhcb analysis comprises a measurement of the differential branching fraction shown in~\reffig{phimumu_LHCb} for which a discrepancy of $3.3\sigma$ between the measured branching fraction and the \SM prediction in the region $1.0  \gevgevcccc < \qsq < 6.0 \gevgevcccc$ is observed~\cite{Aaij:2015esa}. This confirms a tension seen in the branching fraction measured by the previous analysis~\cite{Aaij:2013aln} of about $3.1\sigma$ when comparing to more recent \SM predictions~\cite{Altmannshofer:2014rta,Straub:2015ica} than was referred to in the publication.

\subsubsection{Sum of exclusive \btosll decays}

\begin{figure}[h]
\centering
\includegraphics[width=0.7\linewidth]{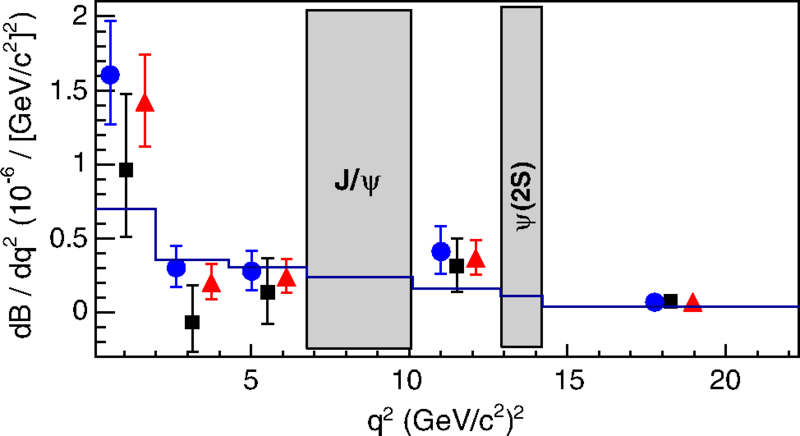}
\caption{Differential branching fraction results for \BXsll decays for electron and muon final states as well as for the flavour-average depending on \qsq with \SM predictions overlaid. Figure from~\cite{Lees:2013nxa}. Due to an incovenient choice of scales, the $\sim 2\sigma$ excess at high \qsq is not visible.}
 \label{fig:Xsll_BF_Babar}
\end{figure}

In contrast to the \lhcb experiment, the \B factories are able to perform inclusive measurements of \eg \BXsll decays. However, inclusive analyses are challenging due to the required full reconstruction of the other \B decay, which significantly reduces the efficiency. Hence, so-called sum-of-exclusive analyses are performed, which use a dominant subset of the corresponding exclusive decays instead. In an analysis of \BXsll decays with $\ell = e, \mu$ on the full \babar dataset, ten different hadronic final states $X_s$ were considered accounting in total for $70\%$ of the fully inclusive rate~\cite{Lees:2013nxa}. In these overall twenty final states, the differential branching fraction of \BXsll decays was measured and by extrapolating this sum-over-exclusive result, the inclusive branching fraction averaged over lepton flavours was determined to be
\begin{align}
\mathcal{B}(\BXsll) = (6.73^{+0.70}_{-0.64} \,^{+0.34}_{-0.25} \pm 0.50)\cdot 10^{-6},
\end{align}
for $\qsq > 0.1 \gevgevcccc$, where the uncertainties are statistical, systematic and originating from the model-dependent extrapolation. A slight excess of $\sim 2\sigma$ is observed in the high \qsq region of the partial branching fraction measurement shown in~\reffig{Xsll_BF_Babar} for both electron and muon final states. From the branching fractions, which are determined for both electron and muon final states as well as the flavour-averaged combination, the direct \CP asymmetry is extracted. By integrating over the \qsq region, the flavour-averaged \CP asymmetry is found to be $\mathcal{A}_{\CP}(\BXsll) = 0.04 \pm 0.11 \pm 0.01$, where the uncertainties are statistical and systematic, respectively~\cite{Lees:2013nxa}, and $\mathcal{A}_{\CP}$ is consistent with the \SM prediction as are the \CP asymmetries in bins of \qsq.\\

\begin{figure}[h]
\centering
\includegraphics[width=0.7\linewidth]{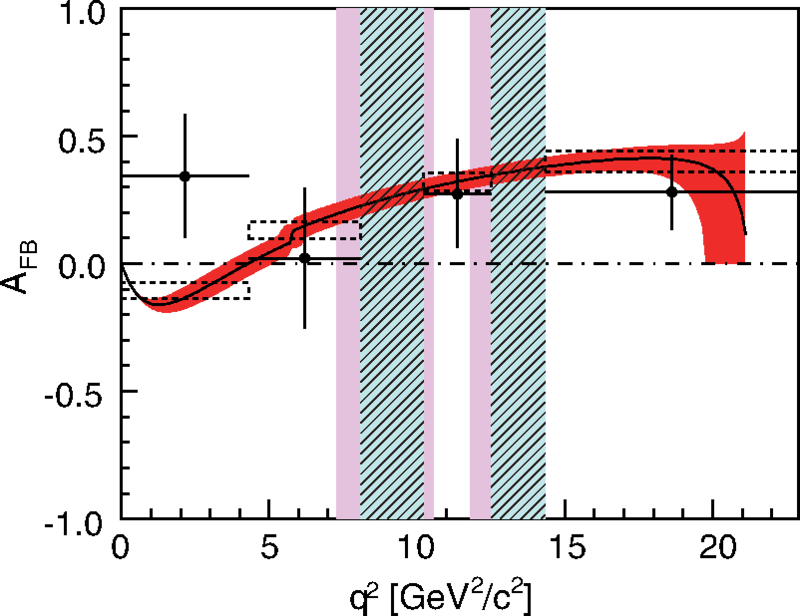}
\caption{Forward-backward asymmetry results for \BXsll decays determined from a sum-of-exclusive approach depending on \qsq with \SM predictions overlaid (red band with black curve and boxes). Figure from~\cite{Sato:2014pjr}.}
 \label{fig:Xsll_AFB_Belle}
\end{figure}

A similar analysis was published by the \belle collaboration on its full dataset, in which the first measurement of $A_{\text{FB}}$ as a function of \qsq was reported~\cite{Sato:2014pjr}, where the reconstruction of the hadronic system extends over ten final states. The sum-of-exclusive results are compatible with \SM predictions as can be seen from~\reffig{Xsll_AFB_Belle}. For $\qsq > 10.2 \gevgevcccc$, $A_{FB} < 0$ is excluded at a level of $2.3\sigma$.

\subsubsection{\Lblll decays}

\begin{figure}[h]
\begin{minipage}[t]{0.49\linewidth}
\centering
\includegraphics[trim={8cm 6cm 0 0}, clip, width=1.1\linewidth]{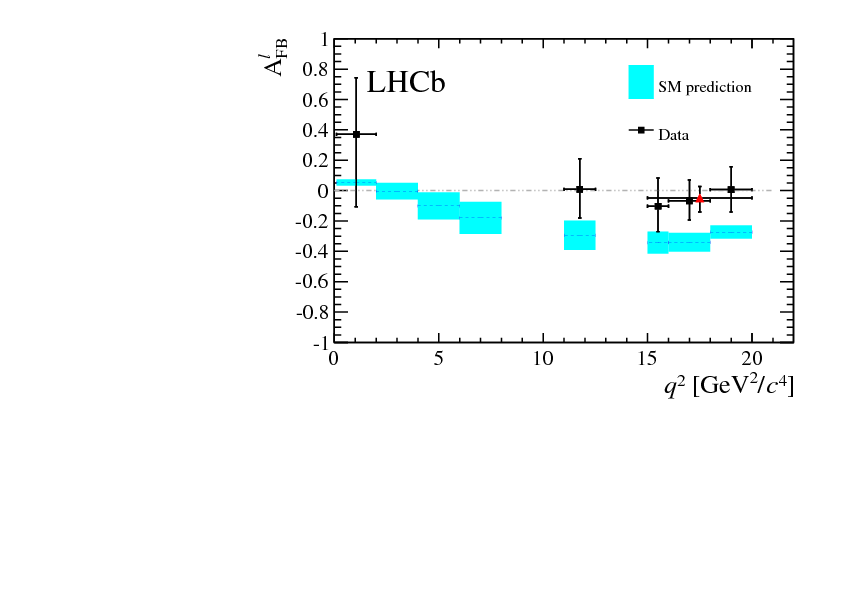}
\end{minipage}
\hspace{\fill}
\begin{minipage}[t]{0.49\linewidth}
\centering
\includegraphics[trim={8cm 6cm 0 0}, clip, width=1.1\linewidth]{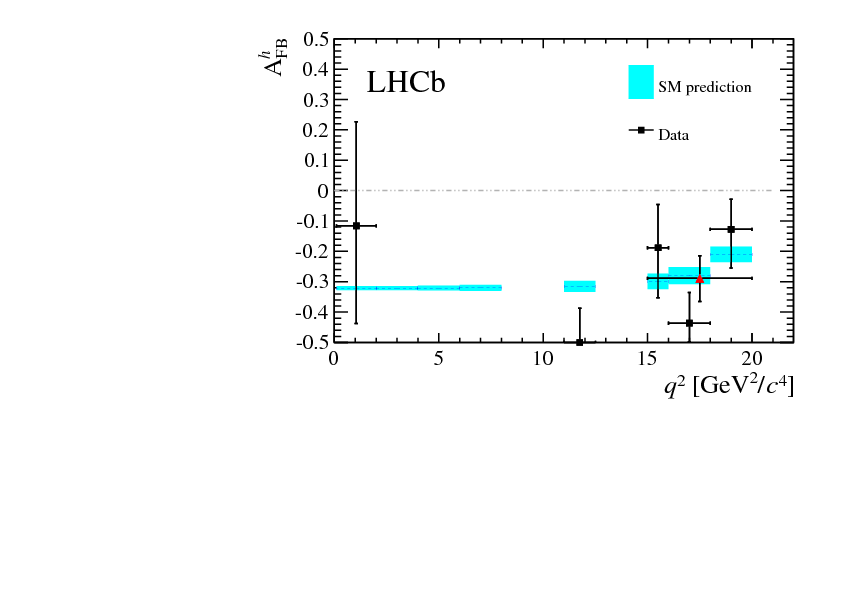}
\end{minipage}
\begin{minipage}[t]{\linewidth}
\centering
\includegraphics[trim={8cm 6cm 0 0}, clip, width=0.55\linewidth]{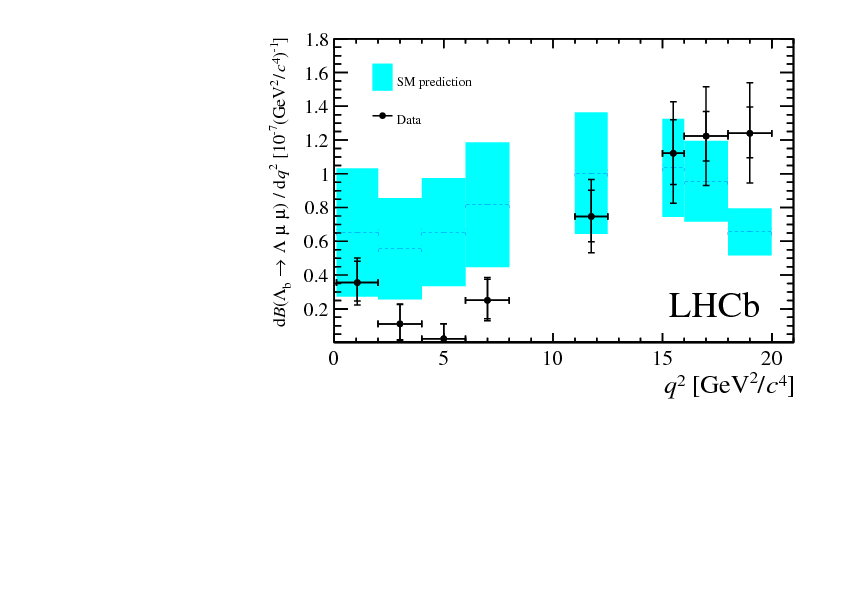}
\end{minipage}
\caption{Results for $A_{\text{FB}}$ of the leptonic system (top left) and hadronic system (top right) and the differential branching fraction (bottom) in \Lbpmunu decays. The \SM predictions~\cite{Meinel:2014wua} are overlaid as band. Figures from~\cite{Aaij:2015xza}.}
 \label{fig:Lmumu_LHCb}
\end{figure}

The \btosll transitions have also been studied in decays of \Lb baryons. Their decay to a ground state \Lz baryons offers a rich phenomenology with ten angular observables in the case of an unpolarised \Lb baryon~\cite{Boer:2014kda}. The \lhcb collaboration has measured the differential branching fraction of  \Lblmumu with \decay{\Lz}{\proton\pim} decays along with the forward-backward asymmetries of the dimuon and the $\proton\pi$ systems denoted by $A_{\text{FB}}^{\ell}$ and $A_{\text{FB}}^{h}$, respectively, as well as the fraction of the longitudinally polarised dimuons, $f_{\text{L}}$~\cite{Aaij:2015xza}. The first significance of a signal is reported in the two regions $0.1  \gevgevcccc < \qsq < 2.0 \gevgevcccc$ and between the charmonium resonances in the range $11.0  \gevgevcccc < \qsq < 12.5 \gevgevcccc$. The angular observables $f_{\text{L}}$ and $A_{\text{FB}}^{\ell}$ are extracted from a fit to the one-dimensional angular distributions as a function of $\cos{\theta_{\ell}}$, which is defined as the angle between the positive (negative) muon and dimuon flight direction in the \Lb (\Lbbar) rest frame. Similarly, the hadronic forward-backward asymmetry is determined from the one-dimensional distribution of $\cos{\theta_{h}}$, where $\theta_h$ is the angle between the proton and \Lz directions. The results on $A_{\text{FB}}^{h}$ are in good agreement with the \SM prediction, whereas the leptonic forward-backward asymmetry lies systematically above the prediction. The results on $A_{\text{FB}}$ along with the differential branching fraction are depicted in~\reffig{Lmumu_LHCb}. The interpretation of these data is challenging and no convincing simultaneous theory explanation of these data and the anomalies in the mesonic modes has been found~\cite{Meinel:2016grj}.

For a subset of the angular observables the impact of all dimension-six \NP contributions
has been worked out in~\cite{Das:2018sms}.
With the future dataset of the \lhcb detector, a novel test of the \SM
could be carried out by studies of polarised versus unpolarised
\Lblll decays~\cite{Blake:2017une}, where the number of angular
observables increases from 10 in the unpolarised case to 34 in the
case of a non-zero production polarisation of the \Lb baryons. 

\subsubsection{Tests of lepton flavour universality}
\label{sec:lfu}

A very clean test for new physics in \btosll decays can be performed by taking ratios of
branching fractions to different lepton species in the final state or by measuring the difference of angular observables across lepton species, \eg in $Q_5$ as has been discussed in~\refsec{Kstll}. Here, the prediction of
lepton flavour universality (LFU) is probed, i.e. the \SM prediction
that weak couplings to all lepton flavours are identical. At the
current experiments, \btosll decays with electrons and muons in the
final state are accessible. For \qsq larger than $1\gev^2$ both muons and electrons are sufficiently relativistic such that uncertainties in the
hadronic form factors cancel to a very good approximation leaving a
\SM prediction with uncertainties well below 1\%~\cite{Hiller:2003js}. Log-enhanced radiative corrections for the decay rates are discussed in~\cite{Bordone:2016gaq} yielding a larger uncertainty $\sim 1\%$. 

In recent years, the interest in lepton flavour universality tests
has increased, mainly driven by two measurements of the LHCb
collaboration: the ratio of \BtoKmumu to \BtoKee, called
\Rk~\cite{Aaij:2014ora}, and the ratio of  \BtoKstmumu to \BtoKstee,
called \Rkst~\cite{Aaij:2017vbb}. The LHCb collaboration uses
basically the same strategy for both analyses, that is discussed here
for general \btosll decays with the corresponding hadron $H$. The lepton flavour universality testing ratio $R(H)$ is then
defined as~\cite{Hiller:2003js}
\begin{align}
R(H) = \frac{\bigintssss \frac{ d\Gamma(B \to H \mumu) }{d\qsq} \, d\qsq}{\bigintssss \frac{ d\Gamma(B \to H\epem) }{d\qsq} \, d\qsq} \, ,
\end{align}
where the differential decay rate is measured in certain \qsq ranges
driven by experimental constraints and theoretical interests. The \qsq range corresponding to
the \jpsi and $\psi (2S)$ is always excluded from the LFU analysis but used as control and normalisation channel, since they are tested to be lepton-flavour universal to very high precision~\cite{Olive:2016xmw}. To cancel experimental uncertainties in
the absolute efficiencies of the measurements, the ratio $R(H)$ is not
measured directly but as double ratio normalising the non-resonant
signal mode to the corresponding high-statistics resonant mode, let us consider a \jpsi resonance in the following. The ratio
$R(H)$ is then measured as 
\begin{align}
R(H) = 
{\frac{\BR(\decay{\B}{H \mumu})}
{\BR(\decay{\B}{H \jpsi(\decay{}{\mumu})})}} 
\bigg{/} 
{\frac{\BR(\decay{\B}{H \epem})}
{\BR(\decay{\B}{H \jpsi(\decay{}{\epem})})}} \, .
\end{align}
A few comments are in order to explain this experimental strategy:
first, this method tests for LFU violations in FCNC decays and it
relies on the conservation of LFU in the corresponding resonant decay
modes. To test this assumption, the ratio of the branching fractions in the resonant channel
\begin{align}
r(\jpsi) = \frac{\decay{\B}{H \jpsi (\to \mumu)}}{\decay{\B}{H \jpsi (\to \epem)}}\, ,
\end{align}
is confirmed to agree with the conservation of lepton flavour
universality. It has to be stressed that this test is more stringent
than required because it cross-checks the absolute ratio of muon to 
electron reconstruction, identification and selection efficiencies
while in the double-ratio of $R(H)$ only relative efficiencies between
non-resonant and resonant channel matter. 
The entire range of \qsq can be tested with this method, if the ratio
$r(\jpsi)$ is measured in bins of the daughter particle momenta.

The most precise measurement of $r(\jpsi)$ has been performed in \lhcb's
analysis of \Rkst, where it was found to be in agreement with unity with a
precision of 4.5\%. Compared to the statistical uncertainties of the
LFU tests of the order of 10\%, this uncertainty is subdominant. For
further tests with enlarged datasets, the precision in the
determination of efficiencies as cross-checked in $r(\jpsi)$ needs to
be studied in greater detail. 

\begin{figure}[h]
\begin{minipage}[t]{0.49\linewidth}
\centering
\includegraphics[width=1\linewidth]{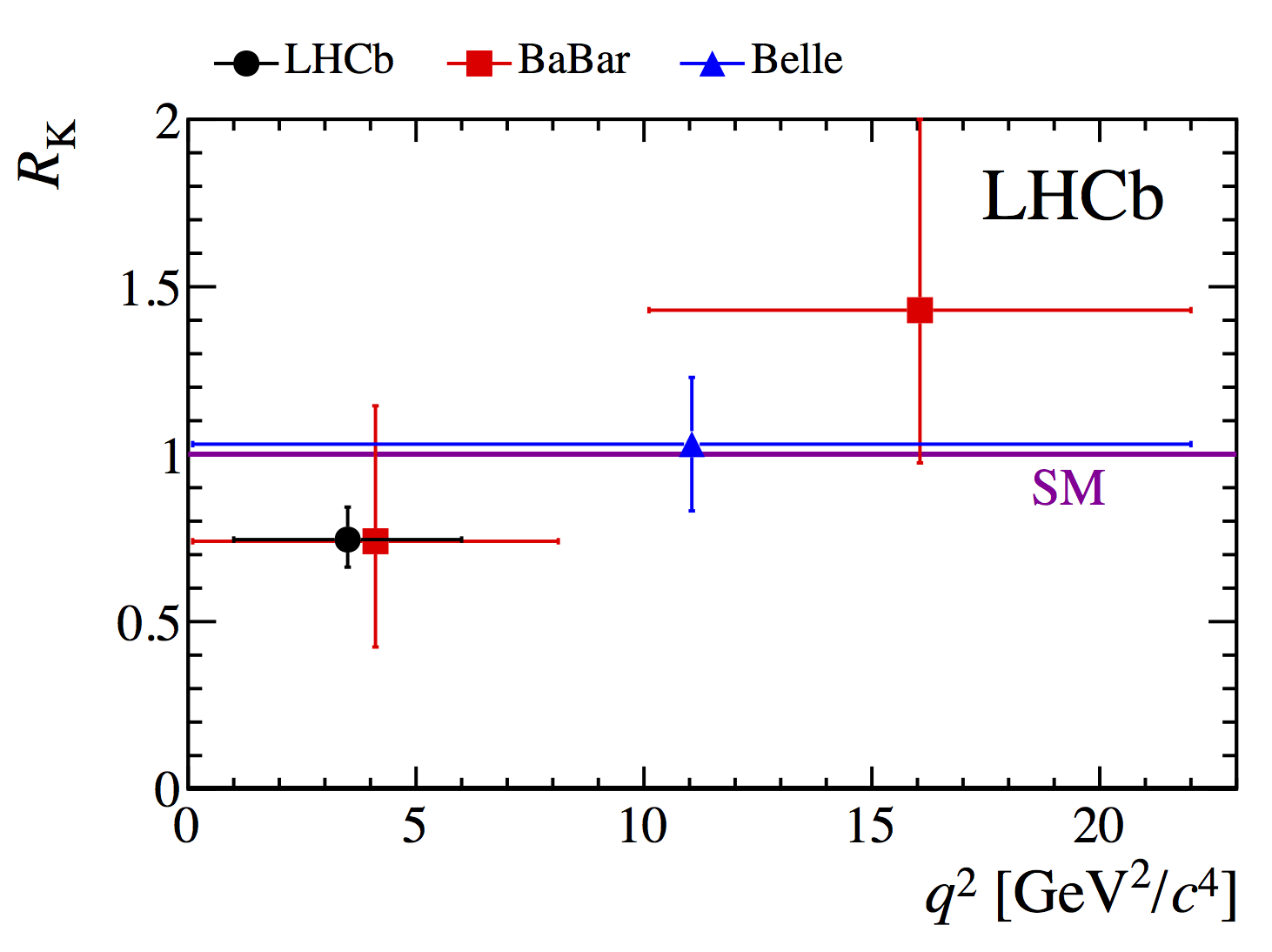}
\end{minipage}
\hspace{\fill}
\begin{minipage}[t]{0.49\linewidth}
\centering
\includegraphics[width=1\linewidth]{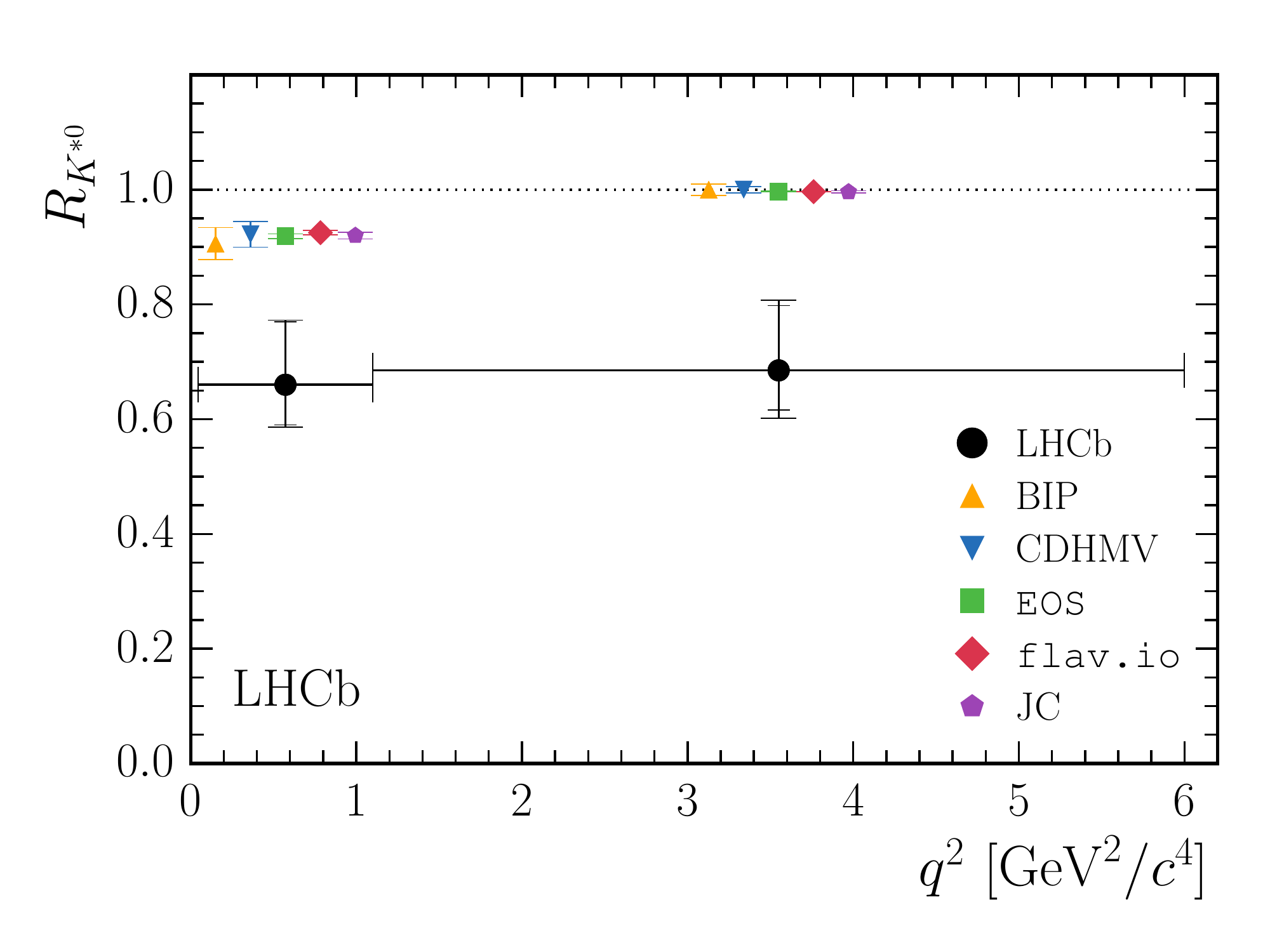}
\end{minipage}
\caption{Tests of LFU in \btosll decays: summary of the measurements of \Rk of the LHCb, \babar and
  \belle experiments with the \SM prediction overlaid as a line at unity (left) and the \lhcb measurement of \Rkst together with several \SM
predictions (right) are shown. Figures from~\cite{Aaij:2014ora,Aaij:2017vbb}.}
 \label{fig:expLFU}
\end{figure}

Among the \btosll decays, the \BtoKll mode is best accessible to the
\lhcb experiment and the \B factories. The former collaboration has
published a measurement using the full Run 1 dataset~\cite{Aaij:2014ora}. 
The value of \Rk is found to be 
\begin{align}
\Rk=0.745^{+0.090}_{-0.074} \pm 0.036,
\end{align}
where the uncertainties are statistical and systematic, and this
measurement constitutes a tension with the \SM
prediction~\cite{Bobeth:2007dw} of 2.6 standard 
deviations. The \babar and \belle experiments have also performed
tests of lepton flavour universality~\cite{Lees:2012tva,Wei:2009zv}
but their analysed dataset is much smaller than the \lhcb dataset and
hence the measurement has significantly increased uncertainties. The
status of all measurements is summarised in~\reffig{expLFU}.   

Further tests of LFU have been carried out with a \Kstarz resonance in the final state; a study in \BToKstzll decays was published recently by the \lhcb collaboration in two bins of \qsq. The results in both \qsq ranges are found below the \SM prediction at
\begin{align}
\Rkst =
\begin{cases}
0.66^{+0.11}_{-0.07}  \pm 0.03     & \textrm{for } 0.045 \gevgevcccc < \qsq < 1.1\gevgevcccc \, , \\
0.69^{+0.11}_{-0.07} \pm 0.05      & \textrm{for } 1.1\phantom{00}\gevgevcccc < \qsq < 6.0\gevgevcccc \, ,
\end{cases}
\end{align}

where the uncertainties are statistical and systematic, respectively. The measurement of \Rkst is shown in~\reffig{expLFU}. The significances of the deviation of the \SM expectation is 2.1-2.3 and
2.4-2.5 standard deviations, respectively. 
The \SM prediction for \Rkst in the lowest bin suffers from an additional source
of theoretical uncertainty due to LFU-violating \SM effects. These effects can for example stem
from almost on-shell hadronic intermediate states that decay at different rates into muons versus
electrons. We refer to~\cite{Bordone:2016gaq} for a discussion.

The data that the \lhcb experiment has collected at the point of this review contains already a
factor three more beauty mesons than in the published LFU analyses on 3\invfb. With the increased sensitivity of the \lhcb detector and the start of the \belle II experiment, the tensions seen in the \Rk and \Rkst measurements will be either confirmed or ruled out in the foreseeable future. 

Additionally to the channels discussed above, LFU can be tested in
$\decay{\Bs}{\phiz \ell^+ \ell^-}$ decays, where a first observation of the
channel $\decay{\Bs}{\phiz \ellp\ellm}$ should be possible already with 3\invfb
of data at \lhcb. Also $\decay{\Bp}{\Kp\pip\pim\ellp\ellm}-$ and
\Lblsll decays are analysed to test for a potential
violation of lepton universality. Combining the already collected
large datasets and the analysis of more channels, the question if LFU
is conserved in the \SM should be conclusively answered in the near
future. A quantitative analysis of the future sensitivities to
discover LFU is discussed in~\refsec{exp-outlook}.

\subsection{A note on \btodll decays}

In contrast to the well tested \btosll decays, that are described
above in~\refsec{btosll}, the decays of the type \btodll are \CKM suppressed by a factor
of about 32~\cite{Olive:2016xmw}. The \btodll system is in principle unconnected to other
systems, wherefore it provides a complementary probe for new physics
effects \eg for the \textit{hypothesis of Minimal Flavour Violation}~\cite{DAmbrosio:2002vsn}, in which these two systems would be connected. Therefore, ratios of
related decays provide a stringent test of this hypothesis.\\

\begin{figure}[h]
\begin{minipage}[t]{0.49\linewidth}
\centering
\includegraphics[width=1\linewidth]{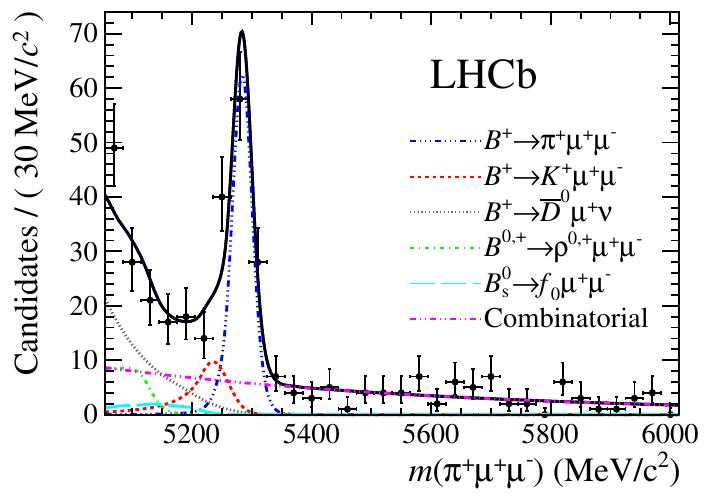}
\end{minipage}
\hspace{\fill}
\begin{minipage}[t]{0.49\linewidth}
\centering
\includegraphics[width=1\linewidth]{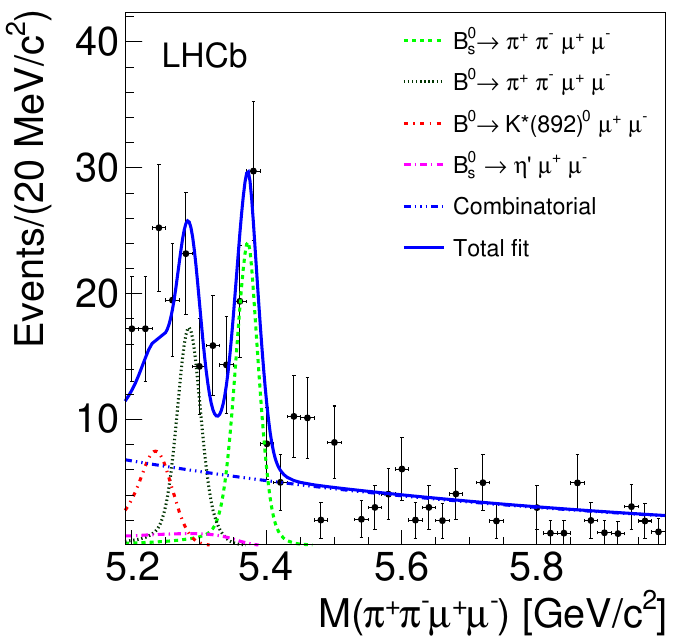}
\end{minipage}
\caption{Fit to the invariant mass distribution of \Butopimumu (left); the separate fit to the charge-conjugated sample is omitted here for reasons of brevity. Fit to the invariant mass distribution of $\decay{\B_{\squark, \dquark}}{\pip\pim\mup\mun}$ candidates (right), where the small contributions from \Bstophimumu and \BToKpmumu are not visible. Figures taken from~\cite{Aaij:2015nea} and~\cite{Aaij:2014lba}, respectively.}
 \label{fig:btodll}
\end{figure}

The first \btodll decay that was observed~\cite{Aaij:2015nea} is
\Butopimumu, where the differential branching ratio was measured with
about one hundred candidates, see~\reffig{btodll}, and the total branching fraction is found to be $\mathcal{B}(\Butopimumu) = (1.83 \pm 0.24 \pm 0.05)\cdot 10^{-8}$, where the uncertainties are of statistical and systematic nature. The publication includes the first test of \CP violation in this decay with the \CP asymmetry being $\mathcal{A}_{\CP}(\Butopimumu) = -0.11 \pm 0.12 \pm 0.01$ where the uncertainties refer to  statistical and systematic uncertainties. Both the differential decay rate and the \CP asymmetry are consistent with  \SM expectations. 
 
The decays $\decay{\B_{\squark, \dquark}}{\pip\pim\mup\mun}$ have also been
measured by the \lhcb collaboration~\cite{Aaij:2014lba}. However, the interpretation of these
decays is more difficult as the hadronically complex structure of the $\pip\pim$ system needs to be considered. The
reconstructed $\pip \pim \mup\mun$ mass distribution shown in~\reffig{btodll}, illustrates this point clearly. In the dipion range of $0.5-1.3 \gevcc$, the first observation of the decay $\decay{\Bs}{\pip\pim\mup\mun}$ with a branching fraction of $\mathcal{B}(\decay{\Bs}{\pip\pim\mup\mun}) = (8.6 \pm 1.5 \pm 0.7 \pm 0.7)\cdot 10^{-8}$ and the first evidence for the decay $\decay{\Bz}{\pip\pim\mup\mun}$ with a branching fraction of $\mathcal{B}(\decay{\Bz}{\pip\pim\mup\mun}) = (2.11 \pm 0.51 \pm 0.15 \pm 0.16)\cdot 10^{-8}$ with the uncertainties being statistical, systematic and originating from the normalisation channel were reported~\cite{Aaij:2014lba}.

An evidence for the decay \BstoKstmumu has recently been seen by the
LHCb collaboration~\cite{Aaij:2018jhg}, using a total
dataset of 4.6\invfb of collision data at $\sqrt s$=7, 8, and 13\tev.
The significance of this evidence is found to be 3.4 standard
deviations. 

In the future, the importance of studies of the decay \btodll will
significantly increase as it allows to measure the same effects as
observed in \btosll decays and thus constitute an independent
verification channel. Furthermore,
ratios of \btodll and \btosll decays provide stringent tests of the
flavour structure of the underlying  interactions and allow to study the hypothesis of minimal flavour
violation~\cite{DAmbrosio:2002vsn}. 

\subsection{Leptonic rare \btoll decays}
\label{sec:btoll}

\begin{figure}
    \centering
    \includegraphics[width=0.85\textwidth]{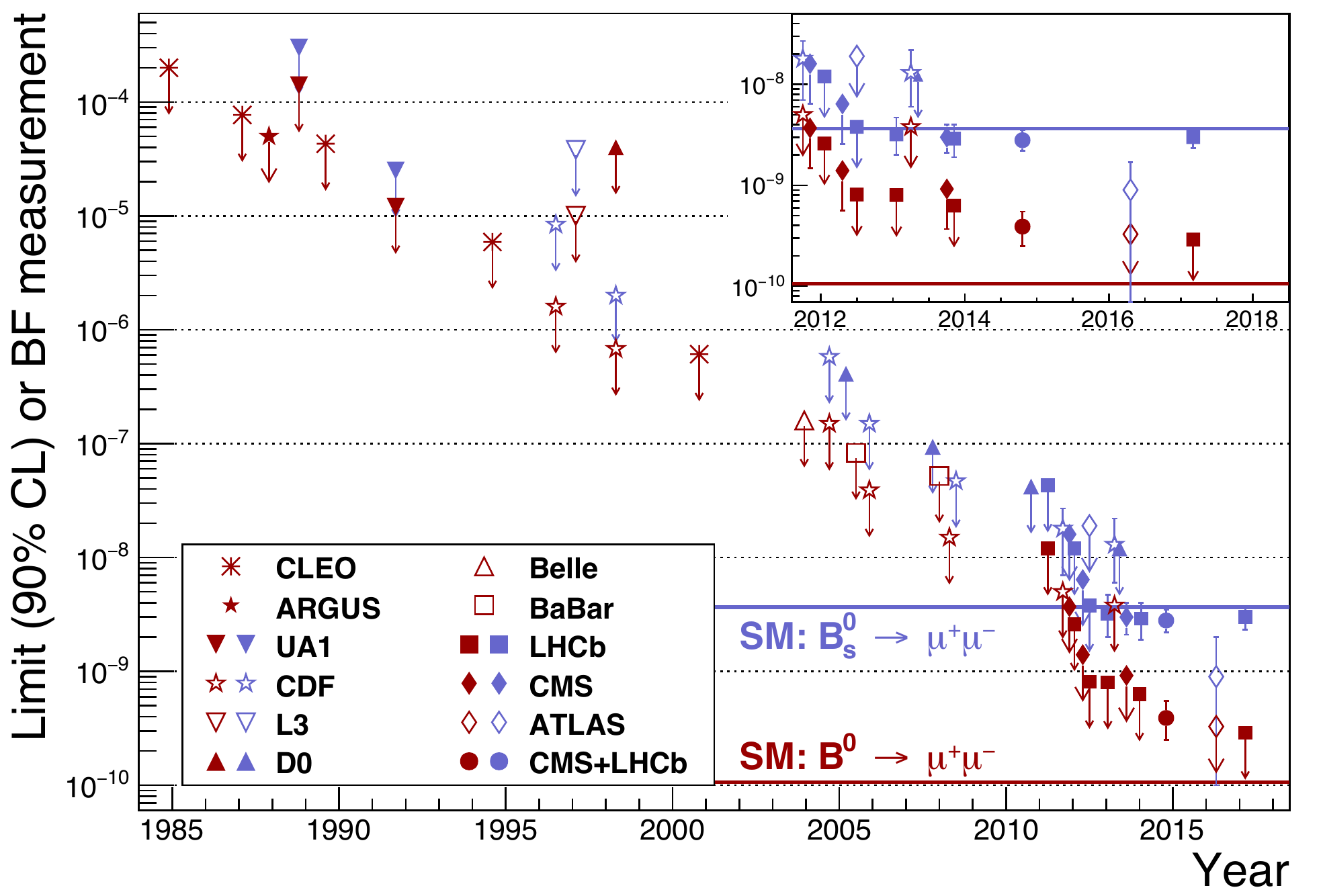}
    \caption{Development of limits and branching fraction measurements
      of the decays \Bsmumu (blue) and \Bdmumu (red) over time. Figure
      from~\cite{CMS:2014xfa}, updated in~\cite{Dettori:2268280}.}
    \label{fig:BsMuMu_history}
\end{figure}

In 1985, the \cleo collaboration published the first limit on \Bdmumu,
whereas UA1 should follow four years later with the first limit in the
\Bsmumu channel; the timeline of limits and measurements is shown in~\reffig{BsMuMu_history} . 
In recent years, the interest in the decays \Bmumu increased as
the \SM rate came into experimental reach. The \SM prediction for the
decay \Bsmm is known with a precision better than 5\% as discussed in
detail in~\refsec{th-had:sm-pred} and lies at
$\mathcal{B}_\text{\SM}(\Bsmumu) = (3.57 \pm 0.17)\cdot
10^{-9}$~\cite{Beneke:2017vpq}. It has to be noted that the
experimentally accessible observable is the time integrated branching
fraction, which differs from the theoretical branching fraction at
$t=0$~\cite{DeBruyn:2012wj} due to \Bs mixing effects. The \Bsll branching fractions discussed in
this chapter are all the time integrated quantities. 
Due to the very small width difference, this effect is negligible for
\Bd decays. 

The corresponding decay \Bdmumu is additionally \CKM-suppressed and therefore
has an \SM rate a factor 32 lower than the \Bs decay.  
Analysing the sensitivity in \NP physics, these modes are
particularly interesting to look for new scalar or pseudo-scalar interactions,
\eg in two Higgs doublet models (2HDM)
type~II~\cite{Cheng:1987rs,DiazCruz:2004tr} large effects are
predicted.

In 2012, the \lhcb collaboration reported a first
evidence of the branching ratio of \Bsmumu, measured to
$\mathcal{B}(\Bsmumu) = (2.9^{+ 
  1.1}_{-1.0})\cdot 10^{-9}$ with a significance of $4.0$ standard
deviations~\cite{Aaij:2013aka}. The first observation of this decay,
exceeding six standard deviations, has been achieved by a joint
analysis of the \cms and \lhcb collaborations~\cite{CMS:2014xfa},
where the branching ratio was determined to be $\mathcal{B}(\Bsmumu) =
(2.8^{+ 0.7}_{-0.6})\cdot 10^{-9}$. 
The \atlas collaboration also published an analysis of 2011 and
2012 data~\cite{Aaboud:2016ire} with a smaller sensitivity than the
\cms or \lhcb collaborations. No significant signal was neither expected nor seen.  
The compatibility of the measurement with the \SM prediction was found
to be at the level of $2.0\sigma$.
The first single
experiment observation of \Bsmumu was reported in 2017 by
the \lhcb collaboration of $\mathcal{B}(\Bsmumu) = (3.0 \pm 0.6 ^{+
  0.3}_{-0.2})\cdot 10^{-9}$ with a significance of 7.8 standard
deviations partly exploiting data from the \lhc's Run
2~\cite{Aaij:2017vad}. In addition to the branching ratio, the first
measurement of the effective lifetime of the \Bs meson, $\tau(\Bsmumu)
= (2.04 \pm 0.44 \pm 0.05)\ps$ was published, which is related to the
parameter  
\begin{align}
A^{\mup\mun}_{\Delta\Gamma} &= -\frac{2\Re{(\lambda)}}{1+|\lambda|^2} \quad \mbox{with} \quad \lambda = \frac{q}{p}\frac{\mathcal{A}(\Bsbmumu)}{\mathcal{A}(\Bsmumu)},
\end{align}
with the complex coefficients $p$ and $q$ connecting mass and flavour
eigenstates in the $\Bs-\Bsb$ system, and $\mathcal{A}$ referring to
the decay amplitudes of the respective process. In certain \NP
models, $A^{\mup\mun}_{\Delta\Gamma}$ ranges between [-1,1]. In the
\SM, only a pseudo-scalar amplitude is allowed and hence
$A^{\mup\mun}_{\Delta\Gamma} \approx 1 - 1.0\cdot 10^{-5}$; the
deviation of unity is caused by power-enhanced QED
corrections~\cite{Beneke:2017vpq}. The effective lifetime $\tau(\Bsmumu)$ is an
orthogonal probe of \NP models compared to the \Bsmumu branching fraction as
it is sensitive to models predicting the branching fraction close to its \SM
value. However, the current experimental sensitivity is insufficient to make a
definitive statement on the validity of the \SM or \NP models, but
$\tau(\Bsmumu)$ will become more important in the future when larger datasets
become available. The \lhcb measurement is consistent
with the hypothesis of $A^{\mup\mun}_{\Delta\Gamma} = +1 (-1)$ at 1.0
(1.4) standard deviations.\\

The experimental results discussed in~\refsec{lfu} hint at a
violation of Lepton Flavour Universality, therefore the non-muonic
decays \Bsee and \Bstautau offer an orthogonal probe for NP. 
Some models predict an enhancement of \Bsee over \Bsmm proportional
to $m_{\mu}/m_e$~\cite{Fleischer:2017ltw}. Also the branching fraction of
the decay \Bstautau could be enhanced with respect to the SM
prediction by several orders of magnitude~\cite{Cline:2015lqp,Becirevic:2016yqi}.


Experimentally, both \Bsdtoee and \Bsdtotautau
final states are more challenging than the muonic mode: when
reconstructing the di-electron final state, bremsstrahlung photons have
to be accounted for, and hence the mass resolution is significantly
degraded and the physical backgrounds are more difficult to
determine. 
In the tauonic decay channel, the decays of
the $\tau$ leptons render the measurement challenging as they involve at least two neutrinos, and hence can only be reconstructed if additional information on the \B momentum is available. 

The current world's best upper limits on
\Bsdtoee stem from the CDF collaboration~\cite{Aaltonen:2009vr}, which
determined the following limits at $90\%$ C.L.  
 \begin{align}
 \mathcal{B}(\Bsee) &< 2.8 \cdot 10^{-7},\\
 \mathcal{B}(\Bdee) &< 8.3 \cdot 10^{-8}.
 \end{align}
The \SM predictions are at the level of $10^{-13}$ (see~\cite{Bobeth:2013uxa}) and hence not
within experimental reach. Searches for \Bsdtoee are therefore a clean
test for effects beyond the Standard Model, as no \SM contribution is
measurable and any observed signal would be a clean signal for New
Physics. A search for \Bsee using the LHCb data is in preparation.\\

 A first upper limit on \Bstautau was recently published by the \lhcb
collaboration~\cite{Aaij:2017xqt} together with a limit on \Bdtautau
improving the previous limits. The limits at $95\%$
C.L. are 
 \begin{align}
 \mathcal{B}(\Bstautau) &< 6.8 \cdot 10^{-3},\\
 \mathcal{B}(\Bdtautau) &< 2.1 \cdot 10^{-3},
 \end{align}
 whereas the \SM predictions are around four to five orders of
 magnitude below the current experimental sensitivity at
 $\mathcal{B}(\Bstautau)=(7.73 \pm 0.49)\cdot 10^{-7}$ and
 $\mathcal{B}(\Bdtautau)=(2.22 \pm 0.19)\cdot 10^{-8}$,
 respectively~\cite{Bobeth:2013uxa}. Therefore, also in \Bsdtotautau
 decays, any observed signal would be a clear sign of New Physics. The
 sensitivity of \Bsdtotautau decays is expected to greatly improve in
 the upgrade of the \lhcb experiment, profiting from the improvements
 in the trigger strategy.

\subsection{Exclusive \btosgamma decays}

Over the past years, the advancements in exclusive \btosgamma decays
have been sparse, and other reviews, see \eg~\cite{Blake:2015tda},
summarise the measurements by the \B factories, wherefore we restrict
this review to the most recent measurements. The \belle collaboration
reported a measurement of the \BtoKstgamma branching fractions,
$\mathcal{B}(\BtoKstzgamma) = (3.96 \pm 0.07 \pm 0.14)\cdot 10^{-5}$
and $\mathcal{B}(\BtoKstpgamma) = (3.76 \pm 0.10 \pm 0.12)\cdot
10^{-5}$~\cite{Horiguchi:2017ntw}, which only place weak constraints
on possible \NP scenarios due to the large form factor
uncertainties. Stronger constraints are obtained by measuring the
direct \CP asymmetry $\Delta\ACP\equiv (\ACP(\BtoKstpgamma) -
\ACP(\BtoKstzgamma)) = (2.4 \pm 2.8 \pm 0.5)\%$ with
$\ACP(\BtoKstzgamma) = (-1.3 \pm 1.7 \pm 0.4)\%$ and
$\ACP(\BtoKstpgamma) = (+1.1 \pm 2.3 \pm 0.3)\%$. The average of the
\CP asymmetries is $\bar{\ACP} = (-0.1 \pm 1.4 \pm
0.3)\%$~\cite{Horiguchi:2017ntw}. This result comprises, for the first
time, evidence of isospin violation (defined similarly to~\refeq{AI}) at a significance of $3.1\sigma$,
$\Delta_{0+} = (+6.2 \pm 1.5 \pm 0.6 \pm 1.2 \%$, where the uncertainties are statistical, systematic and the third uncertainty arises from the
production ratio of charged to neutral \bbbar
pairs~\cite{Horiguchi:2017ntw}; this result is consistent with the \SM
expectation.
We recall that measurements of isospin asymmetries in the decays crucially depend on
accurate determinations of isospin asymmetries in the production. The consequences
of a bias-free determination of the production asymmetry and its consequences on
$\Delta_{0+}$ are discussed in~\cite{Jung:2015yma}.
From the measured \BtoKstzgamma branching fraction,
comprising solely the charged $\decay{\Kstarz}{\Kp\pim}$ channel in
order to reduce systematic uncertainties, the ratio  
\begin{align}
\frac{\mathcal{B}(\BtoKstzgamma)}{\mathcal{B}(\Bstophigamma)} = 1.10 \pm 0.16 \pm 0.09 \pm 0.18,
\end{align}
was determined, where the uncertainties are statistical, systematic
and the third uncertainty comes from the $\B_s^{(*)0}\Bb_s^{(*)0}$ production in \FiveS decays~\cite{Horiguchi:2017ntw}. Here, the $f_s$
related uncertainty arises from the measurement of $\mathcal{B}(\Bstophigamma) = (3.6 \pm 0.5 \pm 0.3 \pm 0.6)\cdot
10^{-5}$~\cite{Dutta:2014sxo}. The results on the \btosgamma branching
fraction ratio are consistent with both the \SM expectations and the \lhcb
result~\cite{Aaij:2012ita}. In \Bstophigamma decays, the photons are
predominantly left-handed with a right-handed component suppressed by
the ratio of the \squark to \bquark masses, where the latter component
could be enhanced \NP scenarios. Information on the polarisation can for example
be obtained from \eg \BToKstzee decays(see~\refsec{btosll} for details), which probe the photon polarisation at low \qsq, or from a measurement of the parameter $A^{\Delta}$. The latter is
a function of the left- and right-handed polarisation amplitudes given
by $A^{\Delta} = \sin (2\Psi)$ with $\tan(\Psi) \equiv
|A(\decay{\Bsb}{\phi\g_R})|/|A(\decay{\Bsb}{\phi\g_L})|$ in the
\SM. The \lhcb collaboration has recently reported the first
time-dependent measurement of a radiative \Bs decay, and has
determined $A^{\Delta} = -0.98^{+0.46}_{-0.52}\, ^{+0.23}_{-0.20}$ in
\Bstophigamma~\cite{Aaij:2016ofv}, which is consistent with the \SM
expectation within two standard deviations.\\ 

In contrast to \Bstophigamma, the decay \Bztophigamma is heavily
suppressed and the current most precise limit is reported by the
\belle collaboration as $\mathcal{B}(\Bztophigamma) < 1.0\cdot
10^{-7}$ at $90\%$ C.L.~\cite{King:2016cxv}. As this decay is not
sensitive to the electromagnetic dipole operator $\Oi{7}$ but to the
QCD-penguin operators $\Oi{3-6}$, we refrain from a more detailed
discussion.

\section{Theory of rare \btosll decays}
\label{sec:th-had}

As discussed in~\refsec{amplitudes}, the description of rare $b$ decays
involves hadronic matrix elements of various local and non-local operators.
We discuss these matrix elements in the following in~\refsec{th-had:local}
and~\refsec{th-had:non-local}. Subsequently in~\refsec{th-had:sm-pred}, we will revisit the theoretical foundations of the observables discussed in~\refsec{exp}.

\subsection{Hadronic matrix elements of local operators}
\label{sec:th-had:local}

For the purely leptonic decays of $B_s$ mesons and at leading order in $\alpha_e$,
the only occuring hadronic matrix element emerges from the axialvector current
\begin{equation}
    \bra{0} \bar{s}\gamma^\mu \gamma_5 b\ket{\bar{B}_s(p)}
        \equiv i p^\mu f_{B_s}\,,
\end{equation}
which introduces the $B_s$ meson decay constant $f_{B_s}$.
The hadronic matrix elements of the operators $\Oi{9,9',7,7'}$ are
suppressed by one power of $\alpha_e$; see~\cite{Beneke:2017vpq}
for a recent study to that effect.\\

Presently, the only ab-initio method to obtain the $B_s$ decay constant is lattice QCD. 
For a discussion of the inherent systematic uncertainties in lattice
determinations of the decay constant we refer to the most recent version of the
FLAG report~\cite{Aoki:2016frl}.  For determinations with $N_f = 2 + 1 + 1$
dynamical light quark flavours, there are presently three results by the European
Twisted Mass collaboration (ETM)~\cite{Carrasco:2013naa}, the HPQCD
collaboration~\cite{Dowdall:2013tga}, and the Fermilab/MILC collaborations (FNAL/MILC)~\cite{Bazavov:2017lyh}.The lattice results read:

\begin{equation}
\begin{aligned}
    f_{B_s} & = 235 \pm 9\,\MeV\,            & \text{(ETM~\cite{Carrasco:2013naa})},\\
    f_{B_s} & = 224 \pm 5\,\MeV\,            & \text{(HPQCD~\cite{Dowdall:2013tga})},\\
    f_{B_s} & = 230.7 \pm 1.2\,\MeV\,        & \text{(FNAL/MILC~\cite{Bazavov:2017lyh})}.
\end{aligned}
\end{equation}

Besides lattice simulations, the decay constants are also accessible through
QCD Sum Rules (QCDSR). The latter are based on an approach based on an \OPE of an inclusive
quantitiy, and relating it to exclusive matrix elements in a dispersive framework.
A large, and difficult-to-quantify systematic uncertainty arises from the use of quark-hadron
duality. The QCDSR results are compatible with the lattice determination, albeit with
uncertainties roughly one order of magnitude larger than the most precise lattice determination:
\begin{equation}
\begin{aligned}
    f_{B_s} & = 234^{+15}_{-11}\,\MeV\,      & \text{(QCDSR~\cite{Gelhausen:2014jea})}.
\end{aligned}
\end{equation}

For the description of the semileptonic $b$-hadron decays, the matrix elements of the local
operators are decomposed in terms of \emph{form factors}: scalar-valued functions of $q^2$, the
square of the momentum transfer to the dilepton system. A commonly used basis consists of three form factors
$f_+$, $f_-$, and $f_T$ for transitions to a single pseudo-scalar meson (see \eg~\cite{Ball:2004ye}), or seven form factors
$V$, $A_{0,1,2}$, $T_{1,2,3}$ for transitions to a single vector meson (see \eg~\cite{Ball:2004rg}).
Transformations of these common bases to the so-called helicity bases~\cite{Bharucha:2010im}
can be convenient for the representation of the theoretical results as is \eg discussed in~\cite{Bobeth:2010wg}. They are also beneficial for the determination of form factor parameters from lattice simulations~\cite{Horgan:2013hoa}. 
In order to parametrize the $q^2$-dependence of the form factors, one relies on their analytic properties~\cite{Boyd:1995sq}.
Their spectral densitites contain potential poles below the crossed-channel
pair production threshold; \eg for the $B \to K$ vector form factor, there is a single pole
contribution from the $B_s^*$ meson with mass $M_{B_s^*}^2 < (M_B + M_K)^2$. Moreover, with the
onset of the pair production the form factors feature a branch cut extending to $q^2 \to +\infty$.
Following~\cite{Boyd:1995sq} and later work, these properties can be incorporated into the
parametrization of the form factors by virtue of a conformal mapping from $q^2$ to a variable $z(q^2; t_+, t_0)$.
The mapping automatically accounts for the first cross-channel branch cut when equating
$t_+ \equiv (M_B + M_K)^2$. A suitable choice of $t_0$ minimizes the value of $|z|$ in the
physical $q^2$ phase space for semileptonic decays~\cite{Bourrely:2008za}.
The common paradigm~\cite{Bharucha:2010im,Straub:2015ica} to parametrize the
form factors is then to first remove the sub-threshold poles through an explicit
factor; and second to expand the remaining functions as a Taylor series around $z = 0$. Our present knowledge of the hadronic local matrix elements can be summarised as follows:

\begin{itemize}
    \item In the case of $B\to K$ transitions, the form factors are available from both lattice simulations
   ~\cite{Bouchard:2013pna}; Light Cone Sum Rules (LCSRs) with kaon distribution amplitudes~\cite{Ball:2004ye};
    and LCSRs with $B$-meson Light Cone Distribution Amplitudes (LCDAs)~\cite{Khodjamirian:2006st}.
    Either of these results, or combinations thereof have been used in phenomenological
    applications. A new analysis using LCSRs with $B$-meson distribution amplitudes
    is pending~\cite{Gubernari:2018}.\\

    \item The case of $B\to K^*$ form factors is more complicated since the $K^*$ is not
    an asymptotic state of QCD. For high-precision analyses, the effects of its finite
    width have to be taken into account. Lattice results in the narrow-width approximation are available from~\cite{Horgan:2013pva}.
    Updated LCSR results for $K^*$-LCDAs in the narrow-width approximation are provided in~\cite{Straub:2015ica}, and a
    combination of lattice QCD and $K^*$-LCSR results have been published in~\cite{Straub:2015ica}. $B$-LCSR~\cite{Khodjamirian:2006st,Cheng:2017smj} results do not reach the same level of sophistication
    as $K^*$-LCSRs, and and updates are expected this year~\cite{Gubernari:2018,Descotes-Genon:2018}.\\

    \item For the case of $B_s\to\phi$ transitions, the full set of form factors is available
    from lattice simulations~\cite{Horgan:2013pva}, and LCSRs with $\phi$-LCDAs~\cite{Straub:2015ica}.\\

    \item The most precise results on $\Lambda_b\to \Lambda$ transition form factors are available
    from lattice simulations~\cite{Detmold:2016pkz}.
    Results on the form factors from LCSRs with $\Lambda$-LCDAs are available from~\cite{Mannel:2011xg}.
    For Soft Collinear Effective Theory Sum Rule results from $\Lambda_b$-LCDAs we refer to~\cite{Feldmann:2011xf},
    with the complete NLO $\alpha_s$ corrections to the leading $\Lambda_b$-LCDAs
    subsequently published in~\cite{Wang:2015ndk}.\\
\end{itemize}

\subsection{Hadronic matrix elements of non-local operators}
\label{sec:th-had:non-local}

Beside the local semileptonic operators $\Oi{9,10}$ and their
counterparts in \NP scenarios, also the operators $\Oi{1\dots 6,8}$ contribute
significantly to the amplitudes. On one hand, the RGE already shows us that
roughly $50\%$ of the value of $\Ci{9}$ at the low scale is generated through
the RGE mixing of the operators $\Oi{1c,2c}$ into $\Oi{9}$. Consequently, this
generates a numerically significant scale dependence of $\Ci{9}$ that can only
be compensated through the scale dependence of the hadronic matrix elements of
these two operators. \footnote{%
    In principle this is also true for the scale dependence of $\Ci{7}$. However,
    here the short-distance mixing of the four-quark operators into the
    electromagnetic operator $\Oi{7}$ are suppressed by $\alpha_s$.
}

The latter arise from a \emph{non-local operator}, the time-ordered product with the electromagnetic current $J_\text{e.m.}$:
\begin{equation}
    \mathcal{H}^\mu = i \int \dd[4] x\, e^{i q\cdot x}
        \bra{H_s(k)} \mathcal{T} \lbrace J_\text{e.m.}^\mu(x),  \sum_{i} \Ci{i} \Oi{i}(0) \rbrace \ket{H_b(p)}\,,
\end{equation}
where the sum runs over $i=1c,2c,1u,2u,3\dots 6, 8$. We shorten the following
discussion by only considering non-local effects of the operators $\Oi{8}$ and $\Oi{1c,2c}$, which enter the amplitudes numerically unsuppressed by either small \CKM matrix elements or small \WC.\\

To leading power in the heavy quark expansion, the
effects due to $O_8$ can be described in the framework of QCD factorisation (QCDF)~\cite{Kagan:2001zk,Beneke:2001at}
as $\order({\alpha_s})$ contributions.
When going to subleading power, QCDF breaks down
due to endpoint-divergent contributions from terms corresponding to photon emissions from the spectator
quark~\cite{Kagan:2001zk}. This problem can presently only be overcome in the frameork of LCSRs~\cite{Dimou:2012un}.
According to~\cite{Dimou:2012un}, the contributions can be written
as generalized and complex-valued form factors for meson transitions. The generalized form factors
are roughly of the same size as the local matrix elements of $\Oi{7}$, however, with large phases.
They enter the amplitudes with a factor two suppression since $\Ci{7}(\mu_b) \simeq 2 \Ci{8}(\mu_b)$.
Additionally, the form factors are further suppressed with increasing $q^2$~\cite{Dimou:2012un}.\\

\begin{figure}
    \centering
    \subfigure[%
        Assuming the absence of a branch cut due to light hadrons, with bound states $J/\psi$
        and $\psi(2S)$.
        \label{fig:4quark-analytic-structure-wo-light-hadron-cut}
    ]{
    \resizebox{.9\textwidth}{!}{%
    \begin{tikzpicture}
        \begin{axis}
        [
            disabledatascaling,
            axis background/.style={%
                postaction={
                    path picture={
                    }
                }
            },
            axis x line=bottom,
            axis y line=left,
            xmin=-4, xmax=25,
            xlabel={$\operatorname*{Re}q^2$ [GeV$^2$]},
            x label style={right,at={(axis description cs:0.5,-0.1)}},
            x=0.4cm,
            ymin=-0.5, ymax=+0.5,
            ylabel={$\operatorname*{Im}q^2$ [GeV$^2$]},
            yticklabel={0}, ytickmin=0, ytickmax=0,
            y label style={above,at={(axis description cs:+0.07,0.5)}},
            y=2.5cm
        ]
            \draw[draw=black!50!white,thick] (axis cs:-4.0, 0.0) -- (axis cs:+25.0, 0.0);

            \addplot[only marks,mark=*] coordinates { ( 9.59,  0.00) }; 
            \addplot[only marks,mark=*] coordinates { (13.59,  0.00) }; 

            \addplot[only marks,mark=o] coordinates { (14.24, -0.30) }; 
            \addplot[only marks,mark=o] coordinates { (16.31, -0.32) }; 
            \addplot[only marks,mark=o] coordinates { (17.31, -0.33) }; 
            \addplot[only marks,mark=o] coordinates { (19.49, -0.27) }; 

            \draw[draw=none,fill=red,fill opacity=1.0]
                (axis cs:+13.89,-0.10) rectangle (axis cs:+25,+0.10);

        \end{axis}
    \end{tikzpicture}
    }
    }\\
    \subfigure[
        In the presence of a branch cut due to light-hadrons, without any bound states.%
        \label{fig:4quark-analytic-structure-full}
    ]{
    \resizebox{.9\textwidth}{!}{%
    \begin{tikzpicture}
        \begin{axis}
        [
            disabledatascaling,
            axis background/.style={%
                postaction={
                    path picture={
                    }
                }
            },
            axis x line=bottom,
            axis y line=left,
            xmin=-4, xmax=25,
            xlabel={$\operatorname*{Re}q^2$ [GeV$^2$]},
            x label style={right,at={(axis description cs:0.5,-0.1)}},
            x=0.4cm,
            ymin=-0.5, ymax=+0.5,
            ylabel={$\operatorname*{Im}q^2$ [GeV$^2$]},
            yticklabel={0}, ytickmin=0, ytickmax=0,
            y label style={above,at={(axis description cs:+0.07,0.5)}},
            y=2.5cm
        ]
            \draw[draw=black!50!white,thick] (axis cs:-4.0, 0.0) -- (axis cs:+25.0, 0.0);

            \addplot[only marks,mark=o] coordinates { ( 9.59, -0.02) }; 
            \addplot[only marks,mark=o] coordinates { (13.59, -0.02) }; 
            \addplot[only marks,mark=o] coordinates { (14.24, -0.30) }; 
            \addplot[only marks,mark=o] coordinates { (16.31, -0.32) }; 
            \addplot[only marks,mark=o] coordinates { (17.31, -0.33) }; 
            \addplot[only marks,mark=o] coordinates { (19.49, -0.27) }; 

            \draw[draw=none,fill=blue!50!white,fill opacity=0.9]
                (axis cs:  0.16,-0.12) rectangle (axis cs:+25,+0.12);

            \draw[draw=none,fill=red,fill opacity=0.9]
                (axis cs:+13.89,-0.10) rectangle (axis cs:+25,+0.10);

        \end{axis}
    \end{tikzpicture}
    }
    }
    \caption{%
        Illustration of the analytic structure of the non-local matrix elements
        of $bsc\bar{c}$ operators, in the complex $q^2$ plane.
        Bound states  are shown as filled circles; resonances are shown as empty circles
        off the real axis, respectively; and the $D\bar{D}$ branch cut is depicted as the red area
        extending from $4 M_D^2$ to $+\infty$.
        The blue shaded area depicts the branch cut due to light hadrons (\eg $\phi$, $K^+K^-$),
        starting at $4 M_\pi^2$.
        These figures originates from auxilliary materials associated with~\cite{Bobeth:2017vxj}.
    }
    \label{fig:4quark-analytic-structure}
\end{figure}
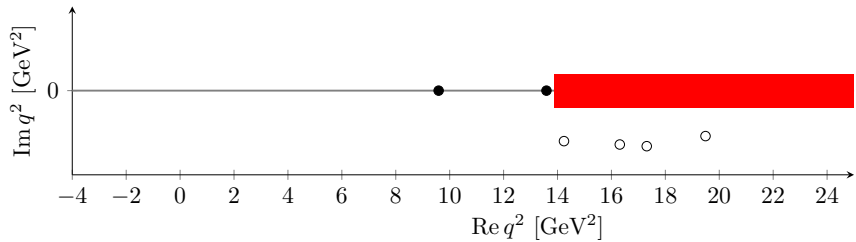
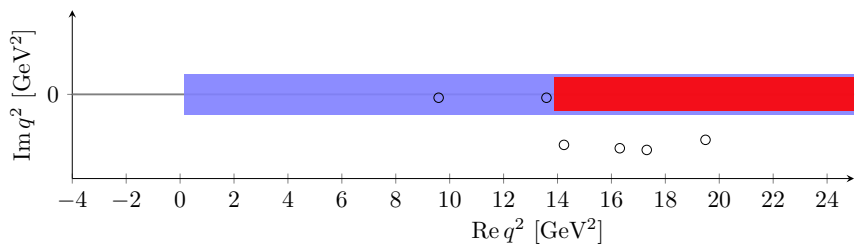

The four-quark operators are both phenomenologically important and conceptually interesting,
since they give rise to hadronic decays of the form $H_b \to H_s V (\to \gamma^* \to \ell^+\ell^-)$.
Here $V$ denotes any state with quantum numbers $J^{PC} = 1^{--}$, which can resonantly
contribute to the decay rate. A special case here are the $\Oi{1c,2c}$ contributions, which
enter with \WC{}s $\Ci{1c,2c} = \order(1)$ and at the same level of the \CKM
Wolfenstein parameter $\lambda$ as the short-distance semileptonic contributions. To illustrate
the problem, consider the measurements of the branching ratio for the decay $\BtoKstmumu$ in the $q^2$ bin
from $1.1\,\GeV^2$ to $6\,\GeV^2$, which yields $\sim 0.3 \cdot 10^{-7}$, while the resonant hadronic decays $\BtoKstJpsi$ and $\BtoKstpsip$ yield $\sim 1.3 \cdot 10^{-3}$
and $\sim 0.6 \cdot 10^{-3}$, respectively. These resonant enhancements are denoted as either poles or resonances
in the complex $q^2$ plane in~\reffig{4quark-analytic-structure}.

Early works on the theoretical description of $\btosll$ decays assumed factorisation of
the $b\to s$ quark current from the $c\bar{c}$ loop -- usually referred to as \emph{naive factorisation}.
A first improvement within the concept of naive factorisation was achieved by Kr\"uger and Sehgal
through replacement of the LO quark-loop function
by a resonant long-distance contribution extracted from $e^+e^- \to \text{hadrons}$ on the $J/\psi$
and $\psi(2S)$ resonances as well as above the $D\bar{D}$ threshold~\cite{Kruger:1996cv}.
The relevant parameters for the description can be inferred from contemporary $R$ ratio measurements~\cite{Olive:2016xmw}, where
\begin{align}
R = \frac{\sigma(\decay{\ep\en}{hadrons})}{\sigma(\decay{\ep\en}{\mup\mun})}\,.
\end{align}

A further theoretical improvement was achieved by embedding the four-quark contributions within
the frame of QCD factorisation~\cite{Beneke:2001at,Beneke:2004dp}, which holds for $q^2 \ll M_{J/\psi}^2$.
The authors were able to show that all contributions to leading power in a combined expansion of local operators
in $1/m_b$ and $1/E$, with $E$ the
hadron energy in the $B$ rest frame, can be expressed either through local hadronic form factors or convolutions of hard-collinear kernels with $B$-meson and light-meson LCDAs. Note here that the restriction on the allowed $q^2$ phase space arises both from the requirements
of the power expansion in QCDF, and from estimating a breakdown of the local \OPE close to the $J/\psi$ pole.
The latter estimate has been questioned more recently in~\cite{Khodjamirian:2010vf,Khodjamirian:2012rm}.
By using a Light-Cone \OPE (LCOPE), the authors estimate that formally power-suppressed soft contributions become
numerically relevant for $q^2 \centernot\ll 4 m_c^2 \simeq 4\,\GeV^ 2$. The leading-power contribution
to the LCOPE corresponds to the local \OPE such that the QCDF results are completely recovered.
At next-to-leading power, corrections to the local \OPE from operators involving one soft gluon field can be
incorporated. The relevant non-local hadronic matrix elements were estimated from LCSRs.
The approximate symmetry of the QCDF results between the two transverse polarization states, arising in the limit of
large $\kaon^*$ energy in the \B-meson rest frame, can be exploited phenomenologically. A large number of
optimised observables have been designed that exhibit a reduced dependence on the local form factors
~\cite{Egede:2008uy,Altmannshofer:2008dz,Egede:2010zc,Becirevic:2011bp}.
The now standard basis of optimised observables is $\lbrace P_{1,2,3}, P'_{4,5,6}\rbrace$~\cite{DescotesGenon:2012zf}.

For the phase space with $q^2 \simeq m_b^2$, the four-quark operators can be
treated in a local \OPE~\cite{Grinstein:2004vb,Beylich:2011aq}, commonly
referred to as the Low Recoil \OPE (LROPE). Due to the universality of the LROPE
results for the various $K^*$ and dimuon polarisation states, it gives rise to a
rich phenomenology and a set of optimised observables~\cite{Bobeth:2010wg,Bobeth:2011gi,Bobeth:2011nj,Bobeth:2012vn,%
Hambrock:2013zya,Hiller:2013cza,Descotes-Genon:2013vna}.
In contrast to the theory approaches below the $J/\psi$, the LROPE results can only be used for sufficiently inclusive
$q^2$-integrated quantities. The small size of the available phase space in $B\to K^{(*)}$ decays
above the open charm threshold bring into question the LROPE predictions (see \eg~\cite{Lyon:2014hpa}).
Possible relief is an item of active research, and includes:
an update to the Kr\"uger/Sehgal model for the resonances including matching onto the LROPE
without~\cite{Lyon:2014hpa} and with~\cite{Brass:2016efg} the use of endpoint relations among
the amplitudes~\cite{Hiller:2013cza}; and
explicit models for the violation of quark-hadron duality in rare $D$ decays~\cite{Feldmann:2017izn}.
Rare $D$ decays might be an interesting laboratory to test our understanding of the long-distance effects,
due to their larger hierarchy between long-distance and short-distance effects when compared to the rare \B decays.\\

Very recently, a new approach to the non-local matrix elements below the $\psi(2S)$ has been proposed
in a proof-of-concept study~\cite{Bobeth:2017vxj}. There the authors parametrize the non-local matrix
elements due to $\Oi{1c,2c}$ in a manner that respects the analytic properties shown in~\reffig{4quark-analytic-structure-wo-light-hadron-cut}.
As consequence, theoretical constraints at $q^2 < 0$ and therefore far below any hadronic threshold can be combined with experimental constraints on the non-leptonic decays $B\to K^* \lbrace J/\psi, \psi(2S)\rbrace$.
It remains to be seen how well this approach can be adapted to other exclusive $b\to s\ell^+\ell^-$ decay
modes.

\subsection{Theory predictions in the Standard Model}
\label{sec:th-had:sm-pred}

For the decay $B_s\to \mumu$, the presently most precise theoretical predictions
include NNLO QCD corrections~\cite{Hermann:2013kca} and NLO electroweak corrections~\cite{Bobeth:2013tba}
at the matching scale, and RGE effects of both QCD and QED in the
evolution to the low scale~\cite{Bobeth:2003at,Huber:2005ig}. Non-local power-enhanced QED effects have been
taken into consideration for the first time in~\cite{Beneke:2017vpq}, which updates
the theory predictions for the observables as follows:
\begin{align}
    \mathcal{B}_{\rm SM}(\Bsmumu)
        & = (3.57 \pm 0.17)\cdot 10^{-9}\,,\\
    A^{\mup\mun}_{\Delta\Gamma}
        & \approx 1 - 1.0\cdot 10^{-5}\,,\\
    C_\lambda
        & \approx \eta_\lambda \cdot 0.6\%\,,\\
    S_\lambda
        & \approx -0.1\%\,.
\end{align}
The time-integrated branching ratio $\mathcal{B}$ and the mixing-induced CP asymmetry $A^{\mup\mun}_{\Delta\Gamma}$
have been introduced in~\cite{DeBruyn:2012wj} and discussed in~\refsec{btoll}. The observables
$C_\lambda$ and $S_\lambda$ are introduced by the time-dependent rate asymmetry~\cite{DeBruyn:2012wk}
Here $\lambda = L,R$ refers to the helicity configuration of the muon pair, and $\eta_{L/R} = \pm 1$.
(Note that the above does not yet include the updated results for the $B_s$ meson decay constant
by the FNAL/MILC collaborations with a substantially reduced uncertainty~\cite{Bazavov:2017lyh}.)\\

Theory predictions for the semileptonic decays cannot be displayed as succinctly and easily
as the ones for the purely leptonic decays. This is due to a large number of choices that need
to be made such as the choice of the form factor inputs and parametrisations as well as the
treatment of the non-local effects.
For $B\to K^{(*)}\ell\ell$ and $B_s \to \phi\ell^+\ell^-$, with either $\ell=e$ or $\ell=\mu$ final states,
there are various predictions that differ in these choices~\cite{Altmannshofer:2014rta,Descotes-Genon:2015uva,Ciuchini:2015qxb}.
However, all of these predictions are mutually compatible at varying degree.\\

For $\Lambda_b\to \Lambda \ell^+\ell^-$, again with $\ell=e,\mu$, there are presently no predictions
for $q^2 < M_{J/\psi}^2$ that systematically take non-local effects into account. For predictions at
$q^2 > M_{\psi(2S)}^2$ we refer to~\cite{Detmold:2016pkz}.\\

For the \SM predictions of lepton-flavour universality ratios, the treatment of
QED effects is very important. For $R_{K^{(*)}}$, this was recently studied in~\cite{Bordone:2016gaq}.
For less inclusive observables, such as the angular observables in $B\to K^*e^+e^-$, the
impact of log-enhanced QED has been discussed in principle in~\cite{Gratrex:2015hna};
in particular their effects on moments of the angular distribution, and higher moments
that do not emerge in the absence of QED effects.\\

We conclude by referring to existing open source software for the prediction of various observables
in exclusive rare (semi-)leptonic $B$ decays. Using both the \texttt{EOS}~\cite{EOS} and \texttt{flavio}~\cite{flavio} software, 
interested parties can gain access to theory predictions within and beyond
the \SM with the benefit of making their own choices of treatment for the relevant hadronic effects.

\section{Phenomenology and New Physics reach}
\label{sec:ph-np}



 


\subsection{Global fits}
\label{sec:global}

The large number of measured observables for exclusive \btosll transitions has
inspired a number of global fits~\cite{Descotes-Genon:2015uva,Altmannshofer:2017fio} of the \WC.
These fits rely on constraints of \qsq-integrated observables from the experimental analysis.
In addition to these global fits, direct fits of the \WC to
the observed events have been proposed~\cite{Hurth:2017sqw,Blake:2017fyh,Chrzaszcz:2018yza,Mauri:2018vbg}.\\
Common to both types of fits are two questions:
\begin{enumerate}
    \item What type of data should be included?
    \item Which \WC should be fitted for, and which can be safely
      set to their \SM values?
\end{enumerate}
The first question has one definite answer: with larger datasets and more complementary
data comes greater confidence in the fit results. In addition, the large number of phase space
bins allows to perform consistency checks amongst the data. It has been shown that the constraints
from individual experiments are in mutual agreement~\cite{Altmannshofer:2017fio}.
Moreover, removing individual $q^2$ bins from the analysis allows to extensively check the
results for stability against mismodelling of the non-local charm effects.
These checks indicate that the results for $C_9$ obtained from subsets of the $q^2$ bin
are mutually compatible, and no explicit $q^2$ dependence of the result is found.
(see \eg~\cite{Descotes-Genon:2015uva,Altmannshofer:2017fio}).
Beside fitting for only the short-distance information encoded in the \WC, it has
also been demonstrated that information on the form factors and non-local matrix elements can be
inferred alongside the \WC~\cite{Beaujean:2012uj,Beaujean:2013soa,Bobeth:2017vxj}.
By fitting an explicit parametrisation of the non-local matrix elements to data, it was
shown that the non-local effects can explain the anomalies, but then fail to explain the
$q^2$ behaviour of the non-local effects as imposed by the dispersion relations
and require corrections to the leading-power theory results beyond what is naively expected
based on power-counting arguments.~\cite{Ciuchini:2015qxb,Hurth:2016fbr}.\\
While fits to the mesonic modes $B\to K^{(*)}\mumu$ and $B_s\to \phi\mumu$ are by now part
of the standard global fits, the inclusion of the baryonic mode \Lblsmumu
is much more complicated and will require future work to be viable. A first study to that
effect concluded that exploitation of the data at low hadronic recoil is possible~\cite{Meinel:2016grj}.
However, the \emph{positive} shift to $C_9$ which emerges in this study is at odds with the consistent
picture of a \emph{negative} shift in all other processes. Moreover, even this positive shift required
large duality-violating effects due to non-local charm effects that are incompatible with our present
understanding of the LROPE~\cite{Meinel:2016grj}.\\

The second question has no definite answer: no amount of data, neither from present nor
from upcoming experiments, will allow to constrain the entire set of \WC simultaneously.
The common paradigm is to restrict the fits to the \WC of operators of mass dimension six. This
is a reasonable restriction, since dimension-eight operators are generically suppressed by factors
$M_B^2 / M_W^2 \sim 0.5\%$.\\

\begin{figure}
    \begin{tabular}{cc}
        \subfigure[Plot taken from~\cite{Descotes-Genon:2015uva}]{%
            \includegraphics[width=.49\textwidth]{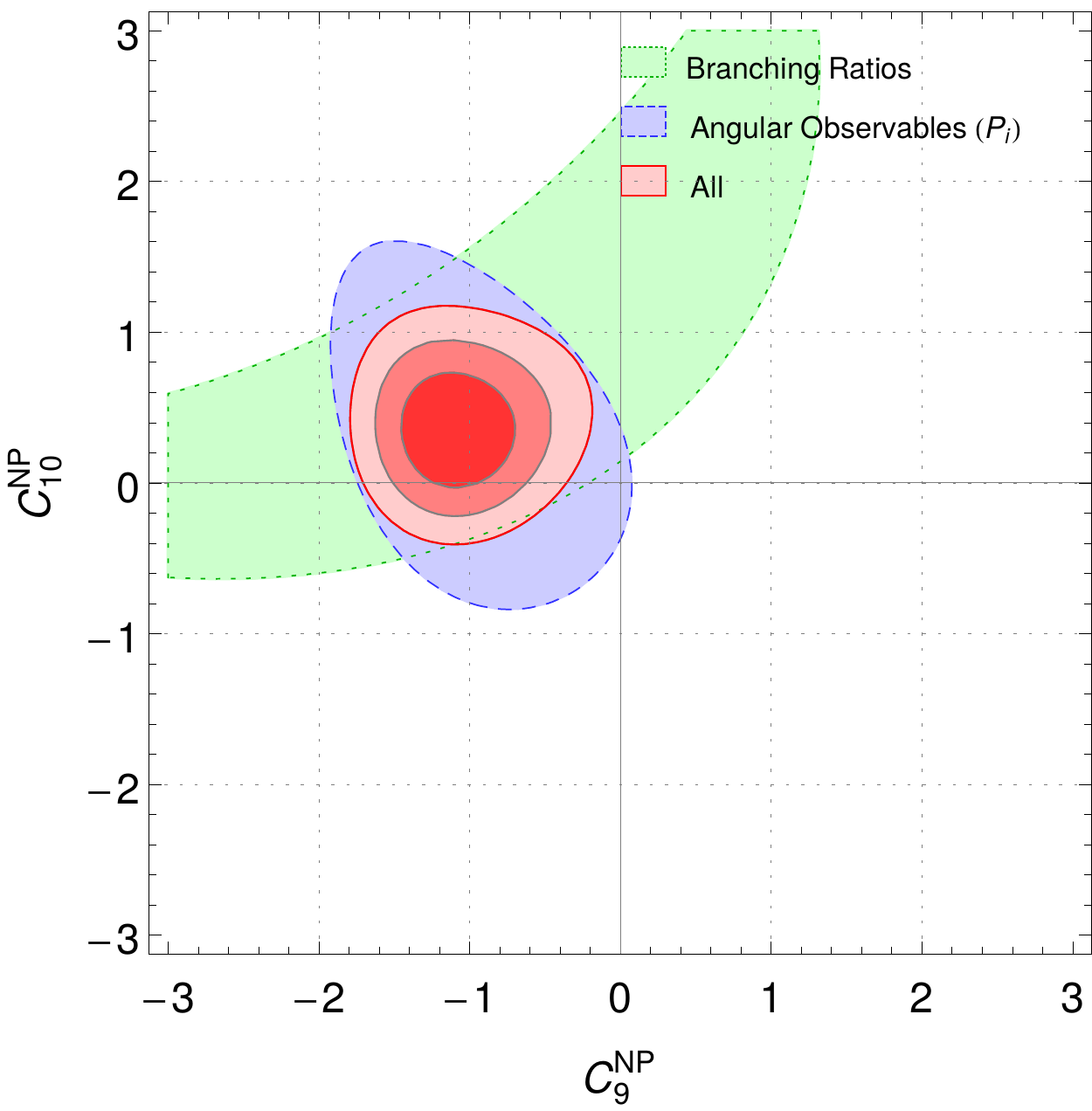}
        }
        &
        \subfigure[Plot taken from~\cite{Altmannshofer:2017fio}]{%
            \includegraphics[width=.49\textwidth]{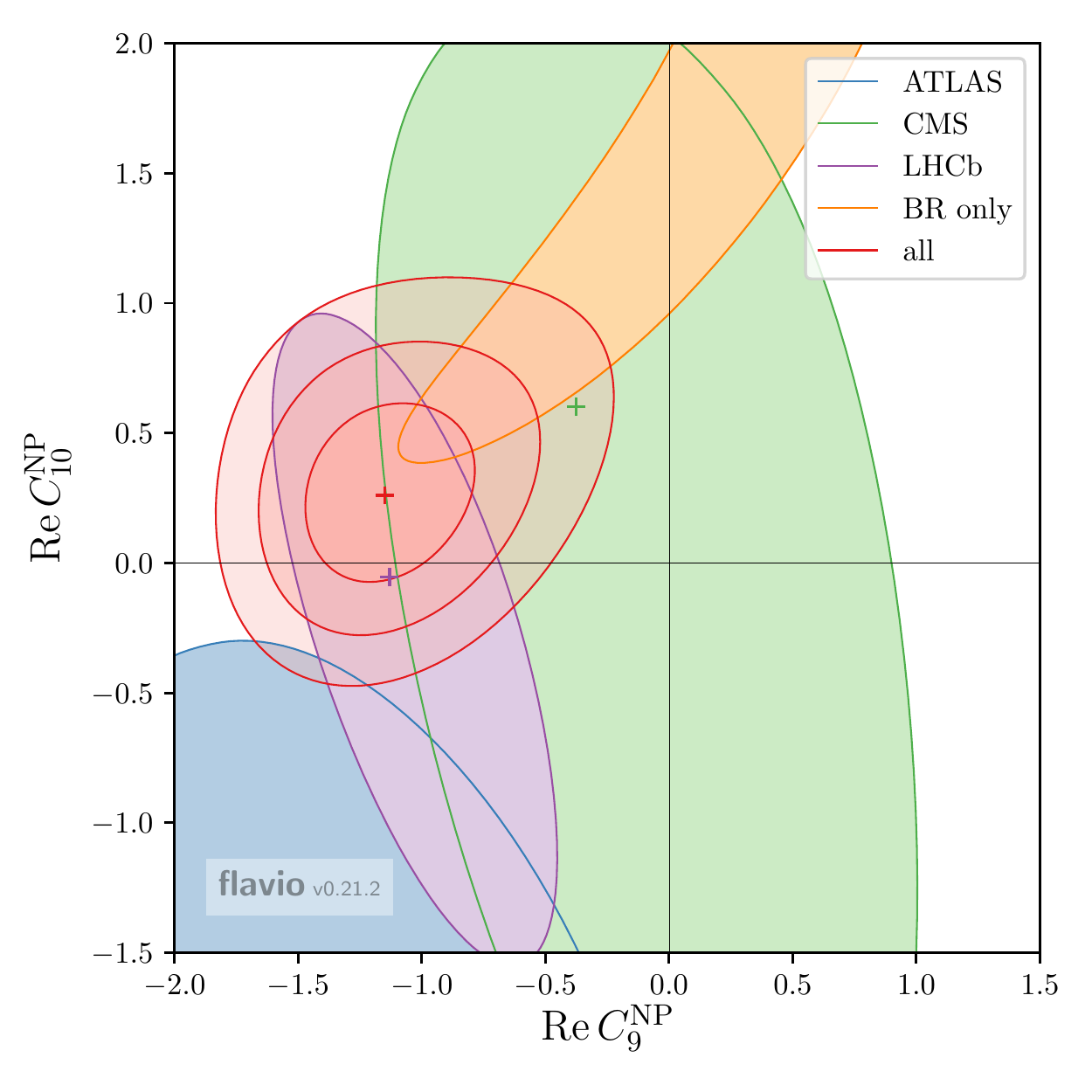}
        }
        \\
        \subfigure[From~\cite{Ciuchini:2017mik}]{%
            \includegraphics[width=.49\textwidth]{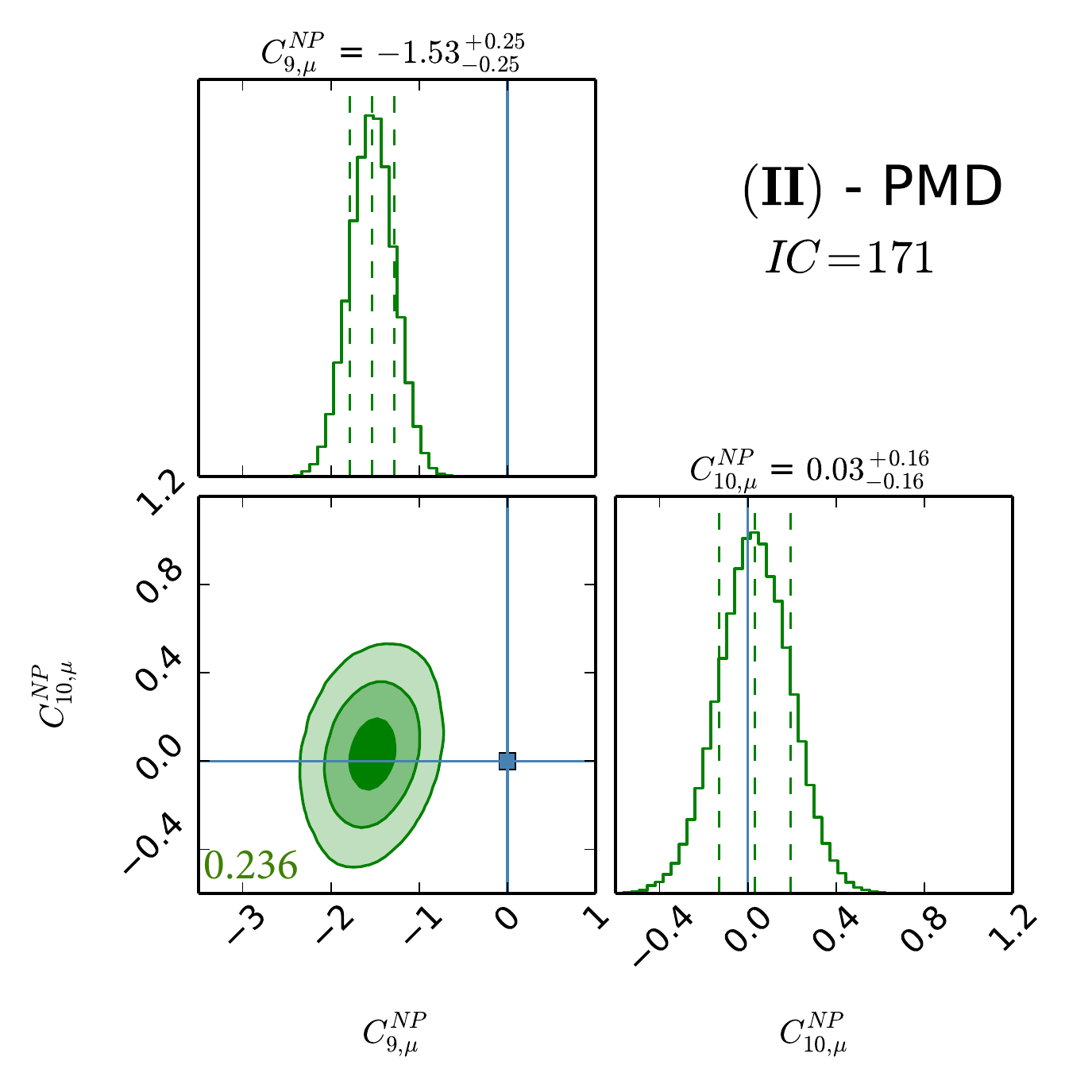}
        }
        &
        \subfigure[From~\cite{Hurth:2017hxg}]{%
            \includegraphics[width=.49\textwidth]{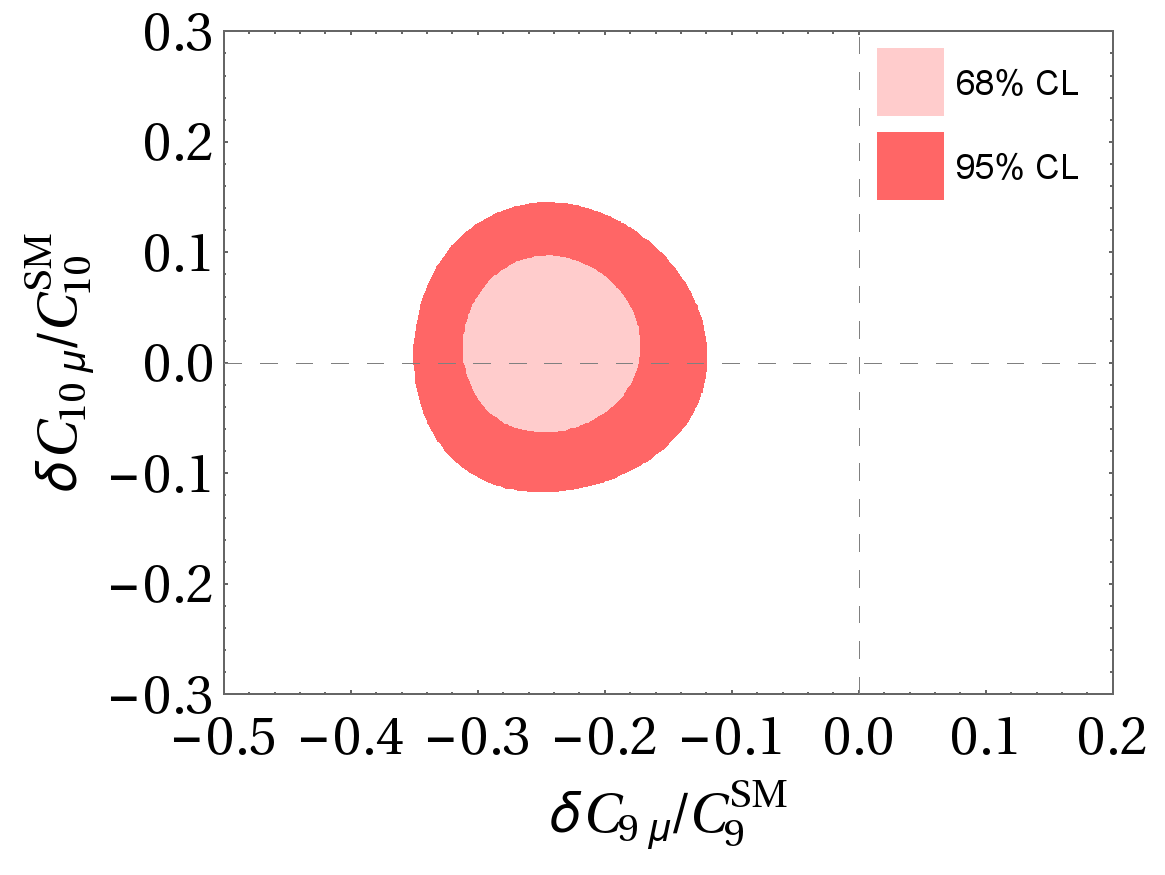}
        }
    \end{tabular}
    \caption{
        Plots of the allowed region in the $C_9$--$C_{10}$ plane based of four
        competing analyses. Even though the four analysis use different approaches
        for the hadronic matrix elements, they all find a substantial tension with
        respect to the \SM value for the parameter $C_9$.
    }
    \label{fig:globals}
\end{figure}

The results for fits to \btosmumu data from the various groups as illustrated in~\reffig{globals} can be summarised as follows~\cite{Descotes-Genon:2015uva,Altmannshofer:2017fio,Ciuchini:2017mik,Hurth:2017hxg}:
\begin{itemize}
    \item the exclusive data can be fitted reasonably well with the
      \SM ansatz, assuming substantial shifts from theory predictions of the
    hadronic matrix elements;
    \item the combination of all \btosmumu data favour a shift
      to the WC $C_9$, which amounts to $25\%$ of its \SM value at $\mu \sim 4.2\GeV$;
    \item there is no clear sign for a $q^2$ dependence of this shift;
    \item from a theory perspective, there is presently no irrefutable evidence that
      this shift is due to \NP;
\end{itemize}
However, in combination with the \btosee data this conclusion changes. Since
the measurements of the LFU observables $R_{K^{(*)}}$ cannot be explained by
hadronic effects within the \SM, all groups find that a \NP interpretation of the data
is strongly favoured. The shift to $C_9$ needed to explain
the \btosmumu data can simultaneously also explain the signs of LFU violation, if \NP
are suppressed or absent in \btosee processes. It is therefore now common to distinguish
between the Wilson coefficients in both processes according to the lepton flavour, and
write $C_9^{(\ell)}$.\\

Sensitivity studies aimed at the dataset sizes expected from the \belle II experiment and the \lhcb Upgrade
indicate that the anomalies can exceed a level of $5\sigma$ if the present central values
remain unchanged~\cite{Albrecht:2017odf}; see~\reffig{outlook} for an illustration.
However, it remains to be shown that the shift to $C_9^{(\mu)}$ is indeed a genuine \NP effect,
and not due to non-local hadronic effects. A promising approach has been recently studied in~
\cite{Mauri:2018vbg} wherein unbinned fits to the angular distributions of \BtoKstmumu
and \BtoKstee are carried out simultaneously, yielding accurate and model-independent results for
the difference $C_9^{(\mu)} - C_9^{(e)}$.
Nevertheless, we emphasise that a better understanding of the non-local hadronic matrix elements
is paramount for our understanding of the individual Wilson coefficients $C_9^{(\mu)}$ and $C_9^{(e)}$
and the potential \NP models that might give rise to LFU violation.

\section{Experimental outlook}
\label{sec:exp-outlook}

The current picture of anomalies in the flavour sector, which have been part of the discussion in this review, promise an interesting future for flavour physics. With the upcoming upgrade of the \lhcb detector in 2019-2020 and the start of the \belle~II experiment later this year, a confirmation or exclusion of those present-day anomalies is to be expected within the next years. The current available datasets of the \lhcb and \belle collaboration correspond to integrated luminosities of $6.7\invfb$ at the beginning of 2018 and $0.7\invab$, respectively. The \belle~II experiment is expected to collect $5\invab$ ($50\invab$) by 2020 (2024). With the \lhcb detector, an increase to 8\invfb for this year before the second long shutdown is foreseen. After the scheduled major upgrade of the \lhcb detector~\cite{Bediaga:2012uyd} datasets corresponding to 22\invfb in 2024 and to 50\invfb by 2029 will be accumulated~\cite{MTP20162020V1}.\\

\begin{figure}[h]
\begin{minipage}[t]{0.49\linewidth}
\centering
\includegraphics[width=1\linewidth]{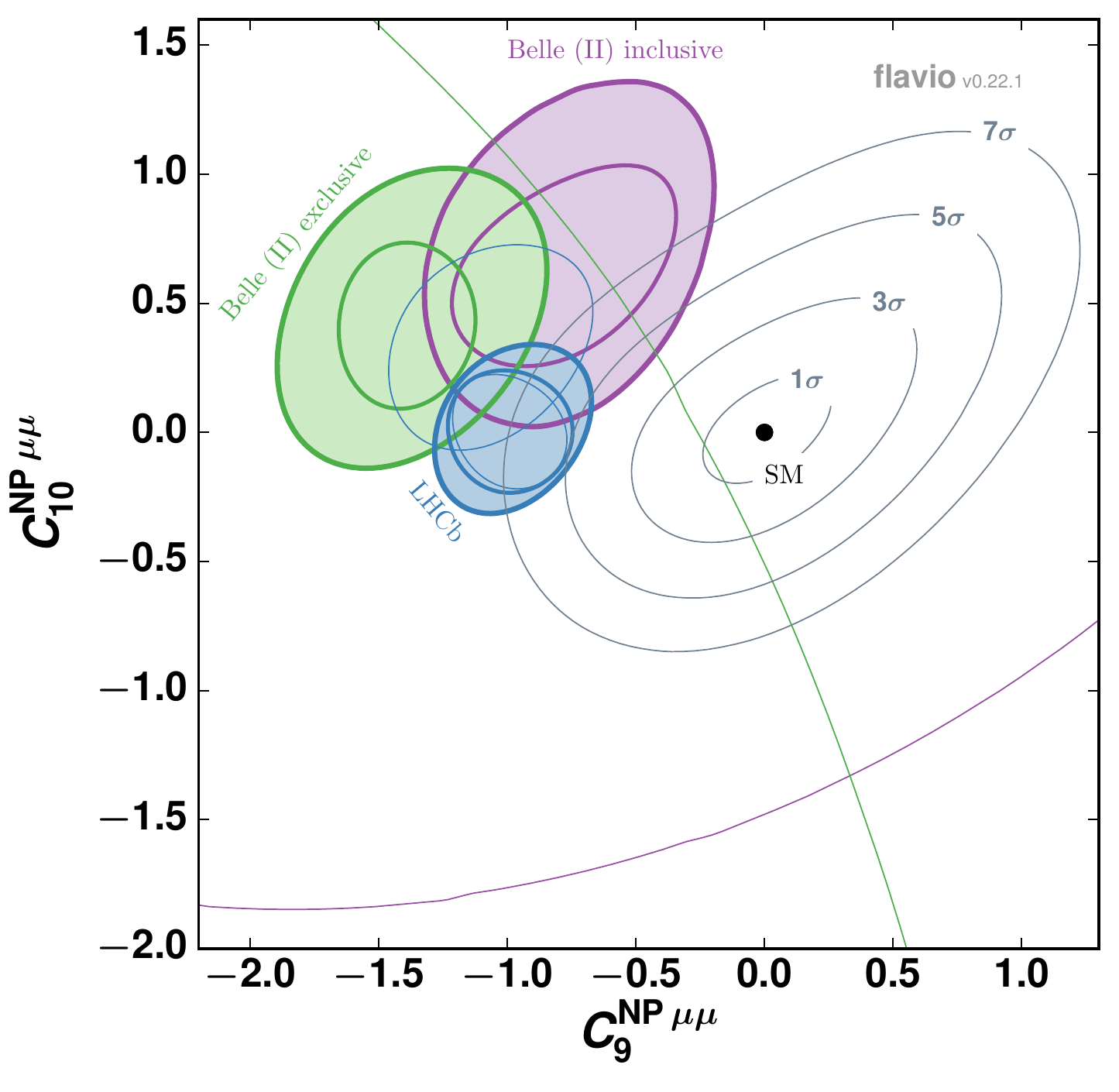}
\end{minipage}
\hspace{\fill}
\begin{minipage}[t]{0.49\linewidth}
\centering
\includegraphics[width=1\linewidth]{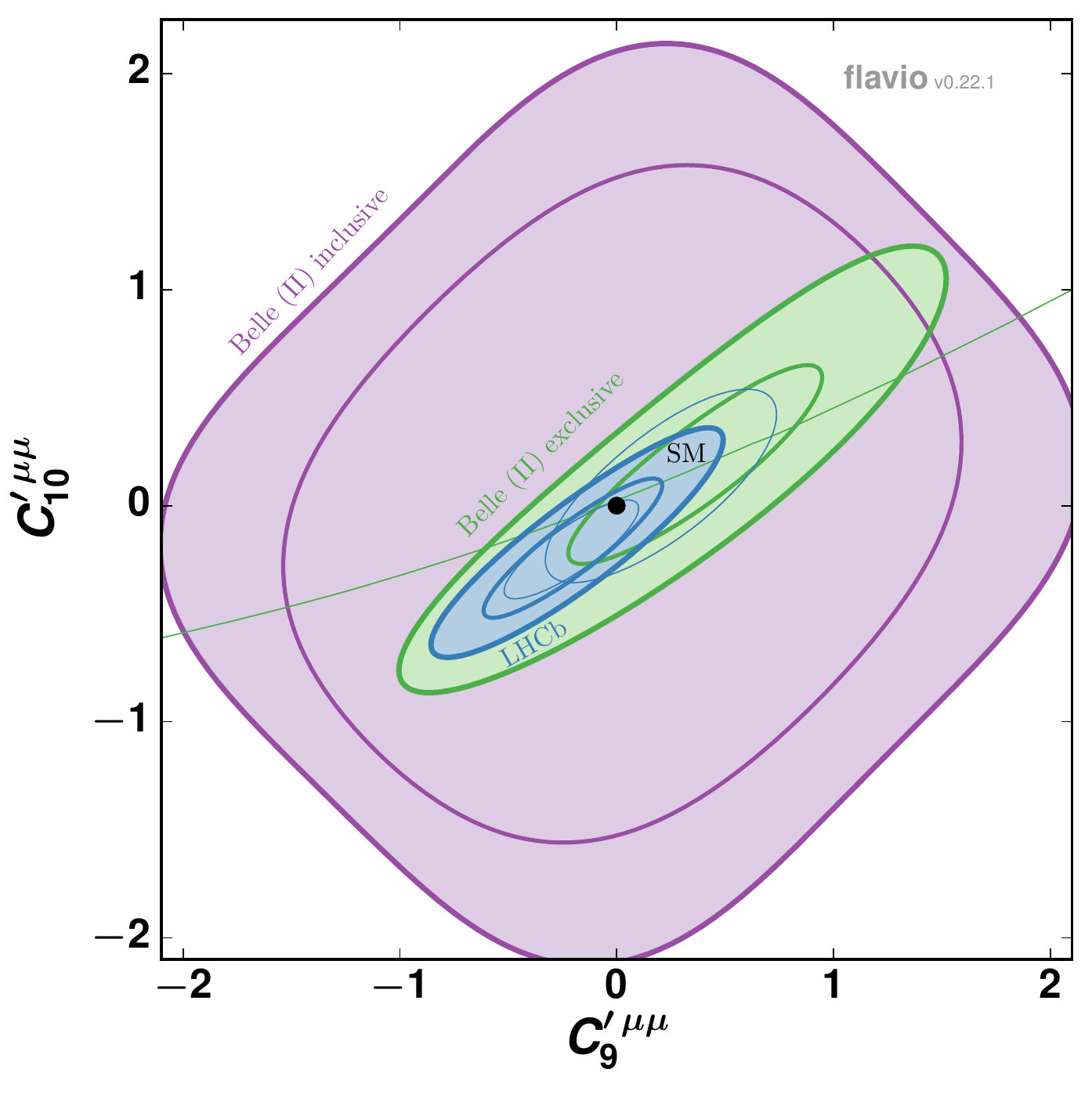}
\end{minipage}
\begin{minipage}[t]{1\linewidth}
\centering
\includegraphics[width=0.49\linewidth]{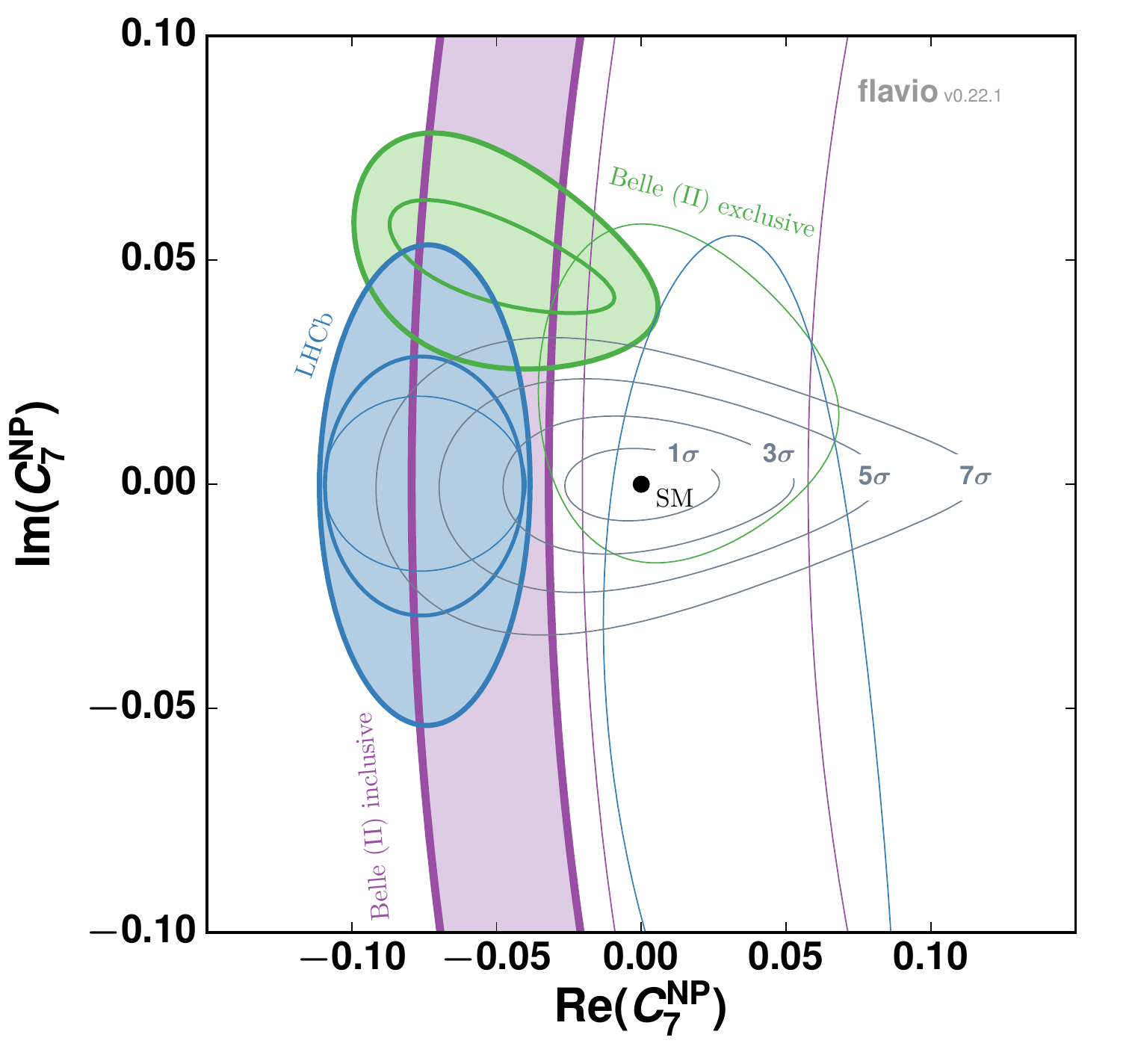}
\end{minipage}
  \caption{Two-dimensional scans of the new physics contribution to \Cnine versus \Cten (top left), \Cninep versus \Ctenp (top right) and \ReCsev versus \ImCsev (bottom). Displayed are the current average (not filled) as well as the extrapolations to future sensitivities (filled) of \lhcb at 8\invfb, 22\invfb and 50\invfb (exclusive decays) and of \belle II at 5\invab and 50\invab (both inclusive and exclusive decays) assuming different \NP models for the three classes of measurements to aid visibility. The \SM prediction (black dot) with the $1\sigma, 3\sigma$, $5\sigma$ and $7\sigma$ exclusion contours with a combined sensitivity of \lhcb's $50\invfb$ and \belle II's $50\invab$ datasets is indicated in light grey. As the primed operators show no tensions with respect to the \SM, no \SM exclusions are provided. Figures from~\cite{Albrecht:2017odf}.}
 \label{fig:outlook}
\end{figure}

In~\cite{Albrecht:2017odf}, the impact of the future sensitivities at those well-defined milestones in 2020, 2024 and 2029 are estimated and their impact on global fits to the Wilson Coefficients is discussed. An illustration of the impact of future sensitivities of the \belle II and \lhcb experiments and their complementarity is shown in~\reffig{outlook}. As can be seen from~\reffig{outlook}, the primed semileptonic operators as well as the electromagnetic dipole operators are currently consistent with the \SM hypothesis. However, the current measurements exhibit a tension from the \SM in $C_9$ (see~\refsec{global}), which seems to be lower than the \SM prediction. No discrepancy to the \SM prediction is observed for $C_{10}$. \\

It is expected that the sensitivities of the upgraded \lhcb experiment and the \belle II detector will allow to either exclude or confirm the currently observed anomalies within the next years. If the current anomalies in \Rk and \Rkst stay at their respective central values, then \lhcb should be able to measure \Rk and \Rkst with a significance exceeding $5\sigma$ already with a dataset corresponding to 8\invfb. The \belle II collaboration will reach the required sensitivities with a dataset of $5\invab$ foreseen to be collected in 2020. Due to the available statistics and improvements to the detector, the origin of the anomalies in the flavour sector is expected to be discovered within the next years and will allow us to gain insight about potential physics beyond the \SM.

\section*{Acknowledgments}

%
J. A. gratefully acknowledges support of the Deutsche
Forschungsgemeinschaft (DFG, Emmy Noether programme: AL 1639/1-1) and
of the European Research Council (ERC Starting Grant: PRECISION
714536).\\
D. v. D. gratefully acknowledges support of the Deutsche
Forschungsgemeinschaft (DFG) within the Emmy Noether programme under
contract DY 130/1-1 and through the DFG Collaborative Research Center
110 ``Symmetries and the Emergence of Structure in QCD''.\\
The authors thank Yasmine Amhis and Javier Virto for proof-reading the manuscript.\\

\appendix

\bibliographystyle{ws-ijmpa}
\bibliography{references}

\end{document}